\documentclass[11pt]{article}
\usepackage[a4paper,hmarginratio=1:1,vmarginratio=2:3,totalwidth=15.6cm,
totalheight=21.20cm]{geometry}
 \usepackage{verbatim} 
\usepackage{latexsym}
\usepackage{graphics} 
\renewcommand{\baselinestretch}{1.155}
\usepackage{amsmath}
\usepackage{amssymb}
\usepackage{amsfonts} 
\usepackage{epsfig} 
\newcommand{\tq}{{\xi_o}}

\newcommand{\osigma}{{\overline\sigma}}

\newcommand{\cD}{{\cal D}}

\newcommand{\cK}{{\cal K}}
\newcommand{\cL}{{\cal L}}
\newcommand{\cM}{{\cal M}}

\newcommand{\cO}{{\cal O}}

\newcommand{\cZ}{{\cal Z}}

\newcommand{\ra}{\rightarrow}
\newcommand{\be}{\begin{equation}}
\newcommand{\ee}{\end{equation}}
\newcommand{\bea}{\begin{eqnarray}}
\newcommand{\eea}{\end{eqnarray}}

\DeclareMathSymbol{\mg}{\mathrel}{symbols}{"1D}

\long\def\symbolfootnote[#1]#2{\begingroup%
\def\thefootnote{\fnsymbol{footnote}}\footnote[#1]{#2}\endgroup} 

\newcounter{oldcounter}
\addtocounter{equation}{1}
\begin{document}
\begin{flushright}
\hfill{CERN-PH-TH/2010-322}\\
\hfill{CPHT-RR 112.1210}
\end{flushright}

\thispagestyle{empty}
\vspace{0.7cm}
\begin{center}
{\Large {\bf Beyond the MSSM Higgs with d=6 effective operators.\\
\vspace{0.3cm}
}}
\end{center}
\vspace{0.5cm}

\begin{center}
{\bf I. Antoniadis$^{\,a, b\,}$,
E. Dudas$^{\,b, c\,}$,
D.~M. Ghilencea$^{\,a,\,b,\,d\,}$, 
P. Tziveloglou$^{\,b,\,}$\footnote{
E-mail addresses:\,\, Ignatios.Antoniadis@cern.ch,
Emilian.Dudas@cpht.polytechnique.fr,\\
$\,\,\,\,$Dumitru.Ghilencea@cern.ch, pantelis.tziveloglou@gmail.com}}\\

 \vspace{0.4cm}
 {\small $^a $Department of Physics, CERN - Theory Division, 1211 Geneva 23,
 Switzerland.}\\[2pt]
 {\small $^b $Centre de Physique Th\'eorique, Ecole Polytechnique, CNRS, 91128
   Palaiseau, France.}\\[2pt]
 {\small $^c $LPT, UMR du CNRS 8627, B\^at 210, Universit\'e de Paris-Sud,
 91405 Orsay Cedex, France.}\\[2pt]
 {\small $^d $DFT, National Institute of Physics and Nuclear Engineering (IFIN-HH)
Bucharest MG-6, Romania.}
\end{center}
\vspace{0.6cm}

\def\baselinestretch{1.1}
\begin{abstract}
\noindent 
We continue a previous study of the MSSM Higgs Lagrangian extended by all effective operators 
of dimension $d=6$  that can be present beyond the MSSM, consistent with its symmetries. By 
supersymmetry, such operators also 
extend the neutralino and chargino sectors, and the corresponding component fields Lagrangian is
 computed onshell.
The corrections to the neutralino and chargino masses, 
due to these operators, are computed analytically in
 function of the MSSM corresponding values. For individual operators, the corrections
 are small, of few GeV for the constrained MSSM (CMSSM) viable parameter space.
We  investigate the correction to the lightest Higgs mass,
which  receives, from {\it individual} operators,  
a {\it supersymmetric} correction of up to $4$ ($6$) GeV above the 2-loop leading-log 
CMSSM value,  from those CMSSM phase space points with:
EW fine tuning $\Delta<200$, consistent with WMAP relic density (3$\sigma$), 
and for a scale of the operators of $M=10\,(8)$ TeV, respectively.
Applied to the CMSSM point of minimal fine tuning ($\Delta=18$), 
such increase gives an upper limit $m_h=120(122)\pm 2$ GeV, respectively.
The increase of $m_h$ from individual operators can be larger 
($\sim 10-30$ GeV) for those CMSSM phase space points with $\Delta>200$; 
these can now be phenomenologically viable, with reduced $\Delta$, and this 
includes those points that would have otherwise violated the LEP2 bound by this 
value.
The neutralino/chargino Lagrangian extended by the effective operators 
can be used in studies of dark matter relic density 
within extensions of the MSSM, by implementing it in  public codes like
micrOMEGAs. 
\end{abstract}

\newpage

\tableofcontents

\section{Introduction}

The physics of the Higgs sector plays 
a central role in the Standard Model  (SM) and its minimal supersymmetric version (MSSM). 
Its discovery would clarify the (electroweak (EW)) gauge symmetry breaking 
mechanism. In  the supersymmetric case it could also provide an insight into the dark matter problem, 
due to its link with the MSSM neutralino sector (higgsinos/gauginos), whose LSP is a dark 
matter candidate. Higgs physics can then be related to  both small and large scale physics
for which EW and dark matter  constraints can be relevant.

The Higgs sector of the MSSM is the minimal that can 
be constructed in a supersymmetric context. Its 
EW vev triggered by radiative EW symmetry breaking is strongly
 related by quantum corrections to 
the scale of supersymmetry breaking and the mass of superpartners.  No discovery of light 
superpartners will indicate a fine tuning \cite{Ellis:1986yg,r3} of 
the EW scale in the MSSM to levels
 phenomenologically unacceptable and will question supersymmetry as a solution to the
hierarchy problem. In constrained MSSM (CMSSM) at 2-loop leading-log (LL),
a Higgs of mass of 120 GeV would mean an EW fine tuning $\Delta=100$ (i.e. 1 part in 100)
\cite{Cassel:2009cx}. Due to quantum corrections (largely QCD ones), 
$\Delta$ grows exponentially, so 
for $m_h=126$ GeV, the fine tuning worsens and becomes $\Delta=1000$. 
Interestingly enough, a minimization of $\Delta$ at 2-loop, with all 
theoretical and experimental constraints, except the LEP2 bound \cite{LEP2} 
on $m_h$ and the WMAP result \cite{wmap}, predicts a value for $m_h$ just 
above the LEP2 bound, 
$m_h=114\!\pm\! 2$ GeV  with an acceptable fine  tuning, 
$\Delta=9$ \cite{Cassel:2009cx}. This 
is only mildly changed when one imposes a saturation of dark  matter 
relic density within 3$\sigma$, 
to $m_h=115.9\pm 2$ GeV, for a fine tuning $\Delta=18$, still an 
acceptable value.  
The question remains though if such results for $m_h$ are stable 
under corrections from new physics 
that may be missed by the otherwise minimal construction of the MSSM higgs sector. 
It is also interesting to investigate what happens if $m_h$ is not found experimentally 
near the  value predicted by minimal $\Delta$ shown above. 
Can one still have a low fine tuning for $m_h$ above these values? 
In other words, a large amount of fine tuning that we mentioned for $m_h>120$ GeV  may 
be taken to indicate that the Higgs sector Lagrangian is not complete, and 
that new physics beyond this  sector can exist, so that its effects could reduce 
$\Delta$ to acceptable values even for $m_h\!>\! 120$ GeV 
(for an example see \cite{Cassel:2009ps}).
If possible, such new physics  can be described, in a model 
independent way, by higher dimensional
operators. These operators respect all the symmetries of the MSSM. For practical 
purposes one can consider operators of dimensions $d=5$ and $d=6$,
and this paper is a continuation of the work in this direction, started in
\cite{Antoniadis:2009rn}.
For studies of effective operators of $d=6$ in the MSSM Higgs sector see  
\cite{Carena:2009gx,Antoniadis:2009rn} and \cite{Antoniadis:2008es}-\cite{Brignole:2010nh}
for studies of effective operators in a related context.

There is only one gauge invariant effective supersymmetric operator of $d=5$  
beyond the MSSM Higgs sector that can depend on Higgs and gauge fields only, 
but the number of similar $d=6$ operators, 
is much larger and their analysis is difficult.
Ignoring the $d=5$ operator, the $d=6$ operators could indicate that the MSSM Higgs, 
and, by  supersymmetry, neutralino  sector, are stable under "new physics"
corrections, since these are strongly suppressed, by  $\sim 1/M^2$ 
($M$ is the scale of new physics).  One would like to clarify if this is true.
However, the extra scale  suppression of $d=6$ operators (relative to $d=5$ ones) 
can be compensated by a large $\tan\beta$, and then   
the $d=5$ and $d=6$ operators can have comparable effects. 
There are stronger motivations to consider  $d=6$ operators.  
New physics beyond MSSM Higgs and neutralino sectors can arise, in the leading order,
as a $d=6$ operator, without any $d=5$ one. For example, integrating a 
massive U(1)$'$ gauge boson generates a $d\!=\!6$ operator in the leading order, but no $d=5$
one. The convergence of the expansion in $1/M$ is another motivation for studying both 
$d\!=\!5$ and $d\!=\!6$ operators, if they are generated by the same physics.

For the effective operators expansion to work the scale of new physics should 
be high enough, to avoid current experimental constraints. Usually EW constraints 
($\rho$ parameter) indicate a value $M\!\sim\! 8$ TeV or larger \cite{Blum:2009na}. 
The expansion parameter  $\tilde m/M$, for $d\!=\!5$ case, and $(\tilde m/M)^2$ for $d\!=\!6$ case, where $\tilde m$ is any low scale of the model (EW vev, $\mu$ parameter, 
$m_0$: Susy breaking scale, $m_{1,2}$ gaugino masses)  should be less than unity. 
If this is not true, the effective approach is unreliable, and
unintegrated states (that generated the effective operators) should be used instead.  

The coefficients of the effective operators can
also be constrained, in a global fit of MSSM plus effective operators,
by dark matter experiments, due to their implications for the 
neutralino sector whose LSP is a dark matter candidate. This  can be translated in 
constraints on the corrections to the Higgs mass. Therefore, the overlap of complementary 
EW and dark matter constraints on new physics would be welcome for model building.
As a first step in this direction,  in this work we compute in component fields, 
for the first time, the most  general extension of the MSSM Lagrangian 
in the neutralino and chargino sectors extended by all allowed $d=5$ and $d=6$ effective 
operators. This is related, by supersymmetry, to  the corresponding MSSM Higgs 
Lagrangian with such effective operators, computed in \cite{Antoniadis:2009rn}.

We then calculate (analytically) 
the corrections to the neutralino and chargino masses, in the leading order, $1/M^2$,
 in function of the MSSM corresponding values. It turns out that the  supersymmetric 
mass corrections to the LSP from {\it individual}
 operators of $d=6$ are in general small, of few GeV and less than $1-2\%$ for a 
scale $M=5$ to $8$ TeV, and as a result, the same is 
true about the change of the LSP composition
relative to the MSSM case. This can change
in cases when all operators of a given order are present simultaneously, and then  their 
combined effect is enhanced. To avoid ambiguities we keep the coefficients 
of all effective operators  as independent, so that one can turn on/off  some of them, 
depending on the details of the model considered. 
The neutralino Lagrangian extended by effective operators  is useful in 
studying the dark matter relic density in  MSSM extensions, 
by implementing it in micrOMEGAs~\cite{Belanger:2006is}.

We also perform a careful investigation of the size of corrections 
to the mass $m_h$ of the lightest MSSM Higgs field, due to effective operators.
In \cite{Antoniadis:2009rn} analytical formulae for these corrections were 
obtained, in $1/M^2$ order, followed by a simple numerical estimate 
in a very special case and under simplifying assumptions.
Here we improve this numerical study by performing a general and accurate 
numerical  analysis of  the corrections to $m_h$, analyzed  separately 
for {\it individual} operators and  including {\it quantum corrections}, not considered before.
 We do so by considering the CMSSM phase space points that 
respect current theoretical and experimental constraints, both electroweak and dark 
matter ones  (except the LEP2 bound on $m_h$ that is not imposed), and treat the 
effective operators corrections as a perturbation 
on this "background". We show that the CMSSM points 
with smallest fine tuning ($\Delta\!<\!200$) are rather stable under (Susy)
 corrections from the effective operators. The correction to the 2-loop 
leading-log CMSSM Higgs mass $m_h$, due to  individual operators of $d\!=\!6$, is 
found to be  in the region of up to: 4 GeV (6 GeV)  for a scale of new physics
 near 10 TeV (8 TeV), respectively. With the above remarks
 on neutralino sector, we could expect that their dark matter relic density
 is unlikely to be affected, but a more careful analysis is needed for this.
Regarding CMSSM phase space points with large EW fine tuning $\Delta\!>\!200$, they
give a larger increase ($\sim\! 10\!-\!30$ GeV) of $m_h$, 
especially for those $m_h$  otherwise  under the LEP2 bound 
(i.e. ruled out in CMSSM), so that the corrected value of $m_h$ 
can be brought above this bound.
For some but not all operators, this value is still close to 120 GeV. 
Therefore points ruled out in the CMSSM by the LEP2 bound or by large EW  
$\Delta$, can become viable phenomenologically and their EW fine tuning
will be reduced, once $m_h$ received a significant classical correction.

The plan of the paper is as follows. Section~\ref{section2} and
\ref{section3} compute the onshell total Lagrangian, of the MSSM Higgs, neutralino 
and chargino sectors, extended by effective operators.
Section~\ref{section4} presents the mass corrections to these fields,  
with phenomenological results  given in Section~\ref{section5}.

\section{The Lagrangian of the model.}\label{section2}

To begin with, consider the MSSM Higgs sector Lagrangian plus all independent
 operators of dimensions $d=5$ and $d=6$ that are allowed in this sector by the MSSM symmetries.
In this section we compute this extended Lagrangian in component fields, in $1/M^2$ order.
Such effective operators parametrize in a model independent way whatever  
new physics may exist in this sector, above  $M\sim$ few TeV. 
All operators are considered here with independent coefficients.  
The Lagrangian is then
\bea
\cL=\int\! d^2\theta
\!\sum_{i=1,2} z_i(S,S^\dagger)\,H_i^\dagger e^{V_i} H_i
+\Big\{
\!\int\!d^2\theta\, \mu (1+B_0 m_0 \theta\theta)\,\, H_1.H_2
+h.c.\Big\}+\cK_0+\sum_{j=1}^8  \cO_j
\eea
where   $z_i(S,S^\dagger)=1-c_i S S^\dagger$, $S=m_0\theta\theta$, $i=1,2$ account for  
Susy breaking in the MSSM Higgs sector, $m_0$ is the Susy breaking scale, 
given by $m_0=\langle F_{hidden}\rangle/M_{Planck}$. $\cK_0$ is the only 
dimension-five operator present  up to non-linear field 
redefinitions \cite{Antoniadis:2008es}, while $\cO_i$ are $d\!=\!6$ operators.
Further:
\begin{eqnarray} 
\mathcal{\cK}_{0}& = & \frac{1}{M}\int d^{2}\theta \,
\,\zeta(S)\,(H_{2}.H_{1})^{2}\!+\!h.c.\\
&=& \zeta_{10}\,\big[2\, (h_2.h_1)(h_2.F_1+F_2.h_1-\psi_2.\psi_1)
\!-\!(h_2.\psi_1+\psi_2.h_1)^2\big]
\!+\!\zeta_{11}\,m_0\,(h_2.h_1)^2\!+\!h.c.\nonumber
\label{dim5}
\end{eqnarray}
where $H_i\equiv (h_i,\psi_i,F_i)$, $h_1.h_2=h_1^0 h_2^0-h_1^- h_2^+$, 
 $(1/M)\,\zeta(S)\equiv \zeta_{10}+\zeta_{11}\,m_0\theta\theta$, so $\zeta_{10},
 \zeta_{11}\sim 1/M$. For 
 the conventions used see Appendix~\ref{appendixA}.
The list of $d=6$ operators is \cite{Antoniadis:2009rn,Carena:2009gx}
 (also \cite{Piriz:1997id})
\bea
\mathcal{O}_{j} &=&
\frac{1}{M^{2}}\int d^{4}\theta \,\,
\mathcal{Z}_{j}(S,S^{\dagger })\,\,
(H_{j}^{\dagger }\,e^{V_{j}}\,H_{j})^{2}, \quad j\equiv 1,2. 
\nonumber\\[-3pt]
\mathcal{O}_{3} &=&
\frac{1}{M^{2}}\int d^{4}\theta \,\,\mathcal{Z} _{3}(S,S^{\dagger
})\,\,
(H_{1}^{\dagger 
}\,e^{V_{1}}\,H_{1})\,(H_{2}^{\dagger }\,e^{V_{2}}\,H_{2}),
\nonumber\\[-3pt]
\mathcal{O}_{4} &=&\frac{1}{M^{2}}\int d^{4}\theta \,\,
\mathcal{Z}_{4}(S,S^{\dagger })\,\,(H_{2}.\,H_{1})\,(H_{2}.\,H_{1})^{\dagger },
\nonumber\\[-3pt]
\mathcal{O}_{5} &=&\frac{1}{M^{2}}\int d^{4}\theta \,\,\mathcal{Z}%
_{5}(S,S^{\dagger })\,\,(H_{1}^{\dagger 
}\,e^{V_{1}}\,H_{1})\,\,H_{2}.\,H_{1}+h.c.
\nonumber\\[-3pt]
\mathcal{O}_{6} &=&
\frac{1}{M^{2}}\int d^{4}\theta \,\,\mathcal{Z}_{6}(S,S^{\dagger
})\,\,
(H_{2}^{\dagger }\,e^{V_{2}}\,H_{2})\,\,H_{2}.\,H_{1}+h.c. 
\nonumber\\[-3pt]
\mathcal{O}_{7} &=&
\frac{1}{M^{2}}\sum_{s=w,y}\frac{1}{16 g^2_s \kappa}
\int d^{2}\theta \,\,
\mathcal{Z}_{7}(S,0)\,\,{\rm Tr}\,( W^{\alpha }\,W_{\alpha } )_s\,
(H_{2}.H_{1})+h.c. \nonumber\\
\nonumber\\[-5pt]
\mathcal{O}_{8} &=&\frac{1}{M^{2}}\int d^{4}\theta
 \,\,\Big[\mathcal{Z}_{8}(S,S^{\dagger
   })\,\,(H_{2}\,H_{1})^{2}+h.c.\Big]
\label{operators18}\eea
where $W^\alpha=(-1/4)\,\overline D^2 e^{-V} D^\alpha\, e^V$
is the chiral field strength
of $SU(2)_L$ or $U(1)_Y$ vector superfields $V_w$ and $V_y$ respectively.
 Also  $V_{1,2}=V_w^a
(\sigma^a/2)+(\mp 1/2)\,V_y$ with the upper (minus) sign for $V_1$.
The expressions of these operators in  component form
are given in Appendix~\ref{appendixA}. The coefficients $\cZ$ are given by
\medskip
\bea
(1/M^2)\,\,\cZ_i(S,S^\dagger)=\alpha_{i0}
+\alpha_{i1}\,m_0\,\theta\theta
+\alpha_{i1}^*\,m_0\,\overline\theta\overline\theta
+\alpha_{i2}\,m_0^2\,\theta\theta\overline\theta\overline\theta,\quad {\textrm{where}}\quad
\alpha_{ij}\sim 1/M^2
\eea

\medskip\noindent
with $\alpha_{ij}$ numerical coefficients, assumed independent;
$\alpha_{j0}$, $\alpha_{j2}$ with $j=1,2,3,4$  are real.

The above equations show only the operators polynomial in fields. 
There are also derivative operators 
\cite{Antoniadis:2009rn} which can be eliminated in the low 
energy effective theory limit, via
 general non-linear field redefinitions or via equations of motion
\cite{Antoniadis:2009rn,Antoniadis:2008es,Antoniadis:2007xc}. For details how 
to eliminate these operators see  \cite{Antoniadis:2008es,Antoniadis:2007xc}. 
To give only two examples of such operators, that can be eliminated, consider 
the D-term 
$(H_1^\dagger e^V {\overline D^2} e^{-V}  D^2 e^V  H_1)\sim (H_1^\dagger \Box H_1)$ 
and  the F-term ${\rm Tr\,}( e^V\,W^\alpha \,e^{-V} D^2 (e^V W_\alpha e^{-V} ))\sim 
{\rm Tr\,}(W^\alpha \Box W_\alpha)$ where $W$ is the supersymmetric field strength\footnote{
These  operators are often generated as one-loop counterterms even in simplest orbifold
 compactifications, see \cite{GrootNibbelink:2005vi,Ghilencea:2005hm},
 after integrating the  Kaluza-Klein
 modes and come multiplied by the compactification volume.}. Such operators can be eliminated up to 
redefinition of the soft masses, wavefunction
 renormalization and $\mu$-term redefinition.

After eliminating the auxiliary fields in $\cL$, one finds the onshell Lagrangian, which is
\medskip
\bea\label{totalL}
\quad 
\cL=\cL_D+\cL_F+\cL_{1}+\cL_{2}
+\cL_{3}+\cL_{4}+\cL_{SSB} 
\qquad \qquad 
\eea

\medskip\noindent
Eliminating the D-dependent terms in $\cL$ one finds, with the notations in
Appendix~\ref{appendixA}, see eqs.(\ref{dd2}) to (\ref{dsq}),   and vector superfields 
notation $V_s=(\lambda_s, V_{s\,\mu},D_s^a/2)$, $s=y,w$:
\bea
\cL_D\!\!\!&=&\!\!\sum_{s=y,w}
-\,(1/2)\,\, D_s^a D_s^a\,\big[ 
1+ 1/2\, (\alpha_{70}^s\,h_2.h_1+h.c.)\big]
=
\Big\{-\frac{g_2^2}{8}
\,\big(\vert h_1\vert^2\!-\!\vert h_2\vert^2\big)
\nonumber\\[-4pt]
&\times&\Big[\big(
1+2\tilde\rho_{1,w}+\frac{1}{2}(\alpha_{70}^w h_2.h_1+h.c.)\big)
\,\vert h_1\vert^2
-
\big(1+ 2\tilde \rho_{2,w}+(1/2)(\alpha_{70}^w\,h_2.h_1+h.c.)
\big)\vert h_2\vert^2
\Big]
\nonumber\\[-2pt]
& - & \!\!\!
(g_2\rightarrow g_1; 
\alpha_{70}^w\rightarrow\alpha_{70}^y;
\tilde\rho_{j,w}\rightarrow\tilde\rho_{j,y})\Big\}\!
- \! \frac{g_2^2}{2}\,\big[ 1\!+\!\tilde\rho_{1,w}\!+\!\tilde\rho_{2,w}\!+\!
({1}/{2})(\alpha_{70}^w\,h_2.h_1\!+\! h.c.)\big]
\vert h_1^\dagger h_2\vert^2
\nonumber\\[3pt]
&+&
g_2/(2\sqrt 2)\,
\big[h_1^\dagger  T^a h_1+h_2^\dagger  T^a h_2\big]
\big[\alpha_{70}^w \,(h_2.\psi_1+\psi_2.h_1)\,\lambda_w^a+h.c.\big]
\nonumber\\
&+&
g_1/(2\sqrt 2)\,
\Big(h_1^\dagger \frac{-1}{2} h_1
+h_2^\dagger \frac{1}{2} h_2\Big)
\big[\alpha_{70}^y \,(h_2.\psi_1+\psi_2.h_1)\,\lambda_y+h.c.\big]
\label{VG}
\eea

\medskip\noindent
Here $h_1.\psi_2=-\psi_2.h_1=h_1^0 \psi_2^0-h_1^- \psi_2^+$, 
$\vert h_1\vert^2=h_1^{0 *} h_1^0+h_1^{- *} h_1^-$, 
$h_1^\dagger h_2=h_1^{0 *} h_2^+ +h_1^{- *} h_2^0$, etc.

Eliminating the $F$-dependent terms in $\cL$ gives $\cL_F$ below (using notation
(\ref{rhos0}), (\ref{rhos}))
\bea
\cL_F&=&\cL_{F,1}+\cL_{F,2}
\nonumber\\
-\cL_{F,1}\!&\equiv&\! \vert F_1\vert^2+
\vert F_2\vert^2
=\vert \mu+2\,\zeta_{10}\,h_1.h_2\vert^2\,\,
\big(\vert h_1\vert^2+\vert h_2\vert^2\big)
\nonumber\\
&+&\!\!\!\!
\Big[\mu\,\Big(
\vert h_1\vert^2\,\rho_{21}+\vert h_2\vert^2\,\rho_{11}
\!+(h_1.h_2)^\dagger\,(\rho_{22}+\rho_{12})
\!+ (\psi_1.h_2)^\dagger \,\rho_{13}
\!+(h_1.\psi_2)^\dagger\,\rho_{23}\Big)
\!+\!h.c.\Big]\nonumber
\label{VF1}\eea
while $\cL_{F,2}$ is due to the nontrivial field  metric in the
 Kahler potential:
\bea
-\cL_{F,2}&=&
\vert \mu\vert^2
\Big[
2\,\big(\alpha_{10}+\alpha_{20}+\alpha_{40}\big)
\vert h_1\vert^2\,\vert h_2\vert^2
+(\alpha_{30}+\alpha_{40})\,\big(\vert h_1\vert^4+\vert h_2\vert^4\big)
\nonumber\\
&& +\,2\,\big(\alpha_{10}+\alpha_{20}+\alpha_{30}\big)
\,\vert h_1.h_2\vert^2+\,\,
\big(\vert h_1\vert^2+2\,\vert h_2\vert^2\big)
\big(\alpha_{50}\,h_2.h_1+h.c.\big)\qquad\qquad\qquad
\nonumber\\
&&+\,\big(2\vert h_1\vert^2+\vert h_2\vert^2\big)
\big(\alpha_{60}\,h_2.h_1+h.c.\big)\Big]
\label{VF2}
\eea
Apart from auxiliary fields contributions, there are also 
terms which contain space-time derivatives, that contribute to the
kinetic terms for Weyl fermions $\psi_{1,2}$, $\lambda^a_{w,y}$ 
when the neutral singlet Higgses $h_{1,2}^0$, components of $h_{1,2}$, acquire a vev:
\medskip
\bea
\cL_{1}&=&\!\!
 \alpha_{10}\,\big[
{i}\, \overline\psi_1\overline\sigma^\mu\cD_\mu \psi_1\,
\vert h_1\vert^2
+{i}\,\overline\psi_1\overline\sigma^\mu\, \psi_1 (\,h_1^\dagger\cD_\mu h_1)
-{i}\, (h_1^\dagger\psi_1)\, \sigma^\mu\overline\psi_1
(\cD_\mu-\overleftarrow \cD_\mu)\,h_1+h.c.\big]
\nonumber\\
&+&
\!\!\!
\alpha_{20}\,\big[
{i}\, \overline\psi_2\overline\sigma^\mu\cD_\mu \psi_2\,
\vert h_2\vert^2
+{i}\,\overline\psi_2\overline\sigma^\mu\, \psi_2 \,(h_2^\dagger\cD_\mu h_2)
-{i}\, (h_2^\dagger\psi_2)\, \sigma^\mu\overline\psi_2
(\cD_\mu-\overleftarrow \cD_\mu)\,h_2+h.c.\big]
\nonumber\\
&+&\!\!\alpha_{30}
\big[
{i}\, \overline\psi_2\overline\sigma^\mu\cD_\mu \psi_2\,
\vert h_1\vert^2
+{i}\,\overline\psi_1\overline\sigma^\mu\, 
\psi_1 \,(h_2^\dagger\cD_\mu h_2)
-{i}\, (h_1^\dagger\psi_1)\, \sigma^\mu\overline\psi_2
(\cD_\mu-\overleftarrow \cD_\mu)\,h_2+h.c.\big](1/2)
\nonumber\\
&+&\!\!\alpha_{30}
\big[
{i}\, \overline\psi_1\overline\sigma^\mu\cD_\mu \psi_1\,
\vert h_2\vert^2
+{i}\,\overline\psi_2\overline\sigma^\mu\, 
\psi_2 \,(h_1^\dagger\cD_\mu h_1)
-{i}\, (h_2^\dagger\psi_2)\, \sigma^\mu\overline\psi_1
(\cD_\mu-\overleftarrow \cD_\mu)\,h_1+h.c.\big](1/2)
\nonumber\\
&+&
\alpha_{40}\big[
{i}\,(\psi_1.h_2+h_1.\psi_2)\,\sigma^\mu\partial_\mu
(\psi_1.h_2+h_1.\psi_2)^\dagger+h.c.\big](1/2)
\nonumber\\
&+&
\big\{\alpha_{50}^*
\big[
i\, h_1^\dagger\, \cD_\mu\psi_1\,\sigma^\mu\,(\psi_1.h_2+h_1.\psi_2)^\dagger
+i\,(h_2.h_1)^\dagger \,
\overline\psi_1\overline\sigma^\mu\cD_\mu\psi_1
\big]+h.c.\big\}
\nonumber\\
&+&
\big\{\alpha_{60}^*
\big[
i\, h_2^\dagger\, \cD_\mu\psi_2\,\sigma^\mu\,(\psi_1.h_2+h_1.\psi_2)^\dagger
+i\,(h_2.h_1)^\dagger \,
\overline\psi_2\overline\sigma^\mu\cD_\mu\psi_2
\big]+h.c.\big\}\nonumber\\
&+&\big\{
({1}/{4})\,\alpha_{70}^w\, (h_2.h_1)
\big[ i\,(\lambda^a_w \sigma^\mu\Delta_\mu\overline\lambda^a_w
  -\Delta_\mu\overline\lambda^a_w \overline\sigma^\mu\lambda^a_w)
\big]+h.c.+(w\rightarrow y)\big\}
\eea

\bigskip\noindent
When the Higgs fields neutral singlets acquire a vev, these terms bring a wavefunction
renormalization of Weyl kinetic terms and a threshold correction to
gauge couplings $g_2, g_1$.

Also, there are terms that contribute to fermions masses, 
when singlet Higgs fields acquire a vev (we denote
$\lambda_{1,2}\equiv g_2 \lambda_w^a\,\sigma^a+g_1\,(\mp1)\lambda_y$,
 with "-" for $\lambda_1$, \,\,$\sigma^a$: Pauli matrices, $a=1,2,3$):
\bea
\cL_2&=&
(\alpha_{10}\,{\sqrt 2}) 
\big[-
 (h_1^\dagger  \lambda_1\psi_1)\vert h_1\vert^2
- (h_1^\dagger\psi_1)\,
h_1^\dagger \lambda_1 h_1\big]
-
\alpha_{11} m_0 (\overline\psi_1 h_1)(\overline\psi_1 h_1)
\nonumber\\
&+& 
(\alpha_{20}\,{\sqrt 2})
\big[-
 (h_2^\dagger  \lambda_2\psi_2)\vert h_2\vert^2
- (h_2^\dagger\psi_2)\,
h_2^\dagger \lambda_2 h_2\big]
-\alpha_{21} m_0 (\overline\psi_2 h_2)(\overline\psi_2 h_2)
\nonumber\\
&+&
\!\! (\alpha_{30}/\sqrt 2)
\big[- (h_2^\dagger  \lambda_2\psi_2)\vert h_1\vert^2
- (h_1^\dagger\psi_1)\,
h_2^\dagger \lambda_2 h_2+(1 \leftrightarrow 2)\big]
-\alpha_{31}^* m_0 ( h_1^\dagger\psi_1)(h_2^\dagger\psi_2)
\nonumber\\
&+&
\!\! (\alpha_{50}/{\sqrt 2})
\big[
 h_1^\dagger\lambda_1 h_1\,(\psi_1.h_2+h_1.\psi_2)
-
(h_2.h_1)(h_1^\dagger \lambda_1\psi_1+
\overline\psi_1\overline\lambda_1 h_1)
\big]
\nonumber\\[1pt]
&+&
\!\!({\alpha_{60}}/{\sqrt 2})
\big[
 h_2^\dagger\lambda_2 h_2\,(\psi_1.h_2+h_1.\psi_2)
-
(h_2.h_1)(h_2^\dagger \lambda_2\psi_2+
\overline\psi_2\overline\lambda_2 h_2)
\big]\nonumber\\
&-&m_0\,(\alpha_{51}^*  \vert h_1\vert^2 
+\alpha_{61}^*\,\vert h_2\vert^2)
\,\psi_2.\psi_1
-m_0\,(\alpha_{51}^*\,h_1^\dagger\psi_1
+\alpha_{61}^*\,h_2^\dagger\psi_2)
(h_2.\psi_1+\psi_2.h_1)
\nonumber\\
&+&
(1/4)\alpha_{71}^w m_0 (h_2.h_1)(\lambda^a_w\lambda^a_w)
+
(1/4)\alpha_{71}^y m_0 (h_2.h_1)(\lambda_y\lambda_y)
+
2\,\alpha_{81}^* \,m_0\,(h_2.h_1)\,(-\psi_2.\psi_1)
\nonumber\\
&+&\alpha_{41}\,m_0\,(h_2.h_1)\,(-\psi_2.\psi_1)^\dagger
+
 \zeta_{10}\big[
2\,(h_2.h_1)(-\psi_2.\psi_1)-(h_2.\psi_1+\psi_2.h_1)^2
\big]
+{\rm h.c.}
\eea

\medskip\noindent
Further, there are some interaction terms
\bea
\cL_3\!\!\!&=&\!\!
-\alpha_{10}\,(\overline\psi_1\psi_1)(\overline\psi_1\psi_1)
-\alpha_{20}\,(\overline\psi_2\psi_2)(\overline\psi_2\psi_2)
-\alpha_{30}\,(\overline\psi_1\psi_1)(\overline\psi_2\psi_2)
+\alpha_{40}(\psi_2.\psi_1)^\dagger(\psi_2.\psi_1)
\nonumber\\[3pt]
&+&\!\!\!
\Big\{\,\,({1}/{4})\,\alpha_{70}^w\,\,\Big[({-1}/{2})\,(h_2.h_1)\,
(F_w^{a\,\mu\nu}F^a_{w\,\mu\nu}+({i}/{2})\,\epsilon^{\mu\nu\rho\sigma}
F^a_{w\,\mu\nu}F^a_{w\,\rho\sigma})
\nonumber\\
&-& \sqrt 2\, (h_2.\psi_1+\psi_2.h_1)\,\sigma^{\mu\nu}\lambda^a_w
F^a_{w, \mu\nu}
- \psi_2.\psi_1\,\lambda^a_w\lambda^a_w\Big]
+(w\rightarrow y)
+h.c.\Big\}
\eea
with $(\overline\psi_1\psi_1)(\overline{\psi_2}\psi_2)=
(\overline{\psi_1^0}\psi_1^0+\overline{\psi_1^-}\psi_1^-)
(\overline{\psi_2^0}\psi_2^0+\overline{\psi_2^+}\psi_2^+)$, etc, where
spinor indices are not shown.

Also, there are $h_{1,2}$ dependent terms that contain space-time derivatives, 
which contribute to the kinetic terms in the Higgs sector, when the singlet
Higgs fields  acquire a vev:
\medskip
\bea
\cL_4&=&
2 \alpha_{1 0}\,
\big[\vert h_1\vert^2 \,\,\vert\cD_\mu h_1\vert^2+
\vert h_1^\dagger  \cD^\mu h_1\vert^2 \big]
+
2 \alpha_{2 0}\,\big[\vert h_2\vert^2\, \vert \cD_\mu h_2\vert^2 +
\vert h_2^\dagger  \cD^\mu h_2\vert^2
\big]\qquad\qquad\nonumber\\
&+&
\alpha_{30}\,\,\big[
 \vert h_1\vert^2\,
\vert \cD_\mu h_2\vert^2
\!+\!(h_1^\dagger \cD_\mu h_1)(h_2^\dagger\overleftarrow \cD^\mu h_2)
\!+\!(1\leftrightarrow 2)\big]
+
\alpha_{40}\,
\vert \partial_\mu (h_2.h_1)\vert^2
\nonumber\\
&+&\big\{
\alpha_{50}
\big[\,
\vert \cD_\mu h_1\vert^2\,(h_2.h_1)
+
(h_1^\dagger \overleftarrow\cD_\mu h_1)\,\partial^\mu(h_2.h_1)\big]
+h.c.\big\}\nonumber\\
&+&
\big\{\alpha_{60}
\big[\,
\vert \cD_\mu h_2\vert^2\,(h_2.h_1)
+
(h_2^\dagger \overleftarrow\cD_\mu h_2)\,\partial^\mu(h_2.h_1)\big]
+h.c.\big\}
\eea

\medskip\noindent
Finally, the Lagrangian contains (F and D-independent)  corrections  
due to supersymmetry breaking, i.e. terms proportional to $m_0$,
 due to spurion dependence 
in the higher dimensional operators (of dimensions $d=5$ and $d=6$) as well as
the usual soft terms of the MSSM. All these together give a final contribution 
to the Lagrangian:
\medskip
\bea
\cL_{SSB}=-V_{SSB}\!\!\!&=& m_0^2\,\big[
\alpha_{12}\,\,\vert h_1\vert^4
+\,\alpha_{22}\,\,\vert h_2\vert^4
+\,\alpha_{32}\,\,\vert h_1\vert^2\,\vert h_2\vert^2
+\,\alpha_{42}\,\,\vert h_2.h_1\vert^2
\\[3pt]
&&+\,\,\big(\alpha_{52}\,\,\vert h_1\vert^2\,(h_2.h_1)+h.c.\big)
+\big(\alpha_{62}\,\,\vert h_2\vert^2\,(h_2.h_1)+h.c.\big)\big]
\nonumber\\[3pt]
&& +\,\,
\big[\,m_0^2\,\alpha_{82}\,(h_1.h_2)^2+\zeta_{11}\,m_0\,(h_2.h_1)^2
+\mu\,B_0\,m_0\,(h_1.h_2)\!+\!h.c.\big]
\nonumber\\[6pt]
&&-\, m_0^2\,(c_1 \vert h_1\vert^2+\!c_2 \vert h_2\vert^2)\nonumber
\label{VSSB}
\eea

\noindent
This concludes the presentation of the full Lagrangian, in $1/M^2$ order.
Additional transformations (fields redefinitions or eqs of motion) can be used 
to eliminate  the non-diagonal kinetic terms of fermions and scalars,
to obtain a canonical form.

\section{The neutralino and chargino Lagrangian.}\label{section3}

From the total Lagrangian of the previous section 
one can obtain the Lagrangian of the neutralino and
 chargino sectors. Since the result is long, its detailed form in component 
fields is provided in Appendix~\ref{appendixB}. In this section we extract 
from this Lagrangian only the terms  that contribute to neutralino 
masses and their  kinetic terms (hereafter $\delta \cL_{D,F,2,1}$). These 
terms originate from $\cL_D$, $\cL_{F,1}$, $\cL_2$, $\cL_1$ of 
 previous section and are present in addition to the MSSM original terms.  
These are detailed below (in component, 
gauge-singlet fields notation):
\bea\label{LD}
\delta \cL_D& = &
- \frac{1}{4\sqrt 2} \big(\vert h_1^0\vert^2-\vert h_2^0\vert^2\big)
\big( g_2\lambda_w^3\alpha_{70}^w -g_1 \lambda_y \alpha_{70}^y\big)
\big( h_2^0\psi_1^0 +\psi_2^0 h_1^0\big)+h.c.\label{m1}
\qquad\qquad\eea
together with
\bea
- \delta \cL_{F,1}&=&
-\frac{\mu}{4}
 \big(\vert h_1^0\vert^2+\vert h_2^0\vert^2\big)
\big( \alpha_{70}^w\lambda_w^3 \lambda_w^3
  +\alpha_{70}^y\,\lambda_y\lambda_y\big)
+\mu\psi_1^0\psi_1^0\,(-2\alpha_{10}\,h_1^{0 *} h_2^0
+\alpha_{50} h_2^{0\, 2})
\nonumber\\
&+&\psi_1^0\psi_2^0\,\, \big[
- (\mu\alpha_{40}+\mu\,\alpha_{30})
(\vert h_1^0\vert^2+\vert h_2^0\vert^2)
+(\alpha_{50}+\alpha_{60})\,(\mu+\mu)\,h_1^0 h_2^0\,\,\big]
\nonumber\\[3pt]
&+&
 \mu \psi_2^0\psi_2^0
(-2\alpha_{20} \,h_1^0 h_2^{0 *} +\alpha_{60} h_1^{0 2})
+h.c.\qquad\label{m2}
\eea
and 
\bea
\delta \cL_2&=&
(g_2 \lambda_w^3-g_1\lambda_y)(\delta_1 \psi_1^0 +\delta_2 \psi_2^0)
+\delta_3\,\psi_1^0\psi_1^0
+\delta_4\,\psi_1^0\psi_2^0
+\delta_5\,\psi_2^0\psi_2^0\qquad\qquad\qquad\quad
\nonumber\\
&-&\frac{1}{4}\, m_0 h_1^0 h_2^0 \,\,\big(
\alpha_{71}^w\lambda_w^3\lambda_w^3
+\alpha_{71}^y\lambda_y\lambda_y\big)+h.c.
\label{m3}\eea
where we introduced the notation:
\bea
\delta_1&=&\!\!\!
-2\sqrt 2 \alpha_{10}\,\vert h_1^0\vert^2\,h_1^{0 *}
+\sqrt 2 \alpha_{50}\,\vert h_1^0\vert^2 \,h_2^0 
+({\alpha_{50}^*}/{\sqrt 2}) \,h_1^{0\, *\, 2} h_2^{0\,*}
-({\alpha_{60}}/{\sqrt 2}) \,\vert h_2^0\vert^2\,h_2^0
\nonumber\\[-3pt]
\delta_2&=& 2\sqrt 2 \alpha_{20}\,\vert h_2^0\vert^2 \,h_2^{0 *}
- \sqrt 2 \alpha_{60}\,\vert h_2^0\vert^2 \,h_1^0 
- ({\alpha_{60}^*}/{\sqrt 2})\,h_1^{0 *} \,h_2^{0 *\, 2}
+({\alpha_{50}}/{\sqrt 2})\,\vert h_1^0\vert^2\,h_1^0
\nonumber\\[-3pt]
\delta_3&=&\!
-\,\alpha_{11}^* \,m_0 h_1^{0 *\,2} +m_0 \alpha_{51}^* \,h_1^{0 *} h_2^0
-\zeta_{10}\,h_2^{0 \,2}
\nonumber\\
\delta_4&=&\!
2m_0 (\alpha_{51}^* \vert h_1^0\vert^2
 +\alpha_{61}^* \vert h_2^0\vert^2)
 -\alpha_{31}^* \,m_0 \,h_1^{0 *} h_2^{0 *}
-2 \alpha_{81}^* m_0  h_1^0\,h_2^0 -\alpha_{41}^* m_0 h_1^{0 *} h_2^{0 *}
-4\zeta_{10}\,h_1^0 \,h_2^0 \nonumber\\
\delta_5&=&\! -\,\alpha_{21}^* m_0 h_2^{0 *\, 2}+ m_0 \,\alpha_{61}^*
\,h_1^0 h_2^{0 *}
-\zeta_{10}\,h_1^{0\,2}
\eea

\medskip\noindent
$\cL_1$ and $\cL_4$ contain non-canonical, non-diagonal
kinetic terms for neutralinos/charginos and neutral higgses, respectively,
and these have  to be carefully considered. 
The terms in $\cL_1$ that generate non-canonical kinetic terms for
neutralinos, are denoted $\delta \cL_1$ and are
\bea
\delta \cL_1&=&
i\,\overline{\psi_1^0} \osigma^\mu \partial_\mu \psi_1^0\, \nu_1
+
i\,\overline{\psi_2^0} \osigma^\mu \partial_\mu \psi_2^0 \,\nu_2
+
i\,\overline{\psi_1^0} \osigma^\mu \partial_\mu \psi_2^0 \,\nu_3
+
i\,\overline{\psi_2^0} \osigma^\mu \partial_\mu \psi_1^0 \,\nu_4
\nonumber\\
&+&
\frac{i}{2}\,\big(\lambda^3_w\sigma^\mu\partial_\mu \overline\lambda^3_w
-\partial_\mu\overline\lambda^3_w \osigma^\mu \lambda^3_w\big)\,\nu_5^w
+
\frac{i}{2}\,\big(\lambda_y\sigma^\mu\partial_\mu \overline\lambda_y
-\partial_\mu\overline\lambda_y \osigma^\mu \lambda_y\big)\,\nu_5^y
+h.c.
\label{mpsi}
\eea
with
\bea
\nu_1&=& 2\alpha_{10}\,\vert h_1^0\vert^2
+(1/2) \,\alpha_{30}\,\vert h_2^0\vert^2
+(1/2)\,\alpha_{40}\,\vert h_2^0\vert^2
-2 \,h_1^{0 *}\,h_2^{0 *}\,\alpha_{50}^*
\nonumber\\
\nu_2&=&
 2\alpha_{20}\,\vert h_2^0\vert^2
+(1/2) \,\alpha_{30}\,\vert h_1^0\vert^2
+(1/2)\,\alpha_{40}\,\vert h_1^0\vert^2
-2 \,h_1^{0 *}\,h_2^{0 *}\,\alpha_{60}^*
\nonumber\\
\nu_3&=& (1/2)\,\alpha_{30}\,h_1^0\,h_2^{0 *} 
+(1/2)\,\alpha_{40}\,h_1^0\,h_2^{0 *}-\alpha_{60}^*\,h_2^{0\, *\, 2}
\nonumber\\
\nu_4&=&
 (1/2)\,\alpha_{30}\,h_2^0\,h_1^{0 *} 
+(1/2)\,\alpha_{40}\,h_2^0\,h_1^{0 *}-\alpha_{50}^*\,h_1^{0\, *\, 2}
\nonumber\\
\nu_5^w &=&- (1/2)\,\alpha_{70}^w \,h_1^0\, h_2^0,\,\qquad
\nu_5^y= - (1/2)\,\alpha_{70}^y \,h_1^0\, h_2^0\,\qquad\qquad\qquad\qquad\qquad
\eea
For our purposes it is useful to re-write this as
\bea
\delta \cL_1 &=&
\frac{i}{2}\,
\nu_{11^*}\,\overline{\psi_1^0}\osigma^\mu\partial_\mu\psi_1^0
+
\frac{i}{2}\,
\nu_{22^*}\,\overline{\psi_2^0}\osigma^\mu\partial_\mu\psi_2^0
+
i\,\nu_{43^*}\,\overline{\psi_2^0}\osigma^\mu\partial_\mu\psi_1^0
\nonumber\\
&+&
\frac{i}{2}\,
\nu_{55^*}^w\,\lambda^3_w\sigma^\mu\partial_\mu
\overline\lambda_w^3+
\frac{i}{2}\,\nu_{55^*}^y\,\lambda_y\sigma^\mu\partial_\mu
\overline\lambda_y
+
\textrm{h.c.}+S_i^\mu\partial_\mu \nu_i
\label{wk}
\eea
 $S_i^\mu$ is a function of fields, not specified here, and its contribution 
is vanishing if $\nu_i$ is a constant, which is indeed the case
when the Higgs fields acquire a vev. Also we introduced: 
\bea
\nu_{ij^*}&\equiv& \nu_i+\nu_{j}^*
\eea
Adding together (\ref{LD}) to (\ref{wk}) and
the original MSSM neutralino/chargino  terms ($\delta \cL_{mssm}$)
we finally have the following part of the neutralino Lagrangian  needed
for the mass spectrum (that is, without interacting terms,
 given in Appendix~\ref{appendixB}):
\bea\label{LN}
\cL_\chi&=&\delta \cL_D+\delta \cL_{F,1}+\delta \cL_1+
\delta \cL_2+\delta \cL_{mssm}\nonumber\\
&=&
\delta \cL_D+\delta \cL_{F,1}+\delta \cL_2\nonumber\\
&+&\Big\{\,\,
\frac{i}{2}\,
(1+\tilde\nu_{11^*})\,\overline{\psi_1^0}\osigma^\mu\partial_\mu\psi_1^0
+
\frac{i}{2}\,(1+\tilde\nu_{22^*})\,\overline{\psi_2^0}\osigma^\mu\partial_\mu\psi_2^0
+
i\,\tilde\nu_{43^*}\,\overline{\psi_2^0}\osigma^\mu\partial_\mu\psi_1^0
\nonumber\\
&+&\frac{i}{2}\,
(1+\tilde\nu_{55^*}^w)\,\lambda^3_w\sigma^\mu\partial_\mu
\overline\lambda_w^3
+
\frac{i}{2}\,(1+\tilde\nu_{55^*}^y)\,\lambda_y\sigma^\mu\partial_\mu
\overline\lambda_y
-\frac{1}{\sqrt 2} (g_2 \lambda_w^3 -g_1\lambda_y)(h_1^{0 *} \psi_1^0
-h_2^{0 *} \psi_2^0)
\nonumber\\
&-&\frac{1}{2}\,m_2\,\lambda_w^3\lambda_w^3
-\frac{1}{2}\,m_1\lambda_y\lambda_y
-\mu\,\psi_1^0\psi_2^0
+\textrm{h.c.}\,\,\Big\}
\eea
where $\tilde\nu_{ij^*}$ are the values of $\nu_{ij^*}$ when the neutral
Higgses acquire a vev:
\bea
\tilde \nu_{ij^*}\equiv\nu_{ij^*}\Big\vert_{h_i^0\ra v_i/\sqrt 2}
\eea
To remove the off-diagonal kinetic terms in $\cL_\chi$, 
we perform a field redefinition\footnote{The off-diagonal
kinetic terms can also be removed by a unitary transformation followed by a (non-unitary)
rescaling of the Weyl fermions. However, the field redefinitions used below 
provide a simpler form for the final result, as they only affect the MSSM
part of $L$, while the unitary transformation (that turns out to be $M$-independent) 
also acts on the rest of the Lagrangian, which results in more
 complicated final expressions. The two approaches are equivalent, 
the eigenvalue problem is not 
changed.}:
\bea\label{redef1}
\lambda_y&=&(1-\tilde\nu_{55^*}^y/2)\lambda_y^{''},\qquad\,
\lambda_w^3=(1-\tilde\nu_{55^*}^w/2)\lambda_w^{3 ''}\nonumber\\
\psi_1^0&=& 
(1-\tilde\nu_{11^*}/2)\,\,\psi_1^{0 ''}- (\tilde\nu_{34^*}/2)\,\, \psi_2^{0 ''}\,
\nonumber\\
\psi_2^0&=& (-\tilde\nu_{43^*}/2)\,\,\, \psi_1^{0''}+(1-\tilde\nu_{22^*}/2)\psi_2^{0 ''}
\eea
which only changes the MSSM part of the Lagrangian $\cL_\chi$, (ignoring
corrections higher than  $1/M^2$).
In  the new basis (double primed) $\cL_\chi$
has  canonical kinetic terms, but it is not yet in the form 
needed to compute the neutralino masses.
This is because there also are non-standard Higgs kinetic terms 
that must be brought to canonical form;  adding the MSSM higgs 
kinetic part, hereafter denoted $\cL_{kt}^{MSSM}$, then 
these terms are \cite{Antoniadis:2009rn}
\bea
\cL_4+\cL^{MSSM}_{kt}\supset 
(\delta_{ij^*}+ g_{ij^*})\,\,
\partial_{\mu}\,h_i^0\,\partial^\mu h_j^{0
  *},\qquad
i,j=1,2.
\eea
where the field dependent metric is:
\bea
g_{11^*}&=& 4\,\alpha_{10}\,\vert h_1^0\vert^2
+(\alpha_{30}+\alpha_{40})\,\vert h_2^0\vert^2
-2\,(\alpha_{50}\, h_1^0\,h_2^0\,+h.c.)
\nonumber\\
g_{12^*}&=&
(\alpha_{30}+\alpha_{40})\,h_1^{0 *}\,h_2^0
-\alpha_{50}^*\,\,h_1^{0 * 2}
-\alpha_{60}\,\,h_2^{0 \, 2},\qquad g_{21^*}=g_{12^*}^*
\nonumber\\
g_{22^*}&=& 4\,\alpha_{20}\,\vert h_2^0\vert^2
+(\alpha_{30}+\alpha_{40})\,\vert h_1^0\vert^2
-2\,(\alpha_{60}\, h_1^0\,h_2^0\,+h.c.)
\eea

\medskip\noindent
The metric $g_{ij^*}$
is expanded about a background value $\langle h_i^0\rangle
=v_i/\sqrt 2$, then v-dependent 
contributions to higgs kinetic terms are generated (plus higher dimensional
interactions for higgs, involving 2 derivatives). 
Higgs field re-definitions can be performed to
obtain canonical kinetic terms for neutral higgs sector;
these bring further corrections
to the scalar potential \cite{Antoniadis:2009rn}  but also shift the
pure MSSM part of $\cL_\chi$ to generate extra $1/M^2$ corrections.
The field re-definitions are then \cite{Antoniadis:2009rn}
\bea\label{redef2}
h_1^0&\ra& h_1^0\,\,\Big(1-\frac{\tilde g_{11^*}}{2}\Big)\,
-\frac{\tilde g_{21^*}}{2}\,h_2^0
\nonumber\\
h_2^0&\ra& h_2^0\,\,\Big(1-\frac{\tilde g_{22^*}}{2}\Big)\,-
\frac{\tilde g_{12^*}}{2}\,h_1^0,\qquad \textrm{where}\qquad
\tilde g_{ij^*}\equiv g_{ij^*}\Big\vert_{h_i^0\rightarrow v_i/\sqrt 2} 
\label{red}
\eea
where $\tilde g_{ij^*}$ are  constants defined above. 
This higgs field redefinition is applied to $\cL_\chi$ in
the doubled-primed basis. The result of applying 
(\ref{redef1}), (\ref{redef2}) to (\ref{LN}) is then, 
after removing the double-primed superscripts:
\bea
\cL_\chi&=&
\delta \cL_D+\delta \cL_{F,1}+\delta \cL_2
\nonumber\\
&+&
\Big\{\,\frac{i}{2}\,\overline{\psi_1^0}\osigma^\mu\partial_\mu\psi_1^0
+
\frac{i}{2}\,\,\overline{\psi_2^0}\osigma^\mu\partial_\mu\psi_2^0
+\frac{i}{2}\,\lambda^3_w\sigma^\mu\partial_\mu\overline\lambda_w^3
+
\frac{i}{2}\,\lambda_y\sigma^\mu\partial_\mu\overline\lambda_y
\nonumber\\
&+&
\frac{\mu}{2}\,\,
\Big[ \,(-2+\tilde\nu_{11^*}+\tilde\nu_{22^*})\,\,\psi_1^0\psi_2^0
+\tilde\nu_{3^*4}\,\psi_1^0\psi_1^0 +\tilde\nu_{34^*}\,\, \psi_2^0\psi_2^0\Big]
\nonumber\\
&+&
\frac{g_2}{2\sqrt{2}}\,\lambda_w^3\,\Big[\,
h_1^{0\, *} \,\big[\, \psi_2^0 \,(\tilde\nu_{34^*}
-\tilde g_{21^*})+\psi_1^0\, (-2+\tilde\nu_{11^*}+\tilde g_{11^*}+\tilde\nu_{55^*}^w)
\big]\nonumber\\
&-&
h_2^{0\, *}\, \big[ \psi_1^0\, \,(\tilde\nu_{3^*4}-\tilde g_{12^*}) +\psi_2^0\,
(-2+\tilde g_{22^*}+\tilde\nu_{22^*}+\tilde\nu_{55^*}^w) \big]\Big]
-(w\ra y, g_2\ra g_1)
\nonumber\\
&-&
\frac{m_1}{2}(1-\tilde\nu_{55^*}^y )\,\lambda_y\lambda_y
-\frac{m_2}{2}(1-\tilde\nu_{55^*}^w )\,\lambda_w^3\lambda_w^3+h.c.\Big\}
\eea
This is the canonical Lagrangian in the neutralino sector that contains the
 mass and kinetic terms.
The interaction terms $\cO(1/M^2)$ can be found in Appendix~\ref{appendixB}; 
they are not affected by the 
above higgs and gaugino/higgsino redefinitions  
since the difference is of order higher than $1/M^2$. Thus, they
 can be simply added to the above $\cL_\chi$. The full Lagrangian 
 can be implemented in micrOMEGAs, to analyze the impact of $\cO_i$, $\cK_0$ 
on dark matter searches.

\section{The spectrum of the Lagrangian.}\label{section4}

\subsection{Corrections to neutralino masses.}\label{neutralinomass}

Using $\cL_\chi$ of the last equation in the previous section, 
we  compute in this section the mass corrections to the neutralino fields, induced by
all effective operators in $1/M^2$ order (i.e. leading order 
in $\alpha_{ij}\sim 1/M^2$, second order in $\zeta_{10}\sim 1/M$).
In the basis $(\lambda_y,\lambda_w^3,\psi_1^0,\psi_2^0)^T$ 
this mass matrix is
\bea
\cM_{11}&=&m_1+\frac{1}{8}\,\,\big[
-2 \,\alpha_{70}^y\,\mu\,v^2+\big(\alpha_{71}^y\,m_0+(\alpha_{70}^y+
\alpha_{70}^{y *})\,\,m_1\big)\,v^2\sin 2\beta\,\big],\qquad
\cM_{12}=0
\nonumber\\
\cM_{13}&=& -\frac{m_Z}{32}\, \sin\theta_w\,\Big[
-4\,\, \big(-8 +(\alpha_{30}+\alpha_{40})\,v^2 \big)
\cos \beta
+v^2\,\,\big[\,4\,(\alpha_{30}+\alpha_{40})\,\cos3\beta
\nonumber\\
&+&
2\sin \beta\,\big(4\alpha_{50}^*+4\alpha_{60}
+\alpha_{70}^y+\alpha_{70}^{y\, *}+
(4\alpha_{50}^*-4\alpha_{60}+3\alpha_{70}^y
+\alpha_{70}^{y\,*})\,\cos2\beta\big)\big]\Big]
\nonumber\\
\cM_{14}&=& \frac{m_Z}{32}\,\sin\theta_w
\Big[32 \sin\beta+2\,v^2\cos\beta\,\big[
4\,\alpha_{50}+4\alpha_{60}^*+\alpha_{70}^y+\alpha_{70}^{y\,*}
\nonumber\\
&+&
\cos 2\beta\,(4\alpha_{50}-4\alpha_{60}^*-3\alpha_{70}^y-\alpha_{70}^{y\,*})
-4\,(\alpha_{30}+\alpha_{40})\,\sin 2\beta\,\big]\Big]
\nonumber\\
\cM_{22}&=&m_2+
\frac{1}{8}\,\,\big[
-2\alpha_{70}^w\,\mu\,v^2+\big(\alpha_{71}^w\,m_0
+(\alpha_{70}^{w}+\alpha_{70}^{w\,*})\,m_2\,\big)\,v^2\sin 2\beta\,\big]
\nonumber\\
\cM_{23}&=&
\frac{m_Z}{32}\,\cos\theta_w\,
\Big[-4\,\big(-8+(\alpha_{30}+\alpha_{40})\,v^2\big)\,\cos\beta+v^2\,\big[\,4
\,(\alpha_{30}+\alpha_{40})\,\cos 3\beta
\nonumber\\
&+&
2\,\big(
4\alpha_{50}^*+4\alpha_{60}+\alpha_{70}^w+\alpha_{70}^{w \,*}
+ (\,4\,\,\alpha_{50}^*-4\alpha_{60}+3\alpha_{70}^w +\alpha_{70}^{w *})
\,\cos 2\beta\,\big)
\sin\beta\,\,\big]\,\Big]\,\,\,\,
\nonumber\\
\cM_{24}&=&
\frac{m_Z}{32}\,\cos\theta_w\,
\Big[
-32 \sin\beta-2\,v^2\,\cos\beta
\,\big[\,\,4\alpha_{50}+4\,\alpha_{60}
+\alpha_{70}^w +\alpha_{70}^{w\, *}
\nonumber\\
&+&(4\,\alpha_{50}-4\,\alpha_{60}^*-3\alpha_{70}^w -\alpha_{70}^{w *}
)\,\cos 2\beta-
4\,(\alpha_{30}+\alpha_{40})\sin 2\beta\,\,\big]\Big]
\nonumber\\
\cM_{33}&=&
\frac{1}{4}\,v^2\,\Big[
4\,m_0\,\cos\beta\,(\alpha_{11}^*\cos\beta-\alpha_{51}^*\,\sin\beta)
+ 4\,\zeta_{10}\,\sin^2\beta
+\mu\,\big[2\alpha_{50}+
\alpha_{50}^*+\alpha_{60}\nonumber\\
&+&(-2\alpha_{50}+\alpha_{50}^*-\alpha_{60})
\,\cos 2\beta-(4\,\alpha_{10}+\alpha_{30}+\alpha_{40})\,\sin 2\beta\big]
\Big]\nonumber\\
\cM_{34}&=&
\frac{1}{4}
\,\Big[
4\,\mu- \Big(2 \,(\alpha_{51}^*+\alpha_{61}^*)\,m_0+
\big(2\alpha_{10}+2\alpha_{20}+3\,(\alpha_{30}+\alpha_{40})\big)
\mu\Big)
\,v^2\nonumber\\
&+&
v^2\,\Big(
2\,\,\big(- (\alpha_{51}^*-\alpha_{61}^*)\,m_0-\mu\,(\alpha_{10}-\alpha_{20})\,\,
\big)\,\cos 2\beta
\nonumber\\
&+&\big[\,(
\alpha_{31}^*+\alpha_{41}^*+2 \,\alpha_{81}^*\,)\,\,m_0
+(3\,\alpha_{50}+\alpha_{50}^*+3\,\alpha_{60}+\alpha_{60}^*)\,\mu+4\zeta_{10}
\big]\,\sin 2\beta\Big)\Big]
\nonumber\\
\cM_{44}&=&
\frac{1}{4}\,v^2\,\Big[
2\alpha_{21}^*\,m_0+(\alpha_{50}+2\,\alpha_{60}+\alpha_{60}^*)
\mu
+\big[-2\alpha_{21}^* m_0+(\alpha_{50}+2\alpha_{60}-\alpha_{60}^*)\,\mu
\big]\cos2\beta\nonumber\\
& +&
4\,\zeta_{10}\,\cos^2\beta
-\big[ 2\alpha_{61}^*\,m_0+\mu\,(4\,\alpha_{20}+\alpha_{30}+\alpha_{40})\,\big]
\sin2\beta\Big]
\eea
with the remaining matrix elements fixed by the symmetry $\cM_{ij}=\cM_{ji}$.

One can find an eigenvalue (denoted $\xi$) of this mass matrix in an analytical approach,
 by a perturbative method, as an expansion about the corresponding MSSM eigenvalue ($\tq$). 
In both cases, the eigenvalues satisfy a characteristic equation
\bea\label{dd9}
 \gamma_l\,\,\xi^l=0,\,\,\,(a) \qquad\qquad
\,\gamma^0_l\,\,(\tq)^l=0,\,\,\,\,(b)
\eea
with sums understood over the repeated index $l=0,1,2,3,4$.
Here (a) refers to the general case and (b) to the MSSM case.
 $\gamma_l$ are coefficients depending on $\alpha_{ij}\sim 1/M^2$ 
and $\zeta_{10}\sim 1/M$ that are found from the mass matrix above 
and $\xi$ denotes any of the four mass eigenvalues in the general case. 
The values of the MSSM counterparts, $\gamma_l^0$ 
for coefficients and $\tq$ for corresponding eigenvalue, are
recovered from the general ones by setting $\alpha_{ij}$ and $\zeta_{10}$ to 0.
Further, any general mass eigenvalue can be written $\xi=\tq+z_1+z_2$
with $z_1\propto \zeta_{10}=\cO(1/M)$ due to the $d=5$ operator, and 
$z_2\propto \sum_{k=1,..8}\sum_{ij} \alpha_{ij}\,\sigma_{kij}
+\beta_2\,\zeta_{10}^2=\cO(1/M^2)$, due to all $d=6$ operators as well as the $d=5$ operator.
From these equations, one computes the difference 
$\xi-\tq$ by consistently retaining the leading order 
approximation in eq.(\ref{dd9}) (a). One finds
\medskip
\bea
\xi=\tq+z_1
-\frac{1}{j\,\gamma_j^0\,\,(\tq)^{j-1}}
\Big[
\gamma_k^{(2)}\,\,(\tq)^{k}+
z_1^2\,C_k^2\,\gamma_k^0\,(\tq)^{k-2}+
z_1\,k\, \gamma_k^{(1)}\,(\tq)^{k-1}\Big]
\eea
where 
\bea
z_1=-\frac{\gamma_k^{(1)}\,(\tq)^k}{j\,\gamma_j^0\,(\tq)^{j-1}}
\eea
with summations understood over indices
 $j,k=0,1,2,3,4$. Here $\gamma_k^{(1)}$ and $\gamma_k^{(2)}$ denote
 the corrections $\cO(1/M)$ and $\cO(1/M^2)$ respectively,
 that are present in $\gamma_k$. Also $C_k^2=k(k-1)/2$. 
Replacing $\tq$ by any of the four values of the neutralino masses in the MSSM,  one
obtains the corresponding neutralino mass in the general case. In particular this 
is true about the LSP mass, when $\xi_o$ is the MSSM corresponding eigenvalue.

We provide below the analytical expression for the neutralino mass corrections,
with the contribution of each operator ($\cO_i$) labelled by the first index 
in $\alpha_{ij}$. 
One can include the effect of a selected set of these 
operators or from all of them, by simply 
adding the {\it corrections} $\delta m_\chi= \xi-\tq$ 
to the MSSM mass eigenvalue ($\tq$), due to the particular 
set of operators considered. 
We found
\bea 
\delta m_{\chi}=\xi-\tq=\sum_{i}\delta m_\chi(\cO_i)+\delta m_{\chi} (\cK_0)
\eea
\bea
\delta m_\chi(\cO_1)
&=&
\frac{2 \alpha_{10}}{\sigma}\,\mu\,v^2\,\cos\beta
\Big[2\,\mu\, (m_1-\tq)\,(\tq-m_2)\,\cos\beta-\big[
(m_1+m_2)\,m_Z^2
\nonumber\\
&+& 2 \,(m_1\,m_2-m_Z^2)\,\tq -
2 \,(m_1+m_2)\,	\tq^2+2\,\tq^3
+(m_1-m_2)\,m_Z^2\,\cos 2\theta_w\big]
\sin\beta\Big]
\nonumber\\
&+&
\frac{\alpha_{11}^*}{4\,\sigma}\,m_0\,v^2\,
\Big[
2\cos^2\beta \big[
(m_1+m_2)\,m_Z^2-2 \,(-2m_1 m_2+m_Z^2)\,\tq-4\,(m_1+m_2)\,\tq^2
\nonumber\\
&+&
4\,\tq^3-m_Z^2\,(m_1+m_2-2\,\tq)\,\cos 2\beta\big]
+(m_1-m_2)\,m_Z^2\,\cos 2\theta_w\sin^2 2\beta\Big]
\eea

\medskip\noindent
where $\sigma$ is defined later on. Further:
\smallskip
\bea
\delta m_{\chi}(\cO_2)\,\,&=&\,\,\delta m_{\chi}(\cO_1)
\Big[\alpha_{10}\ra \alpha_{20},\alpha_{11}^*\ra \alpha_{21}^*, 
\beta\ra \pi/2-\beta\Big]
\nonumber\\[4pt]
\delta m_\chi(\cO_3)&=&
\frac{\alpha_{30}}{4\sigma}\,v^2\,
\Big[
-12 m_1\,m_2\,\mu^2+(m_1+m_2)(12 \mu^2+m_Z^2)\,\tq
-2 \,(6\mu^2+m_Z^2)\,\tq^2
\nonumber\\
&+& m_Z^2\,\tq\,(2\,\tq-m_1-m_2)\cos4\beta+ 2\sin 2 \beta\,
\,\big[
\mu\,\,\big(-3\,(m_1+m_2)\,m_Z^2-2\,m_1\,m_2\,\tq
\nonumber\\
&+& 6 \,m_Z^2\,\tq+2\,(m_1+m_2)\,\tq^2-2 \,\tq^3\big)
+
m_Z^2\,(m_1-m_2)\cos 2\theta_w\,(\tq \sin 2\beta-3\,\mu)\big]\Big]
\nonumber\\
&+&\!\!\!\!
\frac{\alpha_{31^*}}{4\sigma}\,m_0\,v^2\,\sin 2\beta\Big[
m_Z^2\,\big(m_1\!+\!m_2-2\,\tq+(m_1-m_2)\,\cos 2\theta_w\,\big)
\sin 2\beta\nonumber\\ &+&
4\,\mu(m_1-\tq)(m_2-\tq)
\Big]\qquad
\eea
and
\bea
\delta m_\chi(\cO_4)&=&
\delta m_\chi (\cO_3)\Big[\alpha_{30}\ra \alpha_{40},
\alpha_{31}\ra \alpha_{41}\Big]\qquad\qquad\qquad
\qquad\qquad\qquad\qquad\qquad\qquad\quad
\eea
\bea
\delta m_\chi(\cO_5)&=&\!\!\!\frac{\alpha_{50}^*}{8\sigma}v^2\cos\beta\Big[
-5\mu\,m_Z^2\,(m_1+m_2-2 \,\tq)\,\cos 3\beta+2\,\sin\beta\,\big[8\,m_1\,m_2\,\mu^2
\nonumber\\
&-& \!\!\!
(m_1\!+\!m_2)(8\mu^2\!+\!m_Z^2)\tq+2 (4\mu^2\!+\!m_Z^2)\tq^2\big]
\!+\!8(m_2\!-\!m_1) m_Z^2 \tq \cos^2\beta\cos2\theta_w\sin\beta
\nonumber\\
&+&
\mu\cos\beta\,\big[5\,(m_1+m_2)\,m_Z^2+2 (4m_1m_2-5m_Z^2)\,\tq
-8 (m_1+m_2)\,\tq^2+8\,\tq^3
\nonumber\\
&+& 
20\,(m_1-m_2)\,m_Z^2\cos 2\theta_w\sin^2\beta\big]
-2 \,m_Z^2\,(m_1+m_2-2\,\tq)\,\tq\sin3\beta \Big]
\nonumber\\
&+&
\frac{\alpha_{50}}{16\sigma}\,v^2\Big[-\mu m_Z^2\cos 4\beta\,\big[
m_1+m_2-2\,\tq +(m_1-m_2)\cos 2\theta_w\big]
\nonumber\\
&+&
4\mu\cos 2\beta\,\big[(m_1+m_2)\,m_Z^2-2
 (m_1 m_2 +m_Z^2)\,\tq+2 (m_1+m_2)\,\tq^2-2 \tq^3 
\nonumber\\
&+&  
(m_1-m_2)\,m_Z^2 \cos 2\theta_w\big]+3\mu\big[
7\,(m_1+m_2)\,m_Z^2+2 (4\,m_1 m_2 -7 m_Z^2)\,\tq
\nonumber\\&-&
8 (m_1+m_2)\,\tq^2+8 \tq^3+7 (m_1-m_2)\,m_Z^2cos 2\theta_w \big]
-4\,\sin 2\beta\big[-12 m_1 m_2 \mu^2
\nonumber\\
&+&(m_1+m_2)(12 \mu^2+m_Z^2)\,\tq
-2 \,(6\mu^2+m_Z^2)\,\tq^2 +(m_1-m_2)\,m_Z^2 \tq\,\cos 2\theta_w\big]
\nonumber\\
&-& 2 m_Z^2 \,\tq\, (m_1+m_2-2 \tq+(m_1-m_2)\,\cos 2\theta_w)\,\sin 4\beta\Big]
+\frac{\alpha_{51}^*}{8\sigma} \,m_0\, v^2
\nonumber\\
&\times &\!\!\! \Big[32 \mu (m_1-\tq)(\tq-m_2)\,\cos^2\beta
-2 \sin 2\beta \big[ 3 (m_1+m_2) m_Z^2 +
4 m_1 m_2 \,\tq-6 m_Z^2 \,\tq\nonumber\\
&-& 4 \,(m_1+m_2)\,\tq^2
+ 4 \tq^3+3 (m_1-m_2)\,m_Z^2\,\cos 2\theta_w\big]
- m_Z^2\,\big[m_1+m_2-2 \,\tq\nonumber\\
&+& (m_1-m_2)\cos 2\theta_w\big]\sin 4\beta\Big]
\eea
Further 
\bea
\delta m_\chi(\cO_6)\,\,=\,\,
\delta m_{\chi}(\cO_5)\Big[\alpha_{50}\ra \alpha_{60},
\alpha_{51}\ra \alpha_{61},\beta\ra \pi/2-\beta\Big]
\qquad\qquad\qquad\qquad\qquad\qquad
\eea

\noindent
The correction due to bino part (indexed  by $y$) of $\cO_7$ is:
\medskip
\bea
\delta m_\chi(\cO_7^y)&=&\!\!
\frac{\alpha_{70}^y}{16\sigma}
\,v^2\,\Big[\mu\,
\big[
m_1\,m_Z^2+8\mu^2 \,\tq+5\,m_Z^2\,\tq-8 \,\tq^3-m_2\,(8\mu^2+m_Z^2-8\,\tq^2)
\big]
\nonumber\\
&+&\mu\,m_Z^2\,\big[
(m_1+m_2+3\,\tq)\,\cos 2\theta_w-\cos 4\beta\,\big(m_1
+3\,m_2-3\,\tq
\nonumber\\
&+&(m_1-3 \,m_2+3\,\tq)\,\cos 2\theta_w\big)
\big] - 2\big[
m_Z^2\,(2\mu^2+(m_2-\tq)\,\tq)+
m_1\,\big(-2 m_2\,\mu^2\nonumber\\
&+&\!
\tq\,(2\mu^2+m_Z^2)
+2 m_2 \tq^2-2\,\tq^3\big)
+m_Z^2\cos 2\theta_w\big(2\mu^2\!+\!\tq(m_1\!-m_2\!+\!\tq)\big)\big]\sin 2\beta
\Big]
\nonumber\\
&+&
\frac{\alpha_{70}^{y\,*}}{8\sigma_1}
v^2\sin2\beta\Big[2 m_1 m_2 \mu^2
-\big(2 m_1\mu^2\!+\!(m_1\!+\! m_2)\! m_Z^2\big)\,\tq
+(-2 m_1 m_2\!+\!m_Z^2)\tq^2\nonumber\\
&+&\!\!
2\,m_1\tq^3\!+\!
m_Z^2\big(\mu(m_1\!+\!m_2\!-\!\tq)\sin 2\beta\!+\!(m_1\!-m_2\!+\tq)\cos 2\theta_w
(-\tq\!+\!\mu\sin 2\beta)\big)\Big]
\nonumber\\
&+&
\frac{1}{8\sigma}
\,\alpha_{71}^y\,m_0\,v^2\sin 2\beta\,\Big[
2\,m_2 (\mu^2-\tq^2)-\tq\,(2\mu^2+m_Z^2-2\,\tq^2)
\nonumber\\
&+&m_Z^2\,(-\tq\,\cos2\theta_w+2 \mu\cos^2 \theta_w\sin2\beta)\,\Big]
\eea

\medskip\noindent
A similar correction exists for $\cO_7^w$:
\bea
\delta m_{\chi}(\cO_7^w)\,\,=\,\,\delta m_{\chi}(\cO_7^y)\Big[
\alpha_{70}^y\ra \alpha_{70}^w, \alpha_{71}^y\ra \alpha_{71}^w, m_1\ra m_2, m_2\ra m_1,
\theta_w\ra \pi/2- \theta_w\Big]
\eea
Further
\bea
\delta m_{\chi}(\cO_8)&=& \frac{\alpha_{81}^*}{2\sigma}
\,m_0\,v^2\sin 2\beta\,\Big[
4\mu (m_1-\xi_o)(m_2-\xi_o)+m_Z^2 [m_1+m_2-2\xi_o
\qquad\qquad\qquad\quad
\nonumber\\
&+&(m_1-m_2)\cos 2\theta_w
]\sin 2\beta\Big]
\eea
Finally
\bea
\delta m_{\chi}(\cK_0)&=&
\frac{\zeta_{10}}{4\sigma}\,v^2\sigma'
+\frac{\zeta_{10}^2}{8\,\sigma^3}\,v^4 \Big[12\,\sigma^2 
\,(m_1-\tq)(m_2-\tq)\,\sin^2 2\beta
+
\sigma^{' 2}\big[-m_1 m_2 +\mu^2
\nonumber\\
&+&m_Z^2
+3 (m_1+m_2)\,\tq-6\,\tq^2\big]
+\sigma\sigma' \,\big[ 4\,m_1\,m_2-5 m_Z^2-8 (m_1+m_2)\,\tq+12 \tq^2
\nonumber\\
& +& m_Z^2 \cos 4\beta-
8\mu(m_1+m_2-2\,\tq)\sin 2\beta\big]\Big]
\eea
In the above equations we introduced the notation
\bea
\sigma&=&
(m_1+m_2)(2\mu^2+m_Z^2)-4\,(-m_1\,m_2+\mu^2+m_Z^2)\,\tq
-6\,(m_1+m_2)\,\tq^2+8\,\tq^3
\nonumber\\
&+&
m_Z^2\,\big[(m_1-m_2)\cos2\theta_w+2\mu\sin 2\beta\big]
\nonumber\\
\sigma'&=&
5\,(m_1+m_2)\,m_Z^2+2 (4\,m_1\,m_2-5 \,m_Z^2)\,\tq - 8\,(m_1+m_2)\,\tq^2
+8\,\tq^3
\nonumber\\
&+&
m_Z^2\big[-\big((m_1+m_2-2\,\tq)\cos 4\beta\big)-
(m_1-m_2)(-5 +\cos 4\beta)\cos 2\theta_w\big]
\nonumber\\
&+&
16\mu\,(m_1-\tq)(m_2-\tq)\sin 2\beta
\eea

\medskip\noindent
In some cases not all operators are present, and even for 
each operator, one may be interested only in the supersymmetric 
correction (labelled by a "zero" second index, 
$\alpha_{j0}$).  By simply setting $\alpha_{j1}$ and $\alpha_{j1}^*$ to 
zero, one can identify only the supersymmetric corrections, when 
the above results simplify considerably.
Moreover, for the constrained MSSM when the gaugino universality 
is present, $m_1(m_Z)
=(5/3) \tan^2\theta_w(m_Z) m_2(m_Z)$ ,
then the results are further simplified. As for the expression of the MSSM mass eigenvalues 
denoted $\tq$, these are known in the literature \cite{ZZ} and can also be evaluated numerically.
The numerical results due to these corrections
will be presented in Section~\ref{section5}.

\subsection{Corrections to the Higgs fields masses.}\label{higgsmass}

The effective operators also affect the spectrum in the Higgs sector.
The exact corrections of these operators
 to lightest Higgs mass $m_h^2$ to order $\cO(1/M^2)$ can be found in 
eq.(36) and Appendix~C of \cite{Antoniadis:2009rn}.
With $m_A\!>\!m_Z$ assumed, the mass correction to $m_h$, in the large $\tan\beta$ limit with
$m_A$ fixed, has a very simple form \cite{Antoniadis:2009rn}:
\bea\label{dd1}
\delta m_{h}^2
\!\!\!&=&
-2\, v^2\,\Big[ \alpha_{22} \,m_0^2+(\alpha_{30}+\alpha_{40})\mu^2
+2 \alpha_{61} \,m_0\,\mu 
- \alpha_{20}\,m_Z^2\Big]
-\frac{(2\,\zeta_{10}\,\mu)^2\,\,v^4}{m_A^2-m_Z^2}
\nonumber\\[-2pt]
&+&\!\!\!\!\frac{v^2}{\tan\beta}
\bigg[\frac{1}{(m_A^2-m_Z^2)}
\Big( 4 \,m_A^2\,\big(\,
(2 \alpha_{21}\!+\!\alpha_{31}\!+\!\alpha_{41}\! +\!2 \alpha_{81})
\,m_0\,\mu\! +\! (2\alpha_{50}\! +\!\alpha_{60})\,\mu^2
+\alpha_{62}\,m_0^2\big)
\nonumber\\[-2pt]
&-&\, (2 \alpha_{60}-3\alpha_{70})\,m_A^2\,m_Z^2
-(2\alpha_{60}+\alpha_{70})\,m_Z^4\Big)
+\frac{8\,(m_A^2+m_Z^2)\,\,(\mu\,m_0\,\zeta_{10}\,\zeta_{11})
\,v^2}{(m_A^2-m_Z^2)^2}
\bigg]\nonumber\\[-2pt]
&+&\cO(1/\tan^2\beta)
\eea
where $\delta m_{h}^2\equiv m_{h}^2-m_{h}^{0\, 2}$, with $m_{h}^0$ the MSSM value.
Similar formulae exist for the heavier neutral Higgs mass $m_H^2$, and pseudoscalar 
 $m_A^2$. With these corrections 
one can examine the effect of combined  dark matter and EW constraints 
on the scale of new physics that  may be present in the MSSM. For  details 
on the Higgs sector corrections to masses see \cite{Antoniadis:2009rn,Carena:2009gx}. In the 
 numerical results  presented in Section~\ref{section5} we use the exact formula 
for $\delta m_h^2$ (i.e. with no expansion in $\tan\beta$ or other parameter).
Eq.(\ref{dd1}) gives however a good indication on the behaviour of the corrections: 
for example $\delta m_h^2$ is increased by negative $\alpha_{30},\alpha_{40}$, 
as we shall see later on, $\alpha_{50}, \alpha_{60}$ have contributions suppressed 
at large $\tan\beta$, Susy breaking corrections ($\alpha_{j1}, \alpha_{j2}$)
 are comparable to Susy ones ($\alpha_{j0}$), etc.

\subsection{Corrections to the chargino masses.}

Following the method presented for the neutralino case,
one can also evaluate the corrections to the chargino fields masses.
To this purpose one uses the Lagrangian presented in Appendix~\ref{appendixB}.
As for the neutralino case,  chargino fields rescaling is required
to ensure canonical kinetic terms for them,  plus Higgs re-definitions as in 
eq.(\ref{redef2}). The chargino mass corrections in order $1/M^2$ are presented in 
Appendix~\ref{appendixC} and can be useful for phenomenological studies,
in a global fit of MSSM with effective operators.

\section{Phenomenological implications}\label{section5}

In this section we analyze the impact of the effective operators on the MSSM Higgs and 
neutralino LSP masses. We perform the analysis separately for each individual operator
considered. This is because not all operators may be present in a particular model. 
The corrections from a set of operators can be readily 
obtained by combining  appropriately those of individual $\cO_i$.
The number of parameters can be further reduced by considering only the impact of the
supersymmetric
corrections (i.e. take $\alpha_{j1}=\alpha_{j2}=0$, $j=1,...8$), and $\alpha_{j0}\not=0$. 
It turns out
that non-Susy corrections are, in absolute value, of a size generically close to that of 
the supersymmetric case. From this analysis one can identify which of these operators 
has the largest impact on phenomenology.

To this purpose we consider the 
CMSSM phase space points that respect all current constraints, both 
theoretical and experimental. These refer to: radiative EWSB, no electric charge or colour breaking, 
LEP sparticle bounds, $b\ra s\gamma$ bounds and dark matter constraints, but no LEP2
 bound on the Higgs mass (that is not imposed, see later). These points are selected 
using SOFTSUSY \cite{Allanach:2001kg} and micrOMEGAs  \cite{Belanger:2006is}
codes, in the context of  CMSSM, as described and used in \cite{Cassel:2009cx}.  
On  these phase space points we impose the constraint
$\tilde m/M<1/2$, where $\tilde m=\mu, m_0$, or $m_{12}$, to ensure that 
our effective expansion parameter which is actually its square $\alpha_{j0} 
\tilde m \sim (\tilde m/M)^2<1/4$. This value is small enough to trust  the results of the effective 
operators expansion. Using these CMSSM phase points we examine the 
effect on the LSP and Higgs masses  from  the corrections  due to each effective operator. 
That is, we treat the effective operators as a perturbation of the 
CMSSM "background", to analyze which of its phase space points are likely to give sizable 
corrections. In this way we investigate if the "best fit" points of CMSSM are 
stable under corrections
 from the effective operators. By "best fit" points here we mean those points that
satisfy the aforementioned theoretical and experimental constraints,
with  WMAP dark matter relic density consistency or saturation within 3$\sigma$, 
plus electroweak fine tuning\footnote{The definition of EW scale fine tuning that 
we are using is 
\begin{equation}
\Delta\equiv\max\big\vert\Delta_{p}\big\vert_{p=\{
\mu_{0}^{2},m_{0}^{2},m_{1/2}^{2},A_{0}^{2},B_{0}^{2}\}},\qquad\Delta_{p}
\equiv\frac {\partial\ln v^{2}}{\partial\ln p}  \label{ft}
\end{equation}
where $p$ are input parameters at the UV scale, in the standard MSSM notation.
For its value at 2-loop see~\cite{Cassel:2009cx}.}
not worse than 1 part in 200~\cite{Cassel:2009cx}. 
This fine tuning constraint enforces that some points have an
expansion parameter less than $1/4$.

A more careful analysis should implement all the new couplings in the Higgs, chargino and 
neutralino  sectors in SOFTSUSY  and  micrOMEGAs codes, to evaluate the impact of these operators. 
One could then find bounds on the effective operators coefficients
 that can be translated into upper bounds on the corrections to the MSSM Higgs mass. 

\subsection{The neutralino sector.}
\begin{figure}[t!] 
\begin{tabular}{cc|cr|} 
\parbox{6.5cm}{\psfig{figure=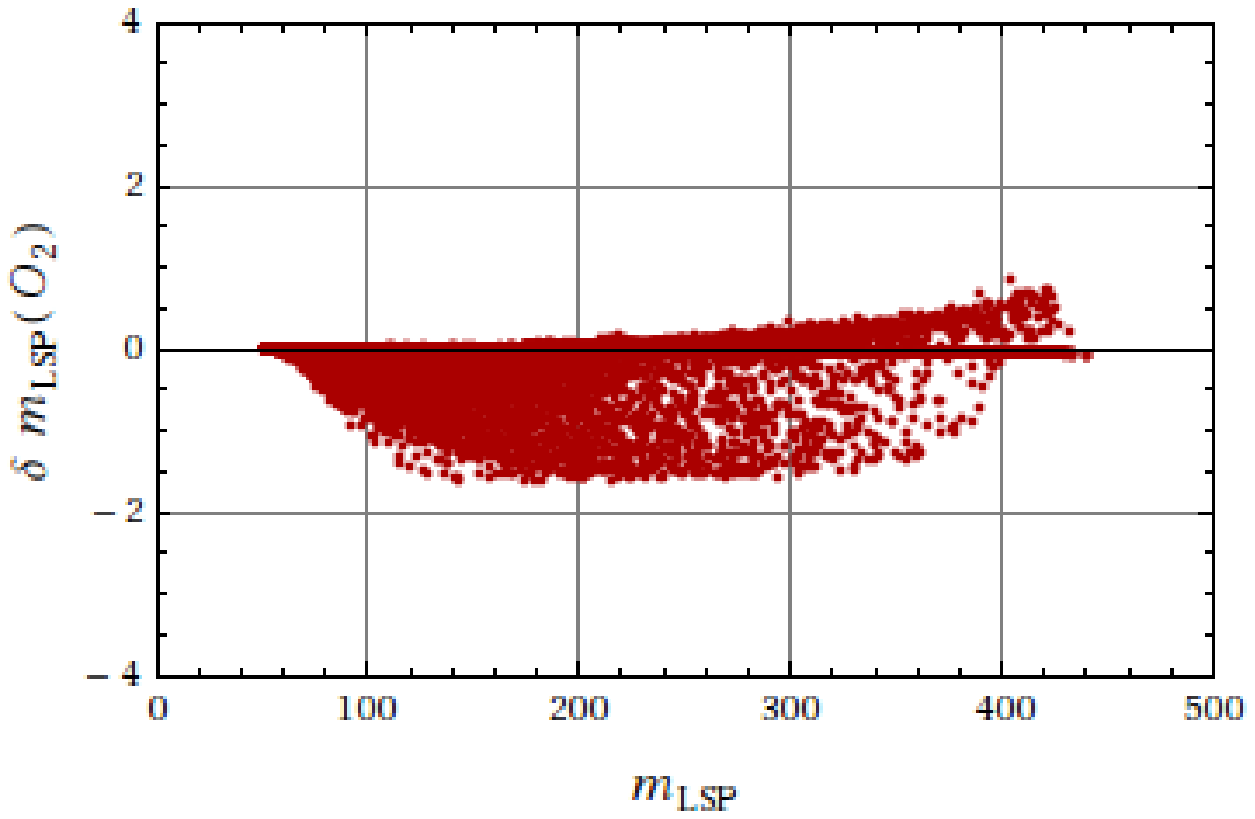,height=4.8cm,width=6.5cm}}
\hspace{1cm}  
\parbox{6.5cm}{\psfig{figure=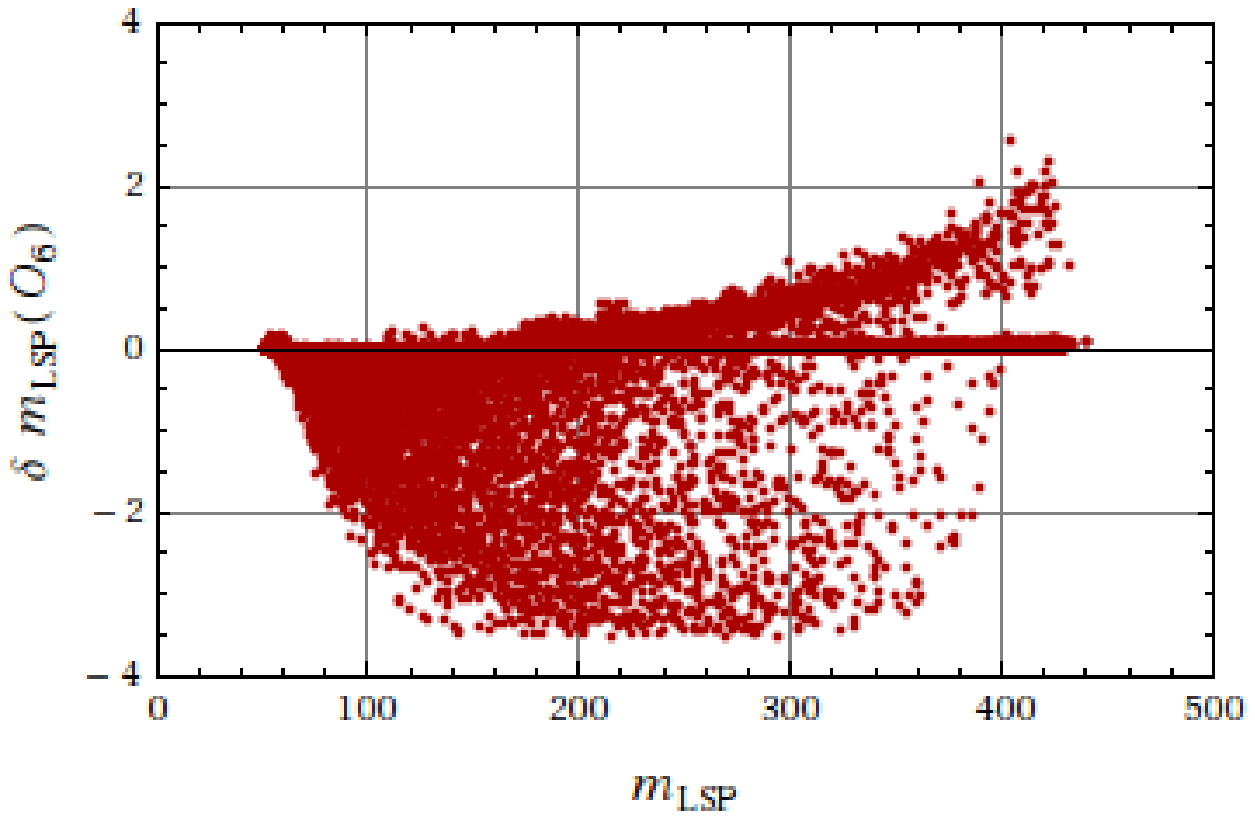,height=4.8cm,width=6.5cm}}
\end{tabular}%
\caption{
\protect\small 
The corrections to the LSP mass, induced by $O_2$ and $O_6$ respectively.
These are the $d=6$ operators that bring the largest corrections $\delta m_{LSP}$ and are 
generated for a scale of new physics of $M=8$ TeV. The corrections are for $\alpha_{j0}=\alpha_{j1}
=-1/M^2$, $j=2,6$, and include Susy breaking effects.}
\label{refmlsp}
\end{figure}

Let us now discuss the mass  corrections to the neutralino LSP. A numerical analysis of the results 
found in Section~\ref{neutralinomass} shows that the mass corrections are actually very small.
In Figure~\ref{refmlsp} we showed the mass corrections induced by operators $\cO_2$ and $\cO_6$ which 
bring the largest corrections, as a function of $m_{LSP}$ in CMSSM for $M=8$ TeV, value 
consistent with $\rho$-parameter constraints \cite{Blum:2009na}.
As shown, the mass corrections to neutralino LSP 
are of the order of few GeV only (including non-Susy corrections), 
while for the remaining operators, these corrections are even smaller. The reason 
for this is that the mass 
of the LSP is suppressed - at large $\mu$ - not only by $\alpha_{ij}$, 
but also by large $\mu$. The mass corrections increase slightly  when non-supersymmetric 
effects are also included, accounted for by $\alpha_{j1}$ in the formulae of
Section~\ref{neutralinomass}. 
Given that the correction to the LSP mass eigenvalue is so small, the LSP 
composition cannot then be significantly changed from its CMSSM value; 
for example the variation of 
$\vert\langle \chi \vert u_i\rangle\vert^2-\vert\langle \chi^0\vert u_i\rangle\vert^2$, 
where $\chi$ ($\chi^0$)  is the neutralino LSP in the presence (absence) of effective operators, and 
$u_i=
\lambda_y,\lambda_w^3,\psi_1,\psi_2$,  does not change by more than $\pm 1\%$ (for all $i$),
 for $M=8$ TeV, which  is very small. The presence of a set or all
 $d=6$ operators can however increase the overall effect on $m_{LSP}$ and LSP 
composition, but this depends on the relative signs of the operators and is not studied here.
For a detailed study of the neutralino sector in the presence of the $d=5$ operator 
see  \cite{Berg:2009mq,Bernal:2009jc,Bernal:2010uf}.

\subsection{The Higgs sector.}

Let us now discuss the corrections to the mass $m_h$ of the
lightest MSSM Higgs field. In \cite{Antoniadis:2009rn} analytical corrections 
to $m_h$ from effective operators were computed in $1/M^2$ order. This was
followed by a simple estimate of the overall size of the correction to 
tree-level $m_h$, in a very special case and under simplifying assumptions
for the coefficients of the operators.
In this section we improve the numerical analysis, to
present a general and accurate numerical investigation 
of the corrections to $m_h$  for {\it individual} operators and
including {\it quantum corrections}, not considered before.

\begin{figure}[t!]
\begin{tabular}{cc|cr|} 
\parbox{6.5cm}{\psfig{figure=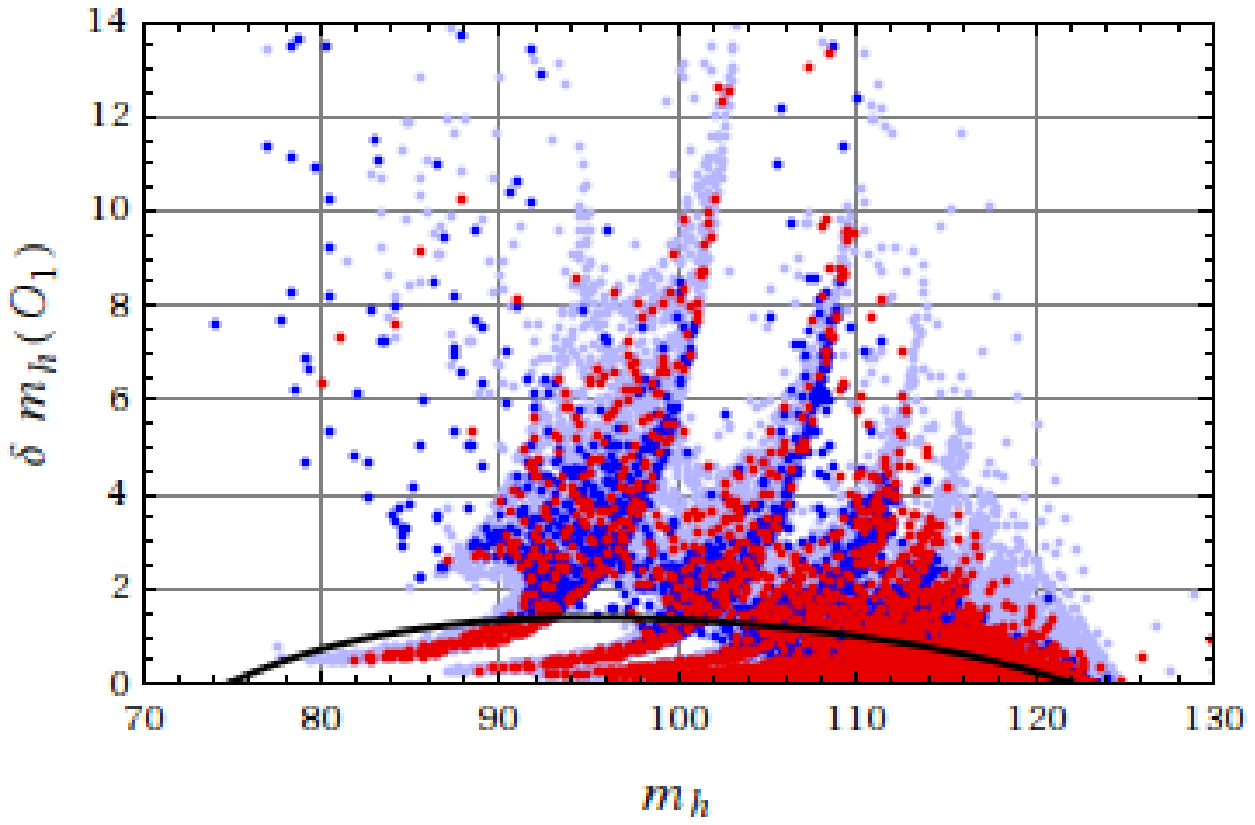, 
height=4.8cm,width=6.5cm}} \hspace{1cm}  
\parbox{6.5cm}{\psfig{figure=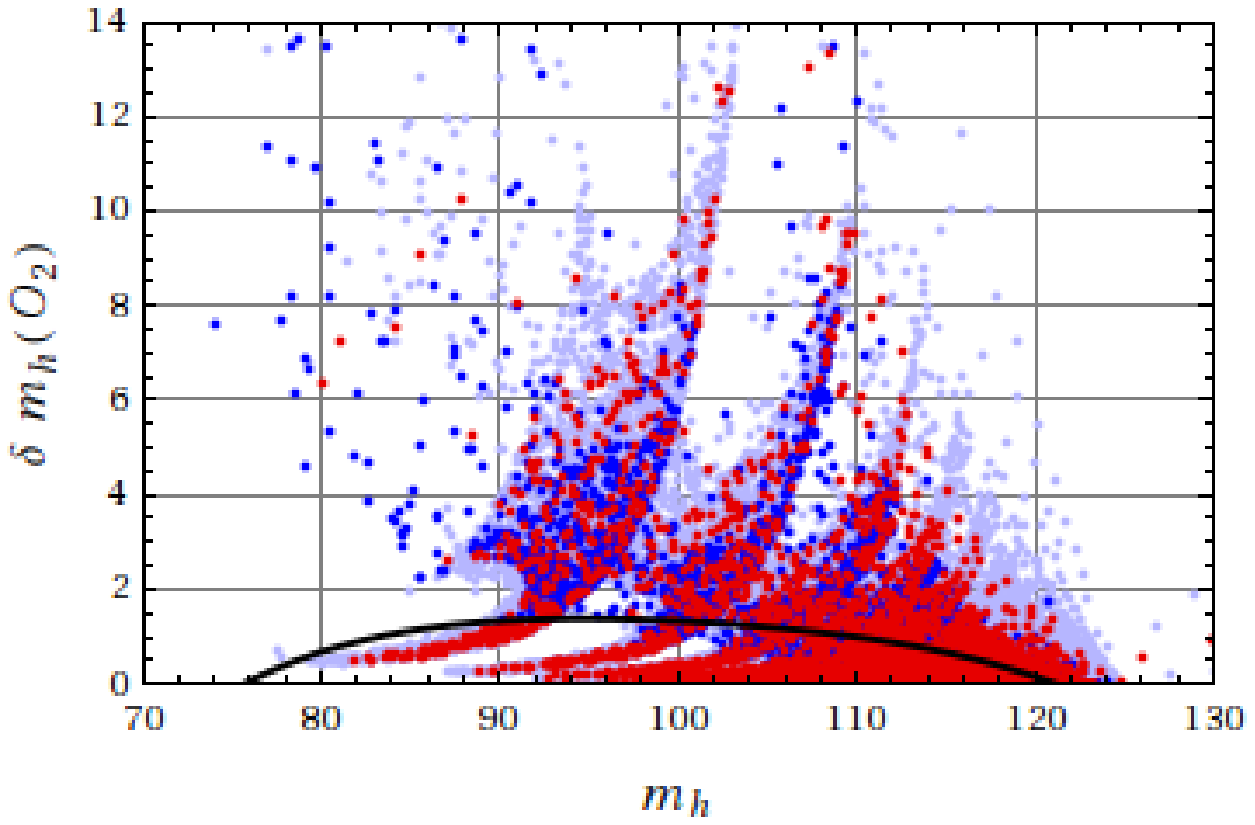, 
height=4.8cm,width=6.5cm}}
\end{tabular}%
\newline
\begin{tabular}{cc|cr|} 
\parbox{6.5cm}{\psfig{figure=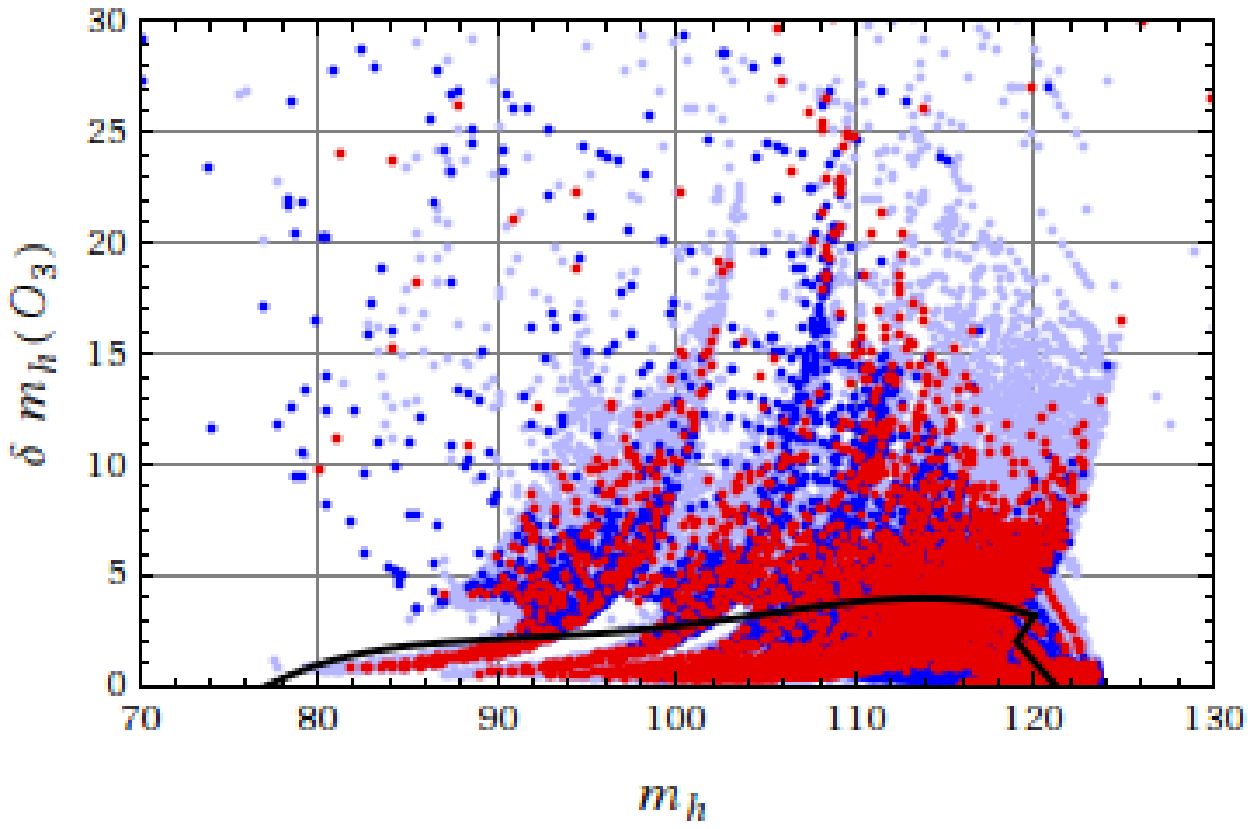,
height=4.8cm,width=6.5cm}} \hspace{1cm}  
\parbox{6.5cm}{\psfig{figure=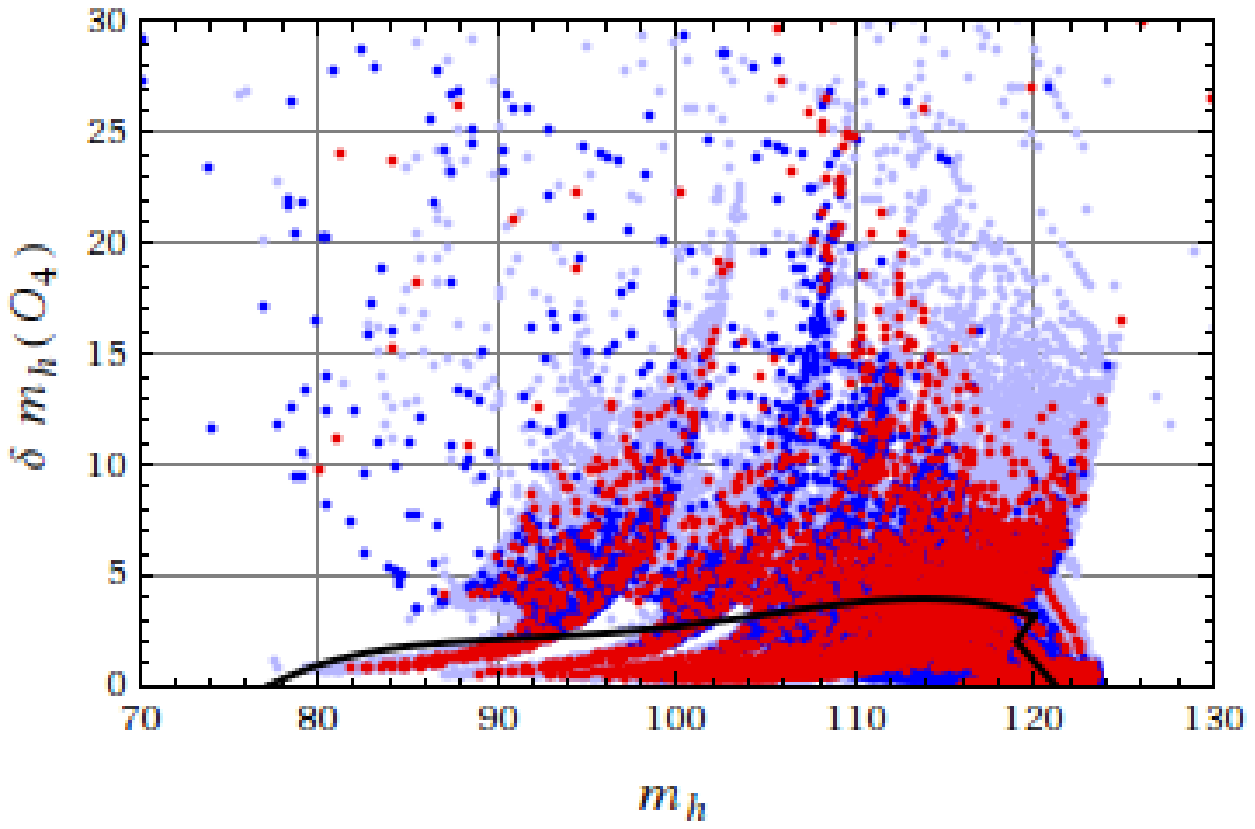,
height=4.8cm,width=6.5cm}}
\end{tabular}%
\newline
\begin{tabular}{cc|cr|} 
\parbox{6.5cm}{\psfig{figure=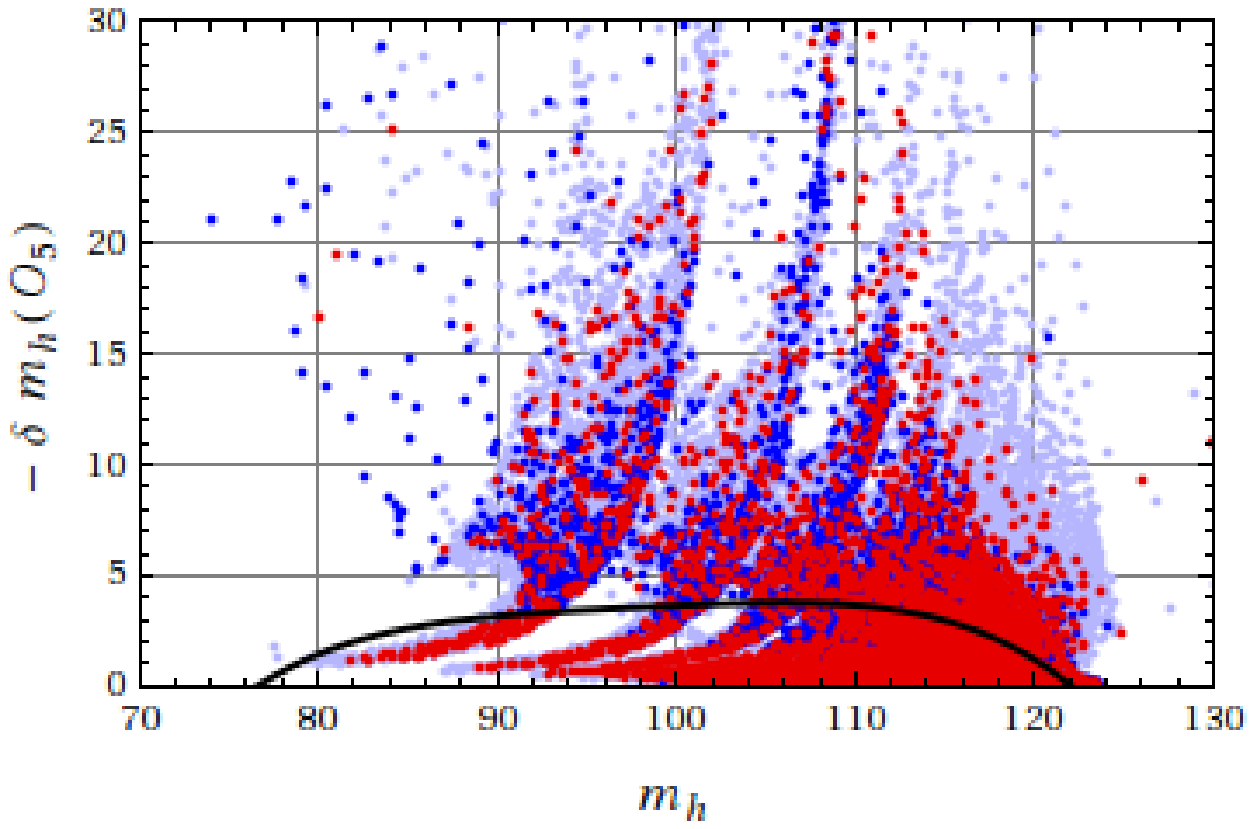,
height=4.8cm,width=6.5cm}} \hspace{1cm}  
\parbox{6.5cm}{\psfig{figure=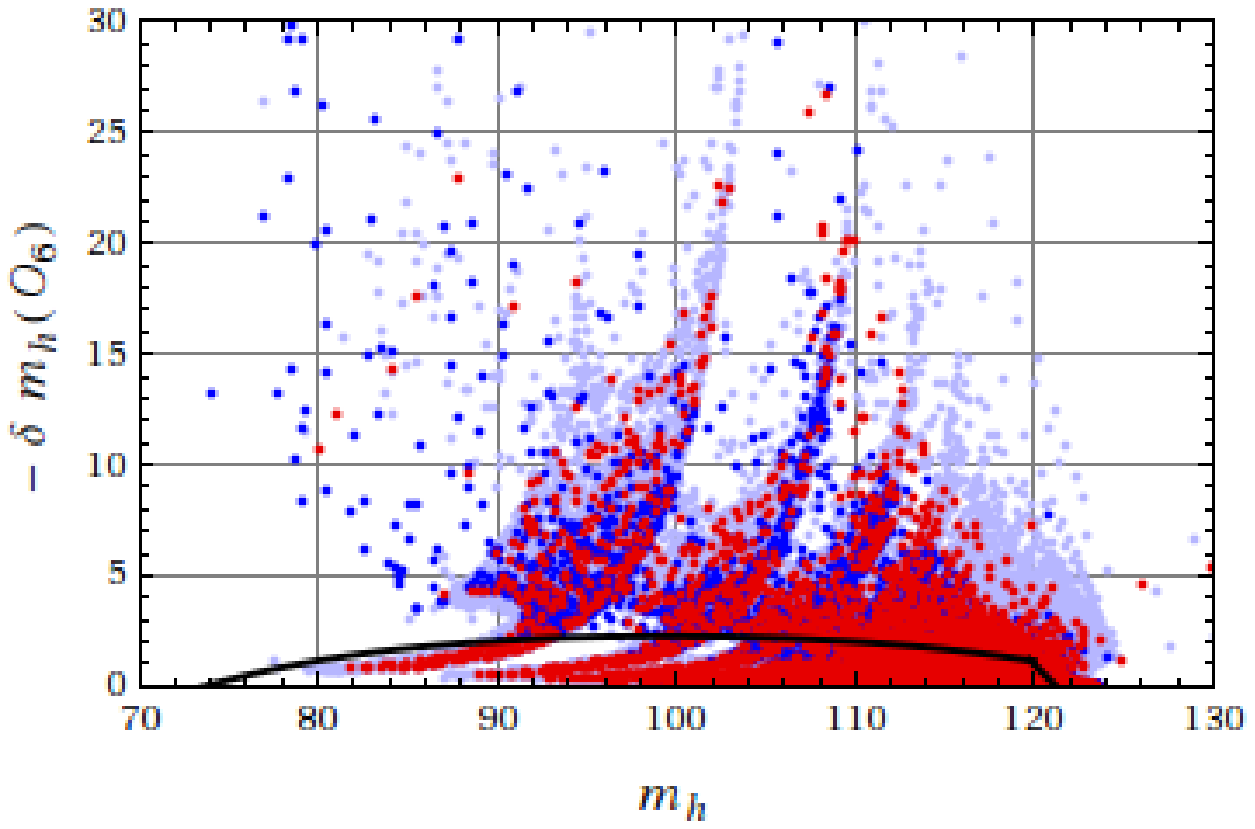,
height=4.8cm,width=6.5cm}}
\end{tabular}%
\caption{\protect\small 
The correction $\delta m_h$ to the lightest MSSM Higgs mass,
due to effective operators, as a function  of the 2-loop (LL) 
CMSSM mass $m_h$, with $M=10$ TeV.
In light blue are CMSSM phase space points with relic density 
$\Omega h^2\!\geq \! 0.1285$; on top, in dark blue, are points with 
$\Omega h^2\!\leq\! 0.0913$ 
(3$\sigma$ deviation) and on top, in red, are MSSM  points that saturate
 WMAP bound within 3$\sigma$: $\Omega h^2=0.1099\pm 0.0186$. 
(WMAP value: $\Omega h^2=0.1099\pm 0.0062$~\cite{wmap}). 
No LEP2 bound on $m_h$ is imposed at any time. The 
corrected value of the Higgs mass is $m_h+\delta m_h$.
The corrections are supersymmetric, generated by $\alpha_{j0}$,
 and can increase/decrease if  Susy-breaking  effects ($\alpha_{j1},\alpha_{j2}$) are also included. 
We assumed $\alpha_{j0}=-1/M^2$, $j=1,2,..6$ and $M=10$ TeV. The points below (above) 
the black continuous line have CMSSM  EW fine-tuning $\Delta<200$ ($\Delta>200$), 
respectively. The points below the continuous line receive a correction of up to 4 GeV.
With  $\alpha_{50}$, $\alpha_{60}<0$,  $\delta m_h(\cO_{5,6})<0$ 
 (note that $ -\delta m_h(\cO_{5,6})$ is plotted).
The gaps ("wedges") in the plots would be
filled in by a better scan of the phase space.}
\label{refmh}
\end{figure}

\begin{figure}[t!]
\begin{tabular}{cc|cr|} 
\parbox{6.5cm}{\psfig{figure=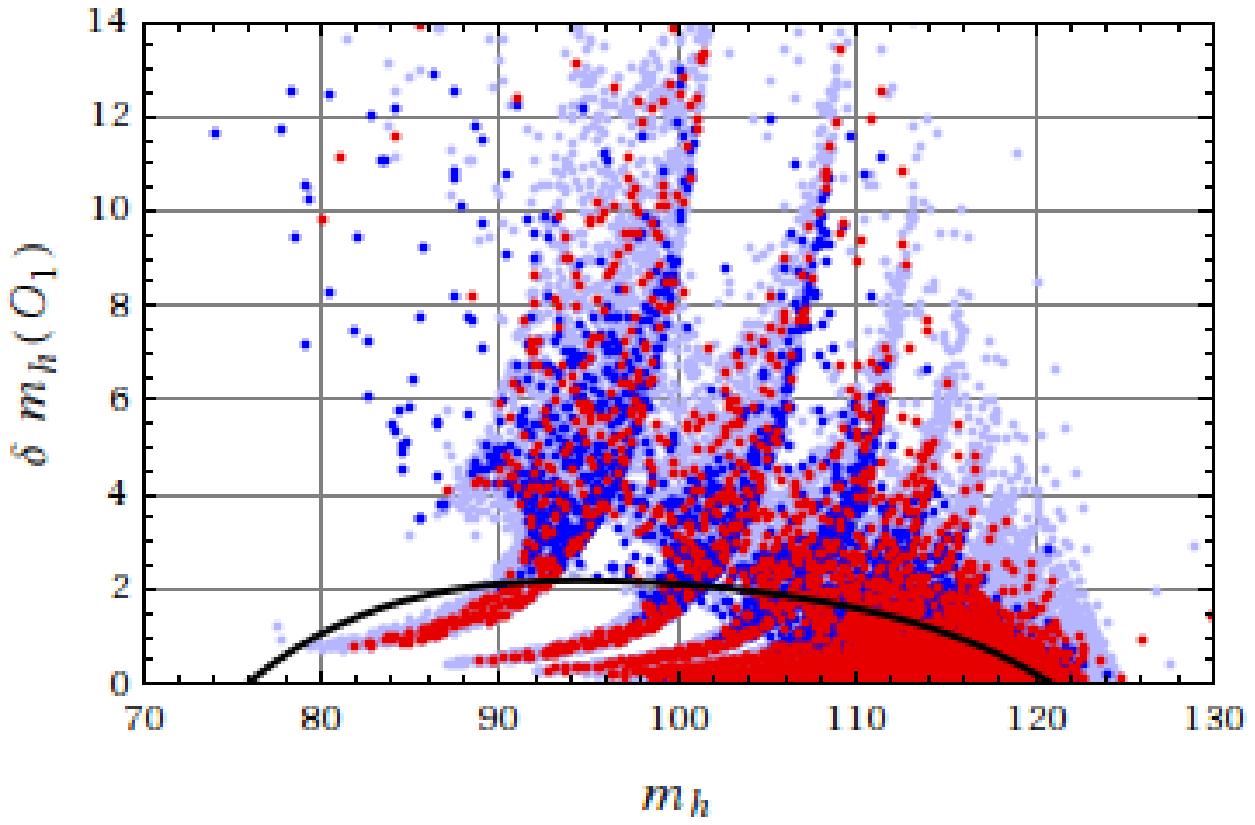, 
height=4.8cm,width=6.5cm}} \hspace{1cm}  
\parbox{6.5cm}{\psfig{figure=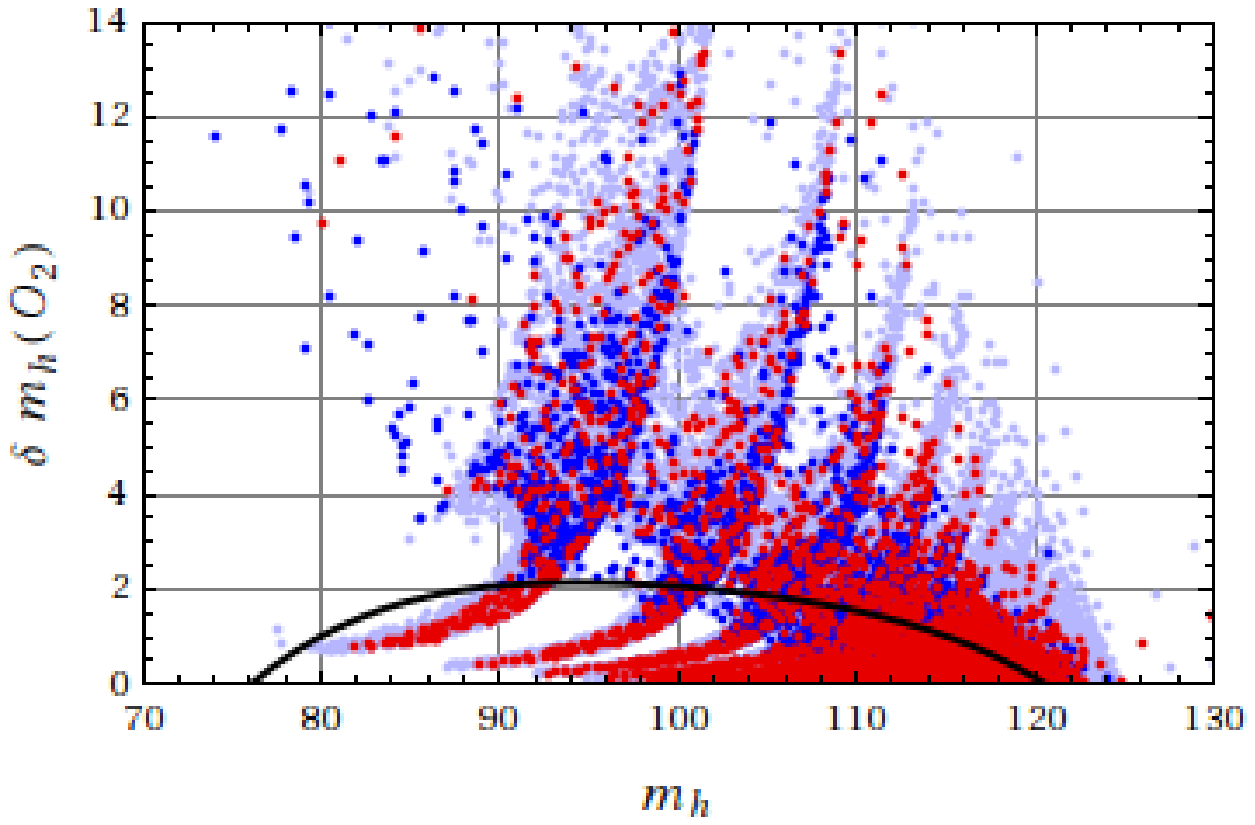, 
height=4.8cm,width=6.5cm}}
\end{tabular}%
\newline
\begin{tabular}{cc|cr|} 
\parbox{6.5cm}{\psfig{figure=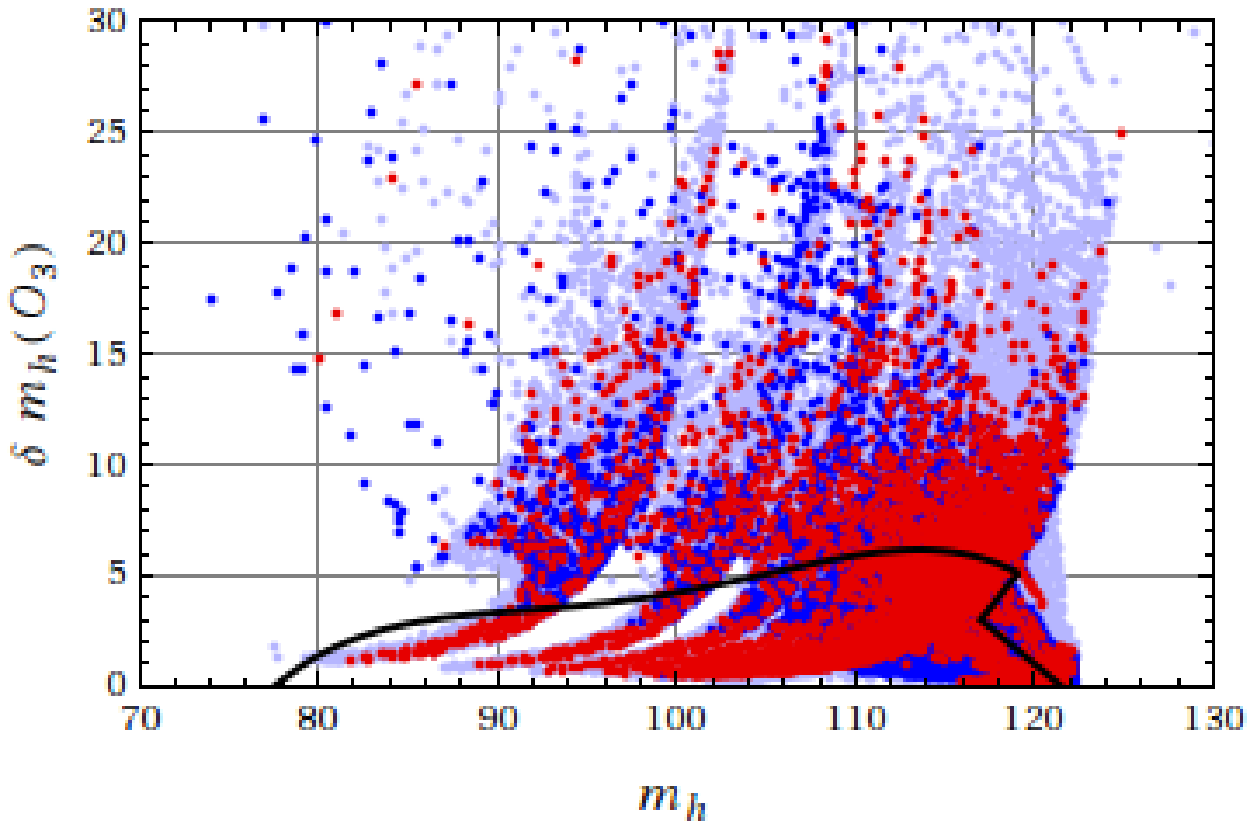,
height=4.8cm,width=6.5cm}} \hspace{1cm}  
\parbox{6.5cm}{\psfig{figure=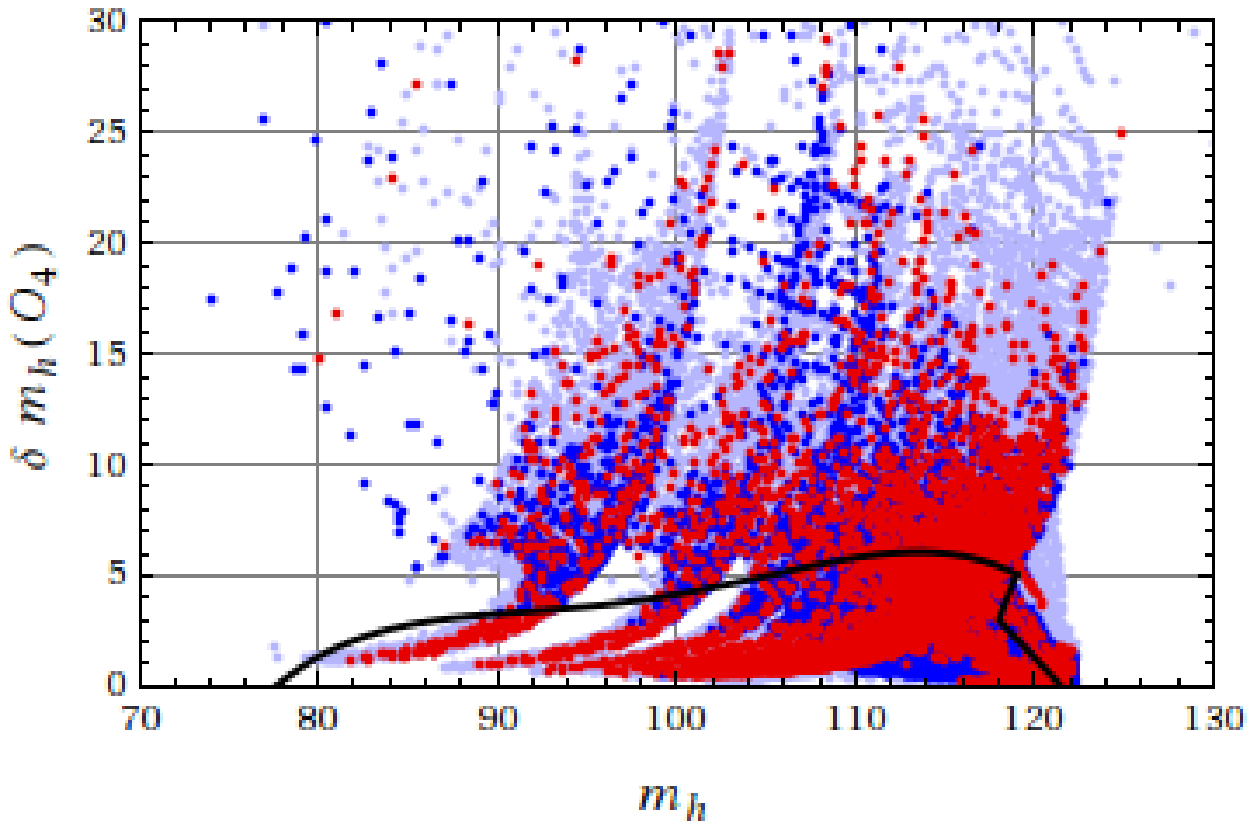,
height=4.8cm,width=6.5cm}}
\end{tabular}%
\newline
\begin{tabular}{cc|cr|} 
\parbox{6.5cm}{\psfig{figure=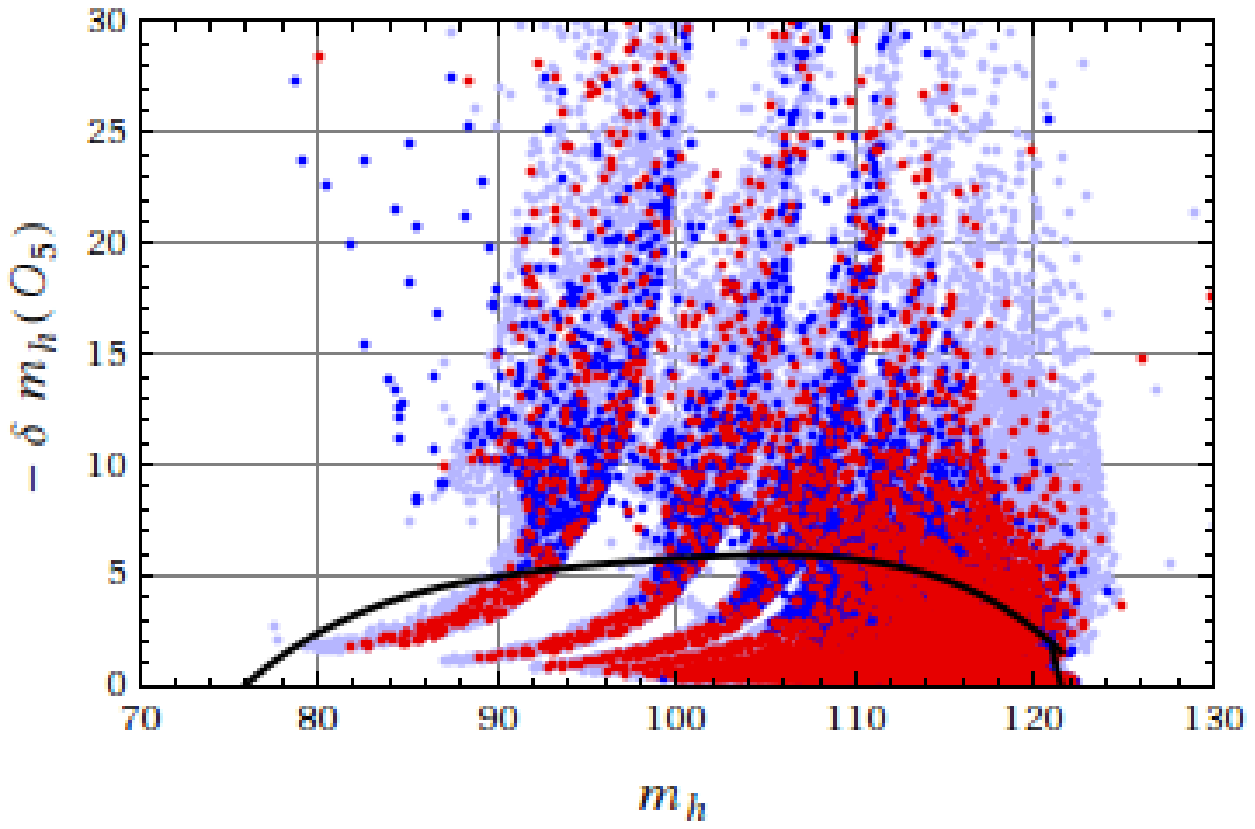,
height=4.8cm,width=6.5cm}} \hspace{1cm}  
\parbox{6.5cm}{\psfig{figure=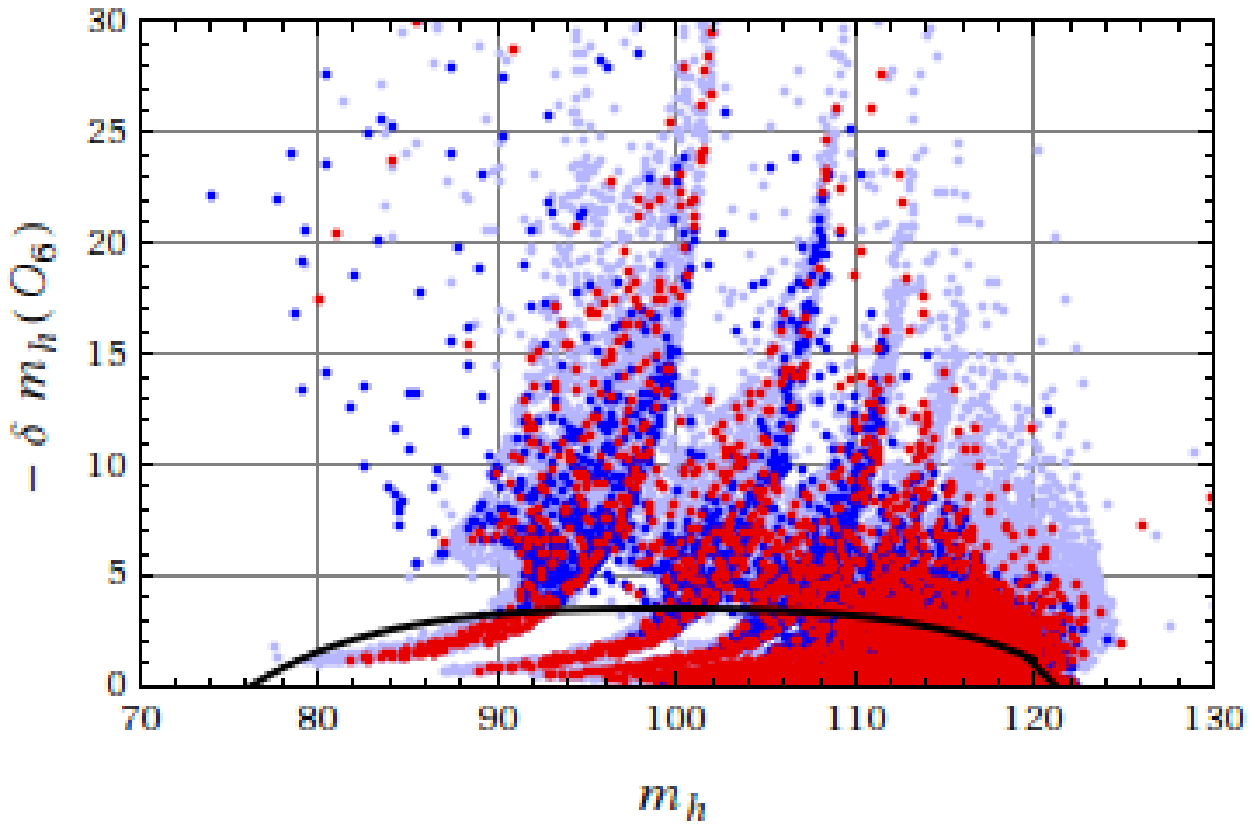,
height=4.8cm,width=6.5cm}}
\end{tabular}%
\caption{\protect\small As for Figure~\ref{refmh}, but with M=8 TeV.
The continuous line of $\Delta=200$ has changed position
 and points under it can bring a $\delta m_h$ up to 6 GeV.}
\label{refmhprime}
\end{figure}

\begin{figure}[t]
\begin{tabular}{cc|cr|} 
\parbox{6.5cm}{\psfig{figure=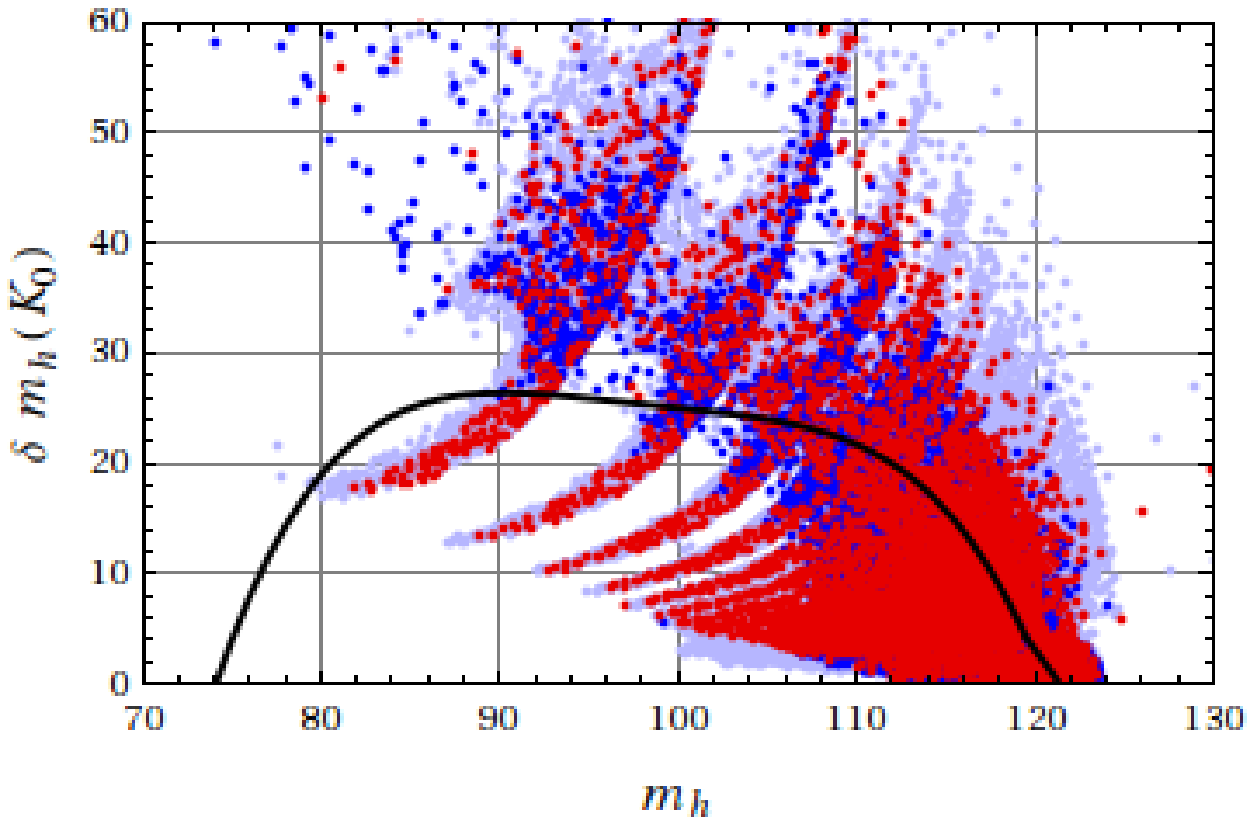,
height=4.8cm,width=6.5cm}} \hspace{1cm}  
\parbox{6.5cm}{\psfig{figure=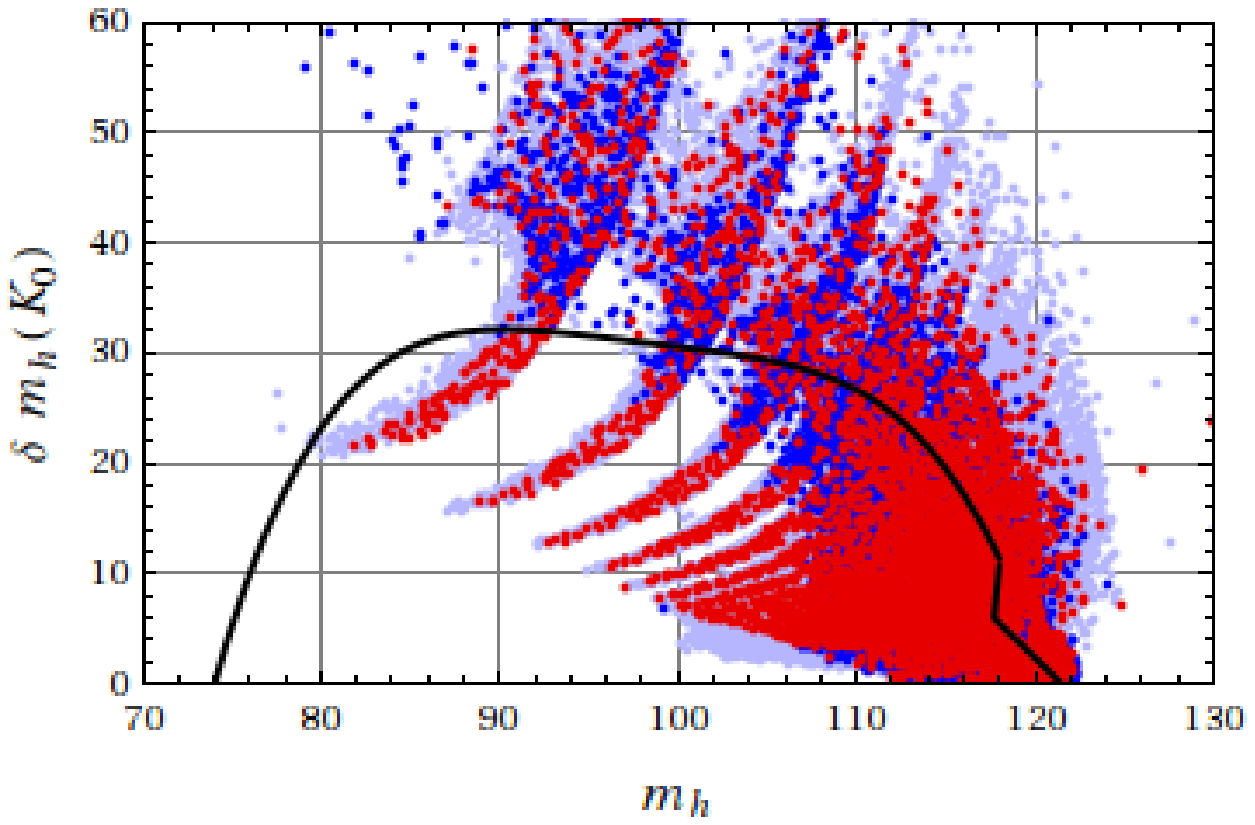,
height=4.8cm,width=6.5cm}}
\end{tabular}%
\caption{\protect\small 
As for Figure~\ref{refmh}, with the correction to the MSSM lightest Higgs mass induced by $\cK_0$,
in function of 2-loop MSSM Higgs mass. Left plot: $M=10$ TeV, right plot: $M=8$ TeV. The corrections
from $d=5$ operators are now larger, due to leading $1/M$ terms present for the $d=5$ operator.}
\label{oz2}
\end{figure}

The results are illustrated in the plots of Figures~\ref{refmh},\ref{refmhprime},\ref{oz2}
where the {\it supersymmetric} correction  $\delta m_h$ is 
shown for each operator $\cO_{1,...6}$ and $\cK_0$ as a function of the 
 CMSSM value for $m_h$ evaluated at 2-loop leading-log (LL), for $M=10$ and $8$ TeV
(values consistent with $\rho$-parameter constraints \cite{Blum:2009na}).
The correction due to $\cO_7$ is very small ($<\!1$GeV)
for all  the parameter space, because it is strongly  suppressed by the small gauge couplings, 
in addition to $\alpha_{ij}$, and it is not shown here. 
The correction $\delta m_h$ shown in the plots as a function of CMSSM value of $m_h$, 
is  defined as
\bea
\delta m_h=\Big[
m_h^2\big\vert_{\rm 2-loop,{\rm MSSM}}+\delta m_h^2\Big]^{1/2}
-m_h\big\vert_{\rm 2-loop,{\rm MSSM}}
=\frac{1}{2} \frac{\delta m_h^2}{m_h\big\vert_{\rm 2-loop,{\rm MSSM}}}+\cO(1/M^4)
\eea
so the total value is then $\delta m_h+m_h\vert_{\rm 2-loop,MSSM}$;
here $m_h\vert_{\rm 2-loop,MSSM}$ is the 2-loop (LL) 
corrected CMSSM value for Higgs mass, while $\delta 
m_h^2$ is the classical correction due to the 
effective operators whose exact expression can be found, 
for exact Susy case, in eq.(36) in \cite{Antoniadis:2009rn}. The large $\tan\beta$ limit of 
$\delta m_h^2$ is given in (\ref{dd1}), including Susy breaking effects; 
 as it can be seen from there,  negative $\alpha_{30}$ and
 $\alpha_{40}$ can bring a positive correction $\delta m_h$,
 and this remains true for all $\tan\beta$ as seen 
in the plots in Figure~\ref{refmh},\,\ref{refmhprime}; 
for $\alpha_{j0}$, $j=1,2,5,6$ the sign of the correction is not clear from (\ref{dd1}).
Finally, in all  plots the points below the 
black continuous line are the CMSSM points with EW 
fine tuning $\Delta<200$, and satisfy all experimental and theoretical  
constraints as explained above, including WMAP constraint within 3$\sigma$  
(in red) or consistent with it (in blue), 
except the LEP2 bound on $m_h$  which is never imposed, 
for reasons that become clear below.

Let us first discuss the correction for points with $\Delta<200$.
From these plots we notice that
CMSSM ("best fit") points below the black continuous line, i.e. 
which respect all constraints  mentioned above plus fine 
tuning $\Delta<200$
and regardless of the LEP2 bound on $m_h$, receive from individual operators
a small change to the Higgs mass $m_h$,  of only few GeV: 
up to 4 GeV for $M=10$ TeV and up to $6$ GeV for $M=8$ TeV. 
This indicates a variation of $\delta m_h$ by about 1 GeV for a 
1 TeV variation of $M$.
These numerical values are even smaller for some operators,
see Figures~\ref{refmh},\ref{refmhprime}.
Note that points which were below the LEP2 bound by this correction 
are now phenomenologically viable. The special 
point of CMSSM of minimal $\Delta=18$  that saturates the 
dark matter relic density  \cite{wmap} 
within  3$\sigma$, and  with $m_h=115.9\pm 2$ GeV \cite{Cassel:2009cx},
could therefore receive a correction $\delta m_h\sim 4$ to $6 $ GeV, so that
$m_h$ can increase to $m_h+\delta m_h=(120-122)\pm 2$ GeV.
Given the relatively small size of the correction $\delta m_h$ 
one can say that these particular CMSSM phase space points and their predictions 
are stable against  the presence of new physics at the scale $M=10$ TeV or $M=8$ TeV.
This is an interesting finding, and can be explained by the fact that
these points generically have a 
light $\mu$ and light $m_{12}$ (focus point region) \cite{Cassel:2009cx},
and thus the supersymmetric corrections $\delta m_h$, 
(generated by $\alpha_{j0}$) are rather suppressed.
 
The corrections $\delta m_h$ can increase or decrease 
if one also includes effects of Susy breaking 
associated with the effective operators and encoded  in $\alpha_{j1}, \alpha_{j2}$, 
by an amount comparable to that 
due to their supersymmetric corrections; for large $\tan\beta$ the size of their effects 
can also be seen from eq.(\ref{dd1}). However the relevance of such corrections for the little
hierarchy problem and for $m_h$ value is questionable, given that these are themselves 
related to supersymmetry breaking. We therefore do not consider such effects further.
Finally, unlike  operators $\cO_i$, in the case of the dimension-five operator, the MSSM 
phase space points of $\Delta<200$ that violate the LEP2 bound
can bring a correction $\delta m_h(\cK_0)$ significantly larger, and these points become 
phenomenologically viable (see Figure~\ref{oz2}, with 
$\delta m_h\sim 20$ GeV for $m_h\sim 110$ GeV, still under 
the black curve, i.e. $\Delta<200$); however, the  final,
 corrected value $m_h+\delta m_h$, while situated above the LEP2 bound now, 
 is still below 125-135 GeV.

Based on our previous results for the corrections to the neutralino LSP mass,
 which turned out to be  significantly smaller, we can say that these "best fit" 
 CMSSM  phase space points ($\Delta<200$) are unlikely to have their
 dark matter constraint changed significantly, and are then rather
stable under "new physics" presence.
This is also supported by the fact that dark matter 
abundance, that depends on the annihilation cross section may not receive large 
corrections since the change of the  LSP composition due to $\cO_i$ 
was small (consistent with a small mass correction). 
However, only a careful implementation of the new couplings in the neutralino sector
into micrOMEGAs and SOFTSUSY can address this issue on  solid grounds. 
Note that to such cross section 
effects all operators $\cO_i$ contribute: some like $\cO_7$ provide a direct
LSP annihilation coupling of the bino ($\propto 1/M^2$) 
but give an otherwise  negligible correction to $m_h$,  while the remaining operators
induce a similar order effect for the LSP, via $\cO(1/M^2)$ mixing
with the MSSM terms.

Let us now discuss the CMSSM points with fine tuning $\Delta>200$ 
i.e. situated above the black continuous line in Figure~\ref{refmh},\ref{refmhprime},\ref{oz2}.
They can bring an increase of $m_h$ which can be 
significant, of 10-30 GeV (larger for $\cK_0$),
 but this depends also on $M$. 
Therefore,  points that in 
the MSSM would be eliminated by the LEP2 mass bound for Higgs  $m_h>114.4$ GeV, 
can now be ruled in as viable points.  For example there are points which for $m_h$ near 
100 GeV can receive corrections of order 20 GeV or so, to now reach and satisfy the LEP2 bound. 
Interestingly, for $\cO_{1,2}$ the Higgs mass increase is such that total
$m_h$ remains close to 120 GeV. In any case, only points that are largely fine 
tuned and have a value for 
$m_h$ significantly below LEP2 bound, are actually receiving the largest corrections to $m_h$.
Thus the phase space of the MSSM is increased and more points which are 
otherwise ruled out on grounds of 
extreme fine tuning and/or LEP2 bound, are "recovered" and can be phenomenologically viable.
The EW fine tuning $\Delta$ of those points can decrease
 by a factor equal to the square of the ratio of the Higgs masses after 
and before adding the correction $\delta m_h$, and this effect can be significant.
For an example of how this works in the presence of the $d=5$
 operator $\cK_0$, see \cite{Cassel:2009ps}, where one sees that $\Delta$ can 
remain acceptable ($\sim 10$), in the presence of $\cK_0$ even for $m_h$ above 120 GeV. 
A similar effect is expected for the case of $d=6$ operators.

\section{Conclusions}

In this paper we considered the extension of the MSSM Higgs 
sector by all possible effective 
operators of dimension-five and dimension-six, allowed by
 the MSSM symmetries. By supersymmetry,  
the same operators 
also provide the most general extension of the neutralino
 and chargino sectors of the MSSM.
The study of such extensions is motivated by the attempts
 to understand better the MSSM higgs sector 
and its stability against corrections from new physics, 
as well as by dark matter studies. 
This is also motivated by the fact that dark matter and
 higgs sectors are intrinsically connected
by supersymmetry.  Complementary constraints from dark
 matter (large length scale physics) 
and electroweak physics (small length scales) can shed
 more light on either of these sectors or on 
both. In this paper we started an analysis in this direction,
 by computing for the first time, in 
component fields,  the Lagrangian in the neutralino and chargino
sectors extended by all effective  operators of dimension $d=5$ and $d=6$,
as well as the corresponding spectrum. The results can be used for studies of
dark matter relic density within extensions of the CMSSM, by implementing this
extended Lagrangian in public codes like micrOMEGAs.
The study also continued our earlier similar calculation of the extended
Lagrangian in the Higgs sector alone. 
The phenomenological impact of the effective operators was then 
studied by analyzing the impact of these operators on the 
CMSSM parameter space, as a perturbation.

We computed the mass corrections to the neutralino and 
chargino fields  and showed that the neutralino LSP 
receives small mass corrections from individual effective operators, 
of few GeV ($<1-2\%$) for a scale of 
the operators at 8 TeV; the sign of the corrections depends on the 
choice for the coefficients for these 
operators. The operators with the largest corrections were identified 
to be $\cO_{2,6}$.

A similar study was done for the Higgs sector, and this continued 
the analysis started in our previous 
work \cite{Antoniadis:2009rn}, where the classical correction to the mass of the
lightest MSSM higgs had been  computed analytically, in the leading order ($1/M^2$).
Using this analytical 
result we performed an accurate numerical investigation of  the size of the correction 
to $m_h$ from individual operators, including quantum corrections, not considered before. 
Using the CMSSM parameter space 
points that satisfied all electroweak and dark matter constraints, except 
the LEP2 bound on $m_h$, we showed that points which would otherwise violate the LEP2 bound or are strongly fine tuned in CMSSM, become viable points, now respect this bound and have a lower fine-tuning.
For such points, with $\Delta>200$,  
the effective operators can bring {\it individual} corrections of $\delta m_h\sim 10-30$ GeV 
with the larger values for $m_h$ further below LEP2 bound, to increase $m_h$ 
just above this bound. Non-Susy effects associated with the effective 
operators can increase or decrease this correction.
The properties of these phase space points need to be analyzed further 
in a global fit of the whole model i.e. MSSM plus effective operators.

An interesting result is that  for the CMSSM phase space points with reduced EW 
fine tuning, $\Delta<200$, and that satisfied the WMAP constraint within 3$\sigma$ or were just 
consistent with this bound. In this case, the {\it supersymmetric} corrections to  
$m_h$ from {\it individual} operators of dimension $d=6$ were small: they were
$<4$ ($<6$) GeV for $M=10$ $(8)$ TeV, respectively (with about a variation of 1 GeV 
for a change of 1 TeV of the scale $M$).
The points below the LEP2 bound by this amount but respecting all 
other experimental and theoretical constraints, become now phenomenologically viable.
The relative smallness of these corrections (for individual operators), suggests that
the CMSSM "best fit" points are rather stable against the effects of "new physics" 
in the Higgs sector that could exist at $10$ ($8$) TeV.
In particular, for the CMSSM point with lowest EW fine tuning ($\Delta=18$)
that saturates the relic density within  3$\sigma$ and predicting $m_h=115.9\pm2$ GeV, 
and considering only corrections from individual operators, 
would bring this value to $m_h=(120-122)\pm 2$ GeV. This could suggest a preference 
for a light $m_h$ even in the presence of "new physics" at $8-10$ TeV.

\section*{Acknowledgments}

This work was supported in part by 
the European Commission under contracts PITN-GA-2009-237920 and ERC Advanced Grant
226371 (``MassTeV''), by the INTAS grant 03-51-6346, by the ANR (CNRS-USAR)
contract 05-BLAN-007901 and by the CNRS PICS 3747 and 4172.
The work of D.G. was supported in part by CNRS research contract 225933 
held at Ecole Polytechnique Paris and ERC Advanced Grant 226371 (``MassTeV'').
We thank Y.~Mambrini (Orsay) for many interesting discussions, and
 S.~Cassel (Oxford) for helping us with an updated set of phase space points 
of the CMSSM, at the 2-loop level, that was used in our numerical estimates.

\newpage
\section{Appendix}

\def\theequation{A-\arabic{equation}}
\def\thesubsection{A}
\setcounter{equation}{0}
\subsection{The expressions of the $d=6$ operators and the auxiliary fields.}
\label{appendixA}

The full component form of the dimension-six operators is
\begin{eqnarray} 
\mathcal{O}_{1} &=&
\frac{1}{M^{2}}\int d^{4}\theta \,\,
\mathcal{Z}_{1}(S,S^{\dagger })\,\,
(H_{1}^{\dagger }\,e^{V_{1}}\,H_{1})^{2} \\
&=& 
2 \alpha_{1 0}\,
\Big[
\vert h_1\vert^2 \,\big[\,\vert\cD_\mu h_1\vert^2
+h_1^\dagger\,\frac{D_1}{2}\,h_1 + \vert F_1\vert^2\big]+
\vert\,h_1^\dagger F_1\vert^2 
+
\vert h_1^\dagger  \cD^\mu h_1\vert^2
\Big]
\nonumber\\
&+&
2 \alpha_{10}\,\Big[
\frac{i}{2}\, \overline\psi_1\overline\sigma^\mu\cD_\mu \psi_1\,
\vert h_1\vert^2
+\frac{i}{2}\,\overline\psi_1\overline\sigma^\mu\, 
\psi_1 \,h_1^\dagger\cD_\mu h_1
-\frac{i}{2}\, (h_1^\dagger\psi_1)\, \sigma^\mu\overline\psi_1
(\cD_\mu-\overleftarrow \cD_\mu)\,h_1+h.c.\Big]
\nonumber\\
&+& 2 \alpha_{10}
\Big[-
\frac{1}{\sqrt 2}\, (h_1^\dagger  \lambda_1\psi_1)\vert h_1\vert^2
-(h_1^\dagger\psi_1) (F_1^\dagger \psi_1)
-\frac{1}{\sqrt 2}\, (h_1^\dagger\psi_1)\,
h_1^\dagger \lambda_1 h_1+h.c.\Big]
\nonumber\\
&-&
\!\!\! \alpha_{10}
(\overline\psi_1\psi_1)(\overline\psi_1\psi_1)
\!+\!\Big[2\alpha_{1 1} m_0 \vert h_1\vert^2\,(F_1^\dagger h_1)
\!-\alpha_{11} m_0 (\overline\psi_1 h_1)(\overline\psi_1 h_1)
\!+\!h.c.\Big]
\!+\!\alpha_{12}\, m_0^2 (\vert h_1\vert^2)^2\nonumber
%
\eea
\bea
\mathcal{O}_{2} 
&=&\frac{1}{M^{2}}\int d^{4}\theta \,\,
\mathcal{Z}_{2}(S,S^{\dagger })\,\,
(H_{2}^{\dagger }\,e^{V_{2}}\,H_{2})^{2} 
\\
&=& 
2 \alpha_{2 0}\,\Big[\vert h_2\vert^2\,\big[\,
\vert \cD_\mu h_2\vert^2
+h_2^\dagger\,\frac{D_2}{2}\,h_2 + \vert F_2\vert^2\,\big]+
\vert  h_2^\dagger F_2\vert^2 
+
\vert h_2^\dagger  \cD^\mu h_2\vert^2
\Big]\nonumber\\
&+&
2\alpha_{20}\,\Big[
\frac{i}{2}\, \overline\psi_2\overline\sigma^\mu\cD_\mu \psi_2\,
\vert h_2\vert^2
+\frac{i}{2}\,\overline\psi_2\overline\sigma^\mu\, 
\psi_2 \,h_2^\dagger\cD_\mu h_2
-\frac{i}{2}\, (h_2^\dagger\psi_2)\, \sigma^\mu\overline\psi_2
(\cD_\mu-\overleftarrow \cD_\mu)\,h_2+h.c.\Big]
\nonumber\\
&+& 2 \alpha_{20}\Big[-
\frac{1}{\sqrt 2}\, (h_2^\dagger  \lambda_2\psi_2)\vert h_2\vert^2
-(h_2^\dagger\psi_2) (F_2^\dagger \psi_2)
-\frac{1}{\sqrt 2}\, (h_2^\dagger\psi_2)\,
h_2^\dagger \lambda_2 h_2+h.c.\Big]
\nonumber\\
&-&\!\!\! \alpha_{20}(\overline\psi_2\psi_2)(\overline\psi_2\psi_2)
\!+\!\Big[2 \alpha_{2 1} m_0  
\vert h_2\vert^2 (F_2^\dagger h_2)
\!-\alpha_{21} m_0 (\overline\psi_2 h_2)(\overline\psi_2 h_2)
\!+\!h.c.\Big]\! +\!\alpha_{22} m_0^2 (\vert h_2\vert^2)^2\nonumber
\eea
\bea
\mathcal{O}_{3} &=&
\frac{1}{M^{2}}\int d^{4}\theta \,\,\mathcal{Z} _{3}(S,S^{\dagger
})\,\,
(H_{1}^{\dagger 
}\,e^{V_{1}}\,H_{1})\,(H_{2}^{\dagger }\,e^{V_{2}}\,H_{2}),
\\
&=&\!\!\!
\alpha_{30}\,\Big[
 \vert h_1\vert^2\,
\big[\vert \cD_\mu h_2\vert^2
+h_2^\dagger\,\frac{D_2}{2}\,h_2 \!+\! \vert F_2\vert^2 \big]
\!+\! (h_1^\dagger  F_1)(F_2^\dagger h_2)
\!+\!(h_1^\dagger \cD_\mu h_1)(h_2^\dagger\overleftarrow \cD^\mu h_2)
\!+\!(1\leftrightarrow 2)\Big]
\nonumber\\
&+&\!\!\alpha_{30}
\Big[
\frac{i}{2}\, \overline\psi_2\overline\sigma^\mu\cD_\mu \psi_2\,
\vert h_1\vert^2
+\frac{i}{2}\,\overline\psi_1\overline\sigma^\mu\, 
\psi_1 \,h_2^\dagger\cD_\mu h_2
-\frac{i}{2}\, (h_1^\dagger\psi_1)\, \sigma^\mu\overline\psi_2
(\cD_\mu-\overleftarrow \cD_\mu)\,h_2+h.c.\Big]
\nonumber\\
&+&\!\!\alpha_{30}
\Big[
\frac{i}{2}\, \overline\psi_1\overline\sigma^\mu\cD_\mu \psi_1\,
\vert h_2\vert^2
+\frac{i}{2}\,\overline\psi_2\overline\sigma^\mu\, 
\psi_2 \,h_1^\dagger\cD_\mu h_1
-\frac{i}{2}\, (h_2^\dagger\psi_2)\, \sigma^\mu\overline\psi_1
(\cD_\mu-\overleftarrow \cD_\mu)\,h_1+h.c.\Big]
\nonumber\\
&+&
\alpha_{30}
\Big[-
\frac{1}{\sqrt 2}\, (h_2^\dagger  \lambda_2\psi_2)\vert h_1\vert^2
-(h_1^\dagger\psi_1) (F_2^\dagger \psi_2)
-\frac{1}{\sqrt 2}\, (h_1^\dagger\psi_1)\,
h_2^\dagger \lambda_2 h_2+(1 \leftrightarrow 2)+h.c.\Big]
\nonumber\\
&-&\!\!\!
\alpha_{30}\,(\overline\psi_1\psi_1)(\overline\psi_2\psi_2)
+\!\!\Big[
\alpha_{31}\,m_0\, \big[\,
\vert h_1\vert^2 (F_2^\dagger h_2)
+\vert h_2\vert^2 (F_1^\dagger h_1)
\big]
-\alpha_{31}^* m_0 ( h_1^\dagger\psi_1)(h_2^\dagger\psi_2)
+h.c.\Big]
\nonumber\\[2pt]
&+&
\alpha_{32}\,m_0^2\,\vert h_1\vert^2 \vert h_2\vert^2
\nonumber
\eea
\bea
\mathcal{O}_{4} &=&\frac{1}{M^{2}}\int d^{4}\theta \,\,
\mathcal{Z}_{4}(S,S^{\dagger })
\,\,(H_{2}\,.\,H_{1})\,(H_{2}\,.\,H_{1})^{\dagger },
\\
&=&
\alpha_{40}\,
\vert \partial_\mu (h_2.h_1)\vert^2
+
\alpha_{40}\Big[
\frac{i}{2}\,(\psi_1.h_2+h_1.\psi_2)\,\sigma^\mu\partial_\mu
(\psi_1.h_2+h_1.\psi_2)^\dagger+h.c.\Big]
\nonumber\\
&+&
\alpha_{40}\,\vert h_2\cdot F_1+F_2\cdot h_1-\psi_2.\psi_1\vert^2
+\Big[\alpha_{41}\,m_0\,(h_2.h_1)\,(h_2.F_1
  +F_2.h_1-\psi_2.\psi_1)^\dagger
 +h.c.\Big]
\nonumber\\
&+&
\alpha_{42}\,m_0^2\,\vert h_2.h_1\vert^2 
\nonumber\eea
\bea
\mathcal{O}_{5} &=&\frac{1}{M^{2}}\int d^{4}\theta \,\,\mathcal{Z}%
_{5}(S,S^{\dagger })\,\,(H_{1}^{\dagger 
}\,e^{V_{1}}\,H_{1})\,H_{2}.\,H_{1}+h.c.\\
&=&
\alpha_{50}
\Big[\big[\,
\vert \cD_\mu h_1\vert^2
+h_1^\dagger\,\frac{D_1}{2}\,h_1 + \vert F_1\vert^2\,\big]\,(h_2.h_1)
+
(h_1^\dagger \overleftarrow\cD_\mu h_1)\,\partial^\mu(h_2.h_1)\Big]
\nonumber\\
&+&
\alpha_{50}^*
\Big[
i\, h_1^\dagger\, \cD_\mu\psi_1\,\sigma^\mu\,(\psi_1.h_2+h_1.\psi_2)^\dagger
+i\,(h_2.h_1)^\dagger \,
\overline\psi_1\overline\sigma^\mu\cD_\mu\psi_1
\Big]
\nonumber\\
&+&
\alpha_{50}\Big[
(\psi_1.h_2+h_1.\psi_2)\,(F_1^\dagger\psi_1+\frac{1}{\sqrt
    2}\, h_1^\dagger\lambda_1 h_1)
-
\frac{1}{\sqrt 2}(h_2.h_1)(h_1^\dagger \lambda_1\psi_1+
\overline\psi_1\overline\lambda_1 h_1)
\Big]\nonumber\\
&+&
\Big[\alpha_{50}(F_1^\dagger h_1) +\alpha_{51}^*\,m_0 \vert
  h_1\vert^2\Big]
(h_2.F_1+F_2.h_1-\psi_2.\psi_1)
-\alpha_{51}^*\,m_0\,
(h_1^\dagger\psi_1)(h_2.\psi_1+\psi_2.h_1)\Big]
\nonumber\\
&+&
m_0\,\Big[\alpha_{51}\,(F_1^\dagger h_1)+\alpha_{51}^*\,
(h_1^\dagger F_1)\Big]\,(h_2.h_1)
+
\alpha_{52}\,m_0^2\,\vert h_1\vert^2\,(h_2.h_1)+{\rm h.c.\,\,\,of \,\,\,all}
\nonumber
\eea
\bea
\mathcal{O}_{6} &=&
\frac{1}{M^{2}}\int d^{4}\theta \,\,\mathcal{Z}_{6}(S,S^{\dagger
})\,\,
(H_{2}^{\dagger }\,e^{V_{2}}\,H_{2})\,\,H_{2}.\,H_{1}+h.c. \\
&=&
\alpha_{60}
\Big[\big[\,
\vert \cD_\mu h_2\vert^2
+h_2^\dagger\,\frac{D_2}{2}\,h_2 + \vert F_2\vert^2\,\big]\,(h_2.h_1)
+
(h_2^\dagger \overleftarrow\cD_\mu h_2)\,\partial^\mu(h_2.h_1)\Big]
\nonumber\\
&+&
\alpha_{60}^*
\Big[
i\, h_2^\dagger\, \cD_\mu\psi_2\,\sigma^\mu\,(\psi_1.h_2+h_1.\psi_2)^\dagger
+i\,(h_2.h_1)^\dagger \,
\overline\psi_2\overline\sigma^\mu\cD_\mu\psi_2
\Big]
\nonumber\\
&+&
\alpha_{60}\Big[
(\psi_1.h_2+h_1.\psi_2)\,(F_2^\dagger\psi_2+\frac{1}{\sqrt
    2}\, h_2^\dagger\lambda_2 h_2)
-
\frac{1}{\sqrt 2}(h_2.h_1)(h_2^\dagger \lambda_2\psi_2+
\overline\psi_2\overline\lambda_2 h_2)
\Big]\nonumber\\
&+&
\Big[\alpha_{60}(F_2^\dagger h_2) +\alpha_{61}^*\,m_0 \vert
  h_2\vert^2\Big]
(h_2.F_1+F_2.h_1-\psi_2.\psi_1)
-\alpha_{61}^*\,m_0\,
(h_2^\dagger\psi_2)(h_2.\psi_1+\psi_2.h_1)\Big]
\nonumber\\
&+&
m_0\,\Big[\alpha_{61}\,(F_2^\dagger h_2)+\alpha_{61}^*\,
(h_2^\dagger F_2)\Big]\,(h_2.h_1)
+
\alpha_{62}\,m_0^2\,\vert h_2\vert^2\,(h_2.h_1)+{\rm h.c.\,\,\,of \,\,\,all}
\nonumber
\eea
\bea
\mathcal{O}_{7} &=&\frac{1}{M^{2}}\sum_{s=w,y}\frac{1}{16 g^2_s \kappa}
\int d^{2}\theta 
\mathcal{Z}_{7}(S,0){\rm Tr}( W^{\alpha } W_{\alpha } )_s
(H_{2}.H_{1})\!+\!h.c. \\
&=&
\!\!\!\!\!\!\sum_{s=w,y}\!\!\!
\frac{\alpha^s_{70}}{4} \Big\{(h_2.h_1)
\Big[ i\,(\lambda^a_s \sigma^\mu\Delta_\mu\overline\lambda^a_s
\!-\!\Delta_\mu\overline\lambda^a_s \overline\sigma^\mu\lambda^a_s)
\!+\! D^a_s D^a_s \!-\!\frac{1}{2}
(F_s^{a\,\mu\nu} F^a_{s\,\mu\nu}\!+\!\frac{i}{2}\epsilon^{\mu\nu\rho\sigma} 
F_{s\, \mu\nu}^a F_{s\, \rho\sigma}^a)
\Big]
\nonumber\\
&-&\!\!\!\sqrt 2\,(h_2.\psi_1+\psi_2.h_1)
(\lambda^a_s D^a_s+\sigma^{\mu\nu}\lambda_s^a F^a_{s\,\mu\nu})
+(h_2.F_1+F_2.h_1-\psi_2.\psi_1)\,\lambda_s^a\lambda_s^a\Big\}
\nonumber\\
&+&
\frac{1}{4}\alpha^s_{71} m_0 (h_2.h_1)(\lambda^a_s\lambda^a_s)
+{\rm h.c.\,\,\,of\,\,\,all},\qquad  
(g_w\equiv g_2;\,\,\, 
g_y\equiv g_1;\,\,\,
w: SU(2);\,\,\,
y: U(1)).
\nonumber
\eea
\bea
\mathcal{O}_{8} &=&
\frac{1}{M^{2}}\int d^{4}\theta
 \,\,\Big[\mathcal{Z}_{8}(S,S^{\dagger })\,\,
\,[(H_{2}\,H_{1})^{2}+h.c.]\Big] 
\\[4pt]
&=&
2\,\alpha_{81}^* \,m_0\,(h_2.h_1)\,(h_2.F_1+F_2.h_1-\psi_2.\psi_1)
+m_0^2\,\alpha_{82}\,(h_2\cdot h_1)^2+{\rm h.c.\,\, of\,\,\, all}
\qquad\qquad\quad\nonumber
\end{eqnarray}%


\noindent
 $W^\alpha$ is the Susy field strength
of $SU(2)_L$ ($U(1)_Y$) vector superfield $V_w$ ($V_y$) of 
auxiliary  component $D_w$ ($D_Y$). Also
\medskip
\bea
(1/M^2)\,\,\cZ_i(S,S^\dagger)=\alpha_{i0}
+\alpha_{i1}\,m_0\,\theta\theta
+\alpha_{i1}^*\,m_0\,\overline\theta\overline\theta
+\alpha_{i2}\,m_0^2\,\theta\theta\overline\theta\overline\theta
\eea

\medskip\noindent
and
$\cD^\mu h_i=(\partial^\mu+i/2\,V^\mu_i)\,h_i$,\,
$h_i^\dagger \overleftarrow\cD^\mu=(\cD^\mu h_i)^\dagger
=h_i^\dagger (\overleftarrow\partial^\mu-i/2 V_i^\mu)
$.

\medskip
Further,  $D_1\equiv \vec D_w\,\vec T+(-1/2)\,\,D_Y$ and
  $D_2\equiv \vec D_w\,\vec T+(1/2)\,\,D_Y$, $T^a=\sigma^a/2$.
Finally, one must rescale in all $\cO_i$ ($i\not=7$ since $\cO_7$ is
  rescaled already):\,\,\,
$V_{w}\ra 2\,g_2\,V_w$, 
$V_{y}\ra 2\,g_1\,V_y$.
Therefore one must replace 
 $V_{1,2}= 2\,g_2\,\vec V_w\,\vec T +2\,g_1\,(\mp 1/2)\,V_y$ with the
upper sign (minus) for $V_1$, where
 $V_{1,2}$ enter above in  the definition of $\cO_{1,2}$.
Similar expressions exist
 for the components of the superfields $V_{1,2}$. For example:
 $\lambda_{1,2}=g_2\,\lambda_w^a\,\sigma^a
+g_1\,(\mp 1)\,\lambda_y$ (minus for $\lambda_1$).

Other notations used above:
$H_1.H_2=\epsilon^{ij}\,H_1^i\,H_2^j$. Also
$\vert h_1\cdot h_2\vert^2
=\vert h_1^i\,\epsilon^{ij}\,h_2^j\vert^2
=\vert h_1\vert^2\,\vert h_2\vert^2
-\vert h_1^\dagger \,h_2\vert^2;$
 $\epsilon^{ij}\,\epsilon^{kj}=\delta^{ik}$;\,\,
$\epsilon^{ij}\,\epsilon^{kl}=
\delta^{ik}\,\delta^{jl}-\delta^{il}\,\delta^{jk}$, $\epsilon^{12}=1$,
with
\bea\label{higgsd}
h_1=\left(\begin{array}{c}
h_1^0 \\[-1pt]
h_1^-\\
\end{array}\right)
\equiv 
\left(\begin{array}{c}
h_1^1 \\[-1pt]
h_1^2\\
\end{array}\right),\,\,Y_{h_1}=-1;
\qquad
h_2=\left(\begin{array}{c}
h_2^+\\[-1pt]
h_2^0
\end{array}\right)
\equiv
\left(\begin{array}{c}
h_2^1\\[-1pt]
h_2^2
\end{array}\right),\,\,\,Y_{h_2}=+1
\eea

\medskip\noindent
In the above expressions for $\cO_{1,2,...8}$, the notations $h_i$, $\psi_i$,
$F_{1,2}$ stand for SU(2) doublets, so for example 
$h_1^\dagger
\psi_1=h_1^{i\, *}\,\psi_1^i$, also $\vert h_1\vert^2=h_1^\dagger
h_1=h_1^{* i} h_1^i$, where the superscript $i$ labels the
SU(2) components, as shown above for the Higgs doublets $h_{1,2}$.
Other notations: $\psi_1.h_2=\psi_1^i \epsilon^{ij} h_2^j$
with a similar notation for the components of the doublet (superscripts).
The derivatives $\cD, (\overleftarrow \cD)$ only act on the first field
to their right (left) respectively.

In $\cO_7$ we used the notation:
\bea
\Delta_\mu\overline\lambda^a=\partial_\mu\overline\lambda^a
-g\,t^{abc}\,V_\mu^b\,\overline\lambda^c,\qquad
\Delta_\mu\overline\lambda
=\partial_\mu\overline\lambda+\frac{i}{2}\,[V_\mu,\overline\lambda],
\eea
where the last equation applies before the rescaling of vector superfield
(in matrix notation). These eqs are considered for 
$\lambda_w$ ($\lambda_y$) with corresponding  $V_{w, \mu}$ ($V_{y,\mu}$).

\medskip
The Lagrangian  with the above $\cO_{1,..,8},\cK_0$  leads to ($q$ is a doublet index)
 \medskip
 \bea\label{arhos}
 F_1^{* q}&=&
 -\big\{\epsilon^{qp}\,h_2^p\, \big[
 \mu+2\,\zeta_{10}\,(h_1.h_2)
 +\rho_{11}\big]+h_1^{* q}\,\rho_{12}+\overline\psi_1^q\,\,\rho_{13}\big\}
 \nonumber\\
 F_2^{* q}&=&
 -\big\{\epsilon^{pq}\,h_1^p \,\big[
 \mu+2\,\zeta_{10}\,(h_1.h_2)
 +\rho_{21}\big]+h_2^{* q}\,\rho_{22}+\overline\psi_2^q\,\,\rho_{23}\big\}
 \eea

\medskip\noindent
where $\rho_{ij}$ are functions of $h_{1,2}$, given in 

\medskip
\bea\label{rhos0}
\rho_{11}
&=&
-(2\alpha_{10}\,\mu+\alpha_{40}\mu+\alpha_{51}^*\,m_0)\vert h_1\vert^2
-(\alpha_{30}\,\mu+\alpha_{40}\mu
+\alpha_{61}^*\,m_0)\,\vert h_2\vert^2
\nonumber\\
&&-\,
(\alpha_{41}^*\,m_0+\alpha_{50}^*\,\mu)\,(h_2.h_1)^*+
\big[\,(\alpha_{60} +2\,\alpha_{50})\,\mu+2\alpha_{81}^*\,m_0\big]\,
(h_1.h_2)
\nonumber\\
&&+\,
\alpha_{40}\,\overline\psi_2.\overline\psi_1
-(1/4)\,\,(\alpha_{70}^w\lambda^a_w\lambda^a_w
+\alpha_{70}^y\lambda_y^2)
\nonumber\\
\rho_{12}
&=&
\,\,\,(2\alpha_{11}^*\,m_0+\alpha_{50}^*\,\mu)\vert h_1\vert^2
+(\alpha_{31}^*\,m_0+\alpha_{50}^*\,\mu)\,\vert h_2\vert^2
\nonumber\\
&&-\big[(2\alpha_{10}+\alpha_{30})\,\mu+\alpha_{51}^*\,m_0\big]\,(h_1.h_2)
+\alpha_{51}^*\,m_0\,(h_2.h_1)^*
-\alpha_{50}^*\,\overline\psi_2.\overline\psi_1
\nonumber\\
\rho_{13}&=&-\big[2\alpha_{10}\,\overline\psi_1 h_1
+\alpha_{30}\overline\psi_2 h_2-
\alpha_{50}^* \,(\psi_1.h_2+h_1.\psi_2)^\dagger
\big]
\eea
and
\bea
\rho_{21}
&=&
 -(2\alpha_{20}\,\mu+\alpha_{40}\mu+\alpha_{61}^*\,m_0)\vert h_2\vert^2
 -(\alpha_{30}\,\mu+\alpha_{40}\mu+\alpha_{51}^*\,m_0)\,\vert h_1\vert^2
\nonumber\\
&&-(\alpha_{41}^*\,m_0+\alpha_{60}^*\,\mu)\,(h_2.h_1)^*
+\big[\,(\alpha_{50} +2\,\alpha_{60})\,\mu+2\alpha_{81}^*\,m_0\big]\,
(h_1.h_2)\nonumber\\
&&
+\,\alpha_{40}\,\overline\psi_2.\overline\psi_1
-(1/4)\,\,(\alpha_{70}^w\lambda^a_w\lambda^a_w
+\alpha_{70}^y\lambda_y^2)
\nonumber\\
\rho_{22}
&=&\,\,\,
(2\alpha_{21}^*\,m_0+\alpha_{60}^*\,\mu)\vert h_2\vert^2
+(\alpha_{31}^*\,m_0+\alpha_{60}^*\,\mu)\,\vert h_1\vert^2
\nonumber\\
&&-\big[(2\alpha_{20}+\alpha_{30})\,\mu+\alpha_{61}^*\,m_0\big]\,(h_1.h_2)
+\alpha_{61}^*\,m_0\,(h_2.h_1)^*
-\alpha_{60}^*\overline\psi_2.\overline\psi_1
\qquad
\nonumber\\
\rho_{23}&=&
-\big[
2\alpha_{20}\,\overline\psi_2 h_2+\alpha_{30}\,\overline\psi_1\,h_1
-\alpha_{60}^*\,(\psi_1.h_2+h_1.\psi_2)^\dagger
\big]
\label{rhos}
\eea
The last line in $\rho_{11}$, $\rho_{21}$ and $\rho_{13}$, $\rho_{23}$
are new contributions, due to fermions only.
Further
\medskip
 \bea\label{ddd}
 D_w^a&=&-g_2\,\Big[\,\,h_1^\dagger T^a\,h_1\,(1+\tilde
 \rho_{1,w})+h_2^\dagger\,T^a\,h_2\,(1+\tilde \rho_{2,w})\,
-\frac{\sqrt 2}{4}\big(\alpha_{70}^w\,(h_2.\psi_1+\psi_2.h_1)
\lambda_w^a\!+\! h.c.\big)\Big]
\nonumber\\
D_Y&=&\!
-g_1\! \Big[h_1^\dagger \frac{-1}{2} h_1 (1+\tilde
\rho_{1,y})+h_2^\dagger\frac{1}{2} h_2 (1+\tilde \rho_{2,y})\,
\!-\!\frac{\sqrt 2}{4}\big(\alpha_{70}^y(h_2.\psi_1+\psi_2.h_1)
\lambda_y^a\!+\! h.c.\big)\Big]\label{dd2}
\eea

\medskip\noindent
with $T^a=\sigma^a/2$, and
\bea
\tilde\rho_{1,w}&=&
2\alpha_{10}\vert h_1\vert^2+\alpha_{30}\vert h_2\vert^2
+\big[(\alpha_{50}-\alpha_{70}^w/2)\,(h_2.h_1)+h.c.\big]
\nonumber\\
\tilde\rho_{2,w}&=&
2\alpha_{20}\vert h_2\vert^2+\alpha_{30}\vert h_1\vert^2
+\big[(\alpha_{60}-\alpha_{70}^w/2)\,(h_2.h_1)+h.c.\big]
\eea
with similar expression for $\rho_{j,y}$ in which one uses instead
   $\alpha_{70}^y$. Therefore
\medskip
\bea
 D_w^a\,D_w^a&=&
 \frac{g_2^2}{4}\,\Big[\,\,\big(\,\,
 (1+\tilde\rho_{1,w})\,\vert h_1\vert^2-
 (1+\tilde\rho_{2,w})\,\vert h_2\vert^2\,\,\big)^2
 +4\,(1+\tilde\rho_{1,w})(1+\tilde\rho_{2,w})\,\vert h_1^\dagger\,h_2\vert^2
\Big]
\nonumber\\
&-&
\frac{\sqrt 2}{2} \, g_2 
\Big[\, h_1^\dagger T^a h_1
+h_2^\dagger T^a h_2\,\Big]
\Big[\,
\alpha_{70}^w\,(h_2.\psi_1+\psi_2.h_1)\,\lambda_w^a+h.c.\Big]
\nonumber\\
D_Y^2&=&
\frac{g_1^2}{4}\,\big(\,\,
(1+\tilde\rho_{1,y})\,\vert h_1\vert^2-
(1+\tilde\rho_{2,y})\,\vert h_2\vert^2\,\big)^2
\nonumber\\&-&
\frac{\sqrt 2}{2} g_1
\Big[ h_1^\dagger \frac{-1}{2} h_1
+h_2^\dagger \frac{1}{2}\, h_2\Big]
\Big[
\alpha_{70}^y\,(h_2.\psi_1+\psi_2.h_1)\,\lambda_y+h.c.\Big]
\label{dsq}
\eea

\def\theequation{B-\arabic{equation}}
\def\thesubsection{B}
\setcounter{equation}{0}
\subsection{The neutralino and chargino Lagrangian.}\label{appendixB}

Here we provide the full neutralino and chargino Lagrangian,
 in component fields (see notation (\ref
{higgsd}) - similar notation for higgsino components), extracted 
from the total Lagrangian of Section~\ref
{section2}.  Below we use the notation:
\bea
\partial_\mu^z f_1&\equiv& (\partial_\mu+(i/2)\, s\,g\,V^z_\mu)f_1
\qquad
\overleftrightarrow\partial_\mu^z f_1 =
(\partial_\mu -\overleftarrow \partial_\mu
+i \,s\,g V_\mu^z)\,f_1
\nonumber\\
\partial_\mu^\gamma f_2&\equiv &
(\partial_\mu-(i/2)\, s\,g\,V^\gamma_\mu) f_2
\qquad
\overleftrightarrow\partial_\mu^\gamma f_2=
(\partial_\mu -\overleftarrow \partial_\mu
-i \,s\,g V_\mu^\gamma)\,f_2
\eea
where $s=+1$ for $f_1=\psi_1^0$, $h_1^0$, $f_2=\psi_1^-, h_1^-$
and $s=-1$ for $f_1=\psi_2^0$, $h_2^0$, $f_2=\psi_2^+, h_2^+$.
Also
\bea
V^{\pm}_\mu&=&V_{w,\mu}^1 \mp i\, V_{w,\mu}^2\equiv \sqrt 2\,W^\pm_\mu,\,\,\quad\,\,\,\,
 g\,V_\mu^z = g_2 V_{w,\mu}^3- g_1 V_{y,\mu}, \nonumber\\
\lambda^\pm_w &=& \lambda_w^1\mp i \,\lambda_w^2\equiv \sqrt 2\, \tilde\lambda_w^\pm
\qquad\qquad\,\,\,
g\,V_\mu^\gamma= g_2 V_{w,\mu}^3+  g_1\,V_{y,\mu},\nonumber\\
 g&=& g_2/\cos\theta_w\,=\,g_1/\sin\theta_w=e/(\sin\theta_w \cos\theta_w)
\eea
where $W_\mu^\pm$, $\tilde\lambda_w^\pm$ denote the charged 
weak bosons and charginos, respectively;
(below we use the $V_\mu^\pm$ and $\tilde\lambda^\pm_w$ notation 
instead, to avoid complicating the 
expressions by the extra $\sqrt 2$ factors).

The neutralino and chargino Lagrangian receives
contributions from $\cL$ of (\ref{totalL}).  
$\cL_D$ gives:
\bea\label{ww1}
\!\!\!\!\!
\cL_{D}\!\!\!\!
&\supset& \!\!\!\!\frac{-1}{4\sqrt 2}
\Big[
g_2\alpha_{70}^w \Big(
\lambda_w^3\,\big[
\vert h_1^0\vert^2\!\!-\vert h_2^0\vert^2\!+
\vert h_2^+\vert^2\!\!-\vert h_1^-\vert^2\big]\!
+\!\lambda_w^+ \,(h_1^{0 *} h_1^- \!\!+\! h_2^{+ *} h_2^0)
\!+\lambda_w^-(h_1^{0 *} h_1^- \!\!+ \! h_2^{+ *} h_2^0)^*\Big)
\nonumber\\
&-&
g_1\,\alpha_{70}^y\,\lambda_y\big[
\vert h_1^0\vert^2-\vert h_2^0\vert^2
+\vert h_1^-\vert^2 -\vert h_2^+ \vert^2\big]
\Big](\psi_1^0 h_2^0\! +\! h_1^0 \psi_2^0\!-\!\psi_1^- h_2^+ \!-\! h_1^- \psi_2^+)
+h.c.\qquad
\eea
$\cL_{F,1}$ contributes the neutralino/chargino terms below, 
given separately for each operator $\cO_i$:
\bea\label{ww2}
\cL_{F,1}&\supset& \sum_{i} R_{\cO_i}\qquad\textrm{where:}
 \eea
 \bea
R_{\cO_1}&=&\!\!\!
2\,\alpha_{10}\,
\mu\,(\psi_1^0 \,h_2^0-\psi_1^- h_2^+)\,
(h_1^{0 *} \psi_1^0 +h_1^{- *}\psi_1^-)
+h.c.
\nonumber\\
R_{\cO_2}&=&\!\!\!
2\,\alpha_{20}
\,\mu\,(\psi_2^0 \,h_1^0-\psi_2^+ h_1^-)\,
(h_2^{0 *} \psi_2^0 +h_2^{+ *}\psi_2^+)
+h.c.
\nonumber\\
R_{\cO_3}&=& \alpha_{30}
\,\mu\,\,\big[
(\psi_1^0 \,h_2^0-\psi_1^- h_2^+)\,
(h_2^{0 *}\psi_2^0 +h_2^{+ *} \psi_2^+)
\nonumber\\
&&\qquad\,+
(\psi_2^0 \,h_1^0-\psi_2^+ h_1^-)\,(h_1^{0 *} \psi_1^0 +h_1^{- *}\psi_1^-)\big]
+h.c.\,\,\,
\nonumber\\
R_{\cO_4}&=&\alpha_{40}
\,\mu\,\,(\psi_1^0\psi_2^0-\psi_1^-\psi_2^+)\,
(\vert h_1^0\vert^2+\vert h_2^0\vert^2+\vert h_1^-\vert^2+\vert h_2^+\vert^2)
+h.c.
\nonumber\\
R_{\cO_5}&=&\!\!\!\!
-\alpha_{50}\,\mu\,\big[
(\psi_1^0 h_2^0 -\psi_1^- h_2^+)\,(\psi_1^0 h_2^0 
+h_1^0 \psi_2^0-\psi_1^- h_2^+ -h_1^- \psi_2^+)
\nonumber\\
&&\qquad+\,
(h_1^0 h_2^0\! -\! h_1^- h_2^+)\,(\psi_1^0\psi_2^0 -\psi_1^-\psi_2^+)\big]+h.c.
\nonumber\\
R_{\cO_6}&=&\!\!\!\!
-\alpha_{60}\,\mu\,\big[
(\psi_2^0 h_1^0 -\psi_2^+ h_1^-)\,(\psi_1^0 h_2^0 
+h_1^0 \psi_2^0-\psi_1^- h_2^+ -h_1^- \psi_2^+)
\nonumber\\
&&\qquad+\,
(h_1^0 h_2^0\! -\! h_1^- h_2^+)\,(\psi_1^0\psi_2^0 -\psi_1^-\psi_2^+)\big]+h.c.
\nonumber\\
R_{\cO_7}&=&
\frac{1}{4}\,\,\mu\,
[\,\alpha_{70}^w \,(\lambda_w^+\lambda_w^-+\lambda_w^3\lambda_w^3)
+\alpha_{70}^y\,\lambda_y\lambda_y\big]
\nonumber\\
&&\qquad\times\,(\vert h_1^0\vert^2+
\vert h_2^0\vert^2+\vert h_1^-\vert^2+\vert h_2^+\vert^2)
+h.c.
\eea
Further, $\cL_1$ generates the following terms, which are 
pairs of higgsino or of gaugino:
\bea\label{ww3}
\cL_1&=&\sum_{i=1}^8 S_{\cO_i}
\eea

\vspace{-0.5cm}
\bea
S_{\cO_1}&=&\!\!
i\,\alpha_{10}\Big[
(\vert h_1^0\vert^2\!+\!\vert h_1^-\vert^2)
\big[
\overline{\psi_1^0}\overline\sigma^\mu \partial_\mu^z
\psi_1^0\! +\!(i/2) g_2 (\overline{\psi_1^0}\overline\sigma^\mu V^+_\mu \psi_1^-
+ \overline{\psi_1^-}\overline\sigma^\mu V_\mu^-\psi_1^0)\!+\!
\overline{\psi_1^-}\overline\sigma^\mu\partial_\mu^\gamma \psi_1^-
\big]
\nonumber\\
&+&
(\overline{\psi_1^0}\overline\sigma^\mu\psi_1^0 + \overline{\psi_1^-}
\overline\sigma^\mu\psi_1^-)
\big[h_1^{0 *}\partial_\mu^z h_1^0+(i/2) g_2 ( h_1^{0 *} V_\mu^+ h_1^-
+ h_1^{- *} V_\mu^- h_1^0)
+ h_1^{- *} \partial_\mu^\gamma \,h_1^-\big]\Big]
\nonumber\\
&-&\!\!\!\!
(h_1^{0 *} \psi_1^0 +h_1^{- *} \psi_1^-)\sigma^\mu
\big[
\overline{\psi_1^0} \overleftrightarrow\partial_\mu^z \,h_1^0+i g_2
(\overline{\psi_1^0} V_\mu^+ h_1^-   \!+\! \overline{\psi_1^-} V^-_\mu h_1^0 )
+\overline{\psi_1^-}\overleftrightarrow\partial_\mu^\gamma\,h_1^-\big]
\Big]\!+\!h.c.
\eea

\vspace{-0.5cm}
\bea
S_{\cO_2}&=&\!\!\!
i\,\alpha_{20}\Big[
(\vert h_2^0\vert^2+\vert h_2^+\vert^2)
\big[
\overline{\psi_2^0}\overline\sigma^\mu \partial_\mu^z
\psi_2^0 +(i/2) g_2 (\overline{\psi_2^0}\overline\sigma^\mu V^-_\mu \psi_2^+
+ \overline{\psi_2^+}\overline\sigma^\mu V_\mu^+\psi_2^0)
+
\overline{\psi_2^+}\overline\sigma^\mu\partial_\mu^\gamma\,\psi_2^+
\big]
\nonumber\\
&+&
(\overline{\psi_2^0}\overline\sigma^\mu\psi_2^0 + \overline{\psi_2^+}
\overline\sigma^\mu\psi_2^+)
\big[h_2^{0 *}\partial_\mu^z h_2^0+(i/2) g_2 ( h_2^{0 *} V_\mu^- h_2^+
+ h_2^{+ *} V_\mu^+ h_2^0)
+ h_2^{+ *} \partial_\mu^\gamma \,h_2^+\big]\Big]
\nonumber\\
&-&\!\!\!
(h_2^{0 *} \psi_2^0 +h_2^{+ *} \psi_2^+)\sigma^\mu
\big[
\overline{\psi_2^0} \overleftrightarrow\partial_\mu^z \,h_2^0+i g_2
(\overline{\psi_2^0} V_\mu^- h_2^+   + \overline{\psi_2^+} V^+_\mu h_2^0 )
\!+\!\overline{\psi_2^+}\overleftrightarrow\partial_\mu^\gamma\,h_2^+\big]
\Big]\!+\!h.c.
\eea

\vspace{-0.5cm}
\bea
S_{\cO_3}&=&\!\!\!
i\,\frac{\alpha_{30}}{2}\Big[
(\vert h_1^0\vert^2\!+\!\vert h_1^-\vert^2)
\big[
\overline{\psi_2^0}\overline\sigma^\mu \partial_\mu^z
\psi_2^0\! +\!(i/2) g_2 (\overline{\psi_2^0}\overline\sigma^\mu V^-_\mu \psi_2^+
\!+\! \overline{\psi_2^+}\overline\sigma^\mu V_\mu^+\psi_2^0)
\!+\!
\overline{\psi_2^+}\overline\sigma^\mu\partial_\mu^\gamma\,\psi_2^+
\big]\!
\nonumber\\
&+&
(\overline{\psi_1^0}\overline\sigma^\mu\psi_1^0 + \overline{\psi_1^-}
\overline\sigma^\mu\psi_1^-)
\big[h_2^{0 *}\partial_\mu^z h_2^0+(i/2) g_2 ( h_2^{0 *} V_\mu^- h_2^+
+ h_2^{+ *} V_\mu^+ h_2^0)
+ h_2^{+ *} \partial_\mu^\gamma \,h_2^+\big]
\nonumber\\
&-&\!\!
(h_1^{0 *} \psi_1^0 +h_1^{- *} \psi_1^-)\sigma^\mu
\big[
\overline{\psi_2^0} \overleftrightarrow\partial_\mu^z \,h_2^0+i g_2
(\overline{\psi_2^0} V_\mu^- h_2^+   + \overline{\psi_2^+} V^+_\mu h_2^0 )
+\overline{\psi_2^+}\overleftrightarrow\partial_\mu^\gamma\,h_2^+\big]
\Big]
\nonumber\\
&+&
(\vert h_2^0\vert^2+\vert h_2^+\vert^2)
\big[
\overline{\psi_1^0}\overline\sigma^\mu \partial_\mu^z
\psi_1^0 +(i/2) g_2 (\overline{\psi_1^0}\overline\sigma^\mu V^+_\mu \psi_1^-
+ \overline{\psi_1^-}\overline\sigma^\mu V_\mu^-\psi_1^0)+
\overline{\psi_1^-}\overline\sigma^\mu\partial_\mu^\gamma\,\psi_1^-
\big]
\nonumber\\
&+&
(\overline{\psi_2^0}\overline\sigma^\mu\psi_2^0 + \overline{\psi_2^+}
\overline\sigma^\mu\psi_2^+)
\big[h_1^{0 *}\partial_\mu^z h_1^0+(i/2) g_2 ( h_1^{0 *} V_\mu^+ h_1^-
+ h_1^{- *} V_\mu^- h_1^0)
+ h_1^{- *} \partial_\mu^\gamma \,h_1^-\big]\Big]
\nonumber\\
&-&\!\!
(h_2^{0 *} \psi_2^0+ h_2^{+ *} \psi_2^+)\sigma^\mu
\big[
\overline{\psi_1^0} \overleftrightarrow\partial_\mu^z h_1^0\!+  i g_2
(\overline{\psi_1^0} V_\mu^+ h_1^-  \! +\! \overline{\psi_1^-} V^-_\mu h_1^0 )
\!+\!\overline{\psi_1^-}\overleftrightarrow\partial_\mu^\gamma h_1^-\big]
\Big]\!+\!h.c.
\eea

\vspace{-0.5cm}
\bea
S_{\cO_4}&=&\!\!
\frac{i\alpha_{40}}{2}
\Big[(\psi_1^0 h_2^0+h_1^0\psi_2^0 -\psi_1^- h_2^+ - h_1^- \psi_2^+)
\sigma^\mu\partial_\mu 
(\psi_1^0 h_2^0+h_1^0\psi_2^0 -\psi_1^- h_2^+ - h_1^- \psi_2^+)^\dagger
+h.c.\Big]\nonumber
\eea

\vspace{-0.5cm}
\bea
S_{\cO_5}&=&
i\alpha_{50}^*\Big[\big[
h_1^{0 *}\partial_\mu^z \psi_1^0 +
(i/2) g_2 (h_1^{0 *} V^+_\mu \psi_1^- +h_1^{- *} V^-_\mu \psi_1^0)
+
h_1^{- *} \partial_\mu^\gamma \psi_1^-\big]
\sigma^\mu(\psi_1^0 h_2^0 +h_1^0 \psi_2^0 
\nonumber\\
&-&
\!\!\!\psi_1^- h_2^+ -h_1^- \psi_2^+)^\dagger
\!-\!(h_1^{0 *} h_2^{0 *} - h_1^{- *} h_2^{+ *})
\big[\overline{\psi_1^0}\overline\sigma^\mu \partial_\mu^z \psi_1^0
\!+\!
(i/2) g_2 (\overline{\psi_1^0} \overline\sigma^\mu V^+_\mu \psi_1^-
\!+\!\overline{\psi_1^-} \overline\sigma^\mu V^-_\mu \psi_1^0)
\nonumber\\
&+&
\overline{\psi_1^-} \overline\sigma^\mu \partial_\mu^\gamma 
\psi_1^-\big]\Big]
+h.c.
\eea

\vspace{-0.5cm}
\bea
S_{\cO_6}&=&
i\alpha_{60}^*\Big[\big[
h_2^{0 *}\partial_\mu^z \psi_2^0 +
(i/2) g_2 (h_2^{0 *} V^-_\mu \psi_2^+ +h_2^{+ *} V^+_\mu \psi_2^0)
+
h_2^{+ *} \partial_\mu^\gamma \psi_2^+\big]
\sigma^\mu(\psi_1^0 h_2^0 +h_1^0 \psi_2^0 
\nonumber\\
&-&
\!\!\!\psi_1^- h_2^+ -h_1^- \psi_2^+)^\dagger
\!-\!(h_1^{0 *} h_2^{0 *} - h_1^{- *} h_2^{+ *})
\big[\overline{\psi_2^0}\overline\sigma^\mu \partial_\mu^z \psi_2^0
\!+\!
(i/2) g_2 (\overline{\psi_2^0} \overline\sigma^\mu V^-_\mu \psi_2^+
\!+\!\overline{\psi_2^+} \overline\sigma^\mu V^+_\mu \psi_2^0)
\nonumber\\
&+&
\overline{\psi_2^+} \overline\sigma^\mu \partial_\mu^\gamma 
\psi_2^+\big]\Big]
+h.c.
\eea

\vspace{-0.5cm}
\bea
S_{\cO_7}\,\,=\,\,
\frac{i}{2} \alpha_{70}^w\big[-h_1^0 h_2^0+h_1^- h_2^+\big]
\,
\lambda_w^a\sigma^\mu\Delta_\mu\overline{\lambda_w^a}
 +
\frac{i}{2}\alpha_{70}^y\,\big[-h_1^0 h_2^0+h_1^- h_2^+\big]
\,\lambda_y\sigma^\mu\partial_\mu\overline{\lambda_y}+h.c.\,\,
 \eea
where
\bea
\lambda_w^a\sigma^\mu\Delta_\mu\overline{\lambda_w^a}
&=&
\frac{1}{2}\lambda_w^+\sigma^\mu\partial_\mu\overline{\lambda_w^+}
+
\frac{1}{2}\lambda_w^-\sigma^\mu\partial_\mu\overline{\lambda_w^-}
+\lambda_w^3\sigma^\mu\partial_\mu\overline{\lambda_w^3}
\nonumber\\
&+&
(i/2)\,g_2\,
\Big[\lambda_w^+\sigma^\mu (V_\mu^- \overline{\lambda_w^3}
-V_w^3 \overline{\lambda_w^+})
+\lambda_w^-\sigma^\mu\,(V_w^3\,\overline{\lambda_w^-}-V_\mu^+
\overline{\lambda_w^3})
\nonumber\\
&-&
\lambda_w^3 \sigma^\mu \,(-V^+_\mu \,\overline{\lambda_w^+}
+V^-_\mu\overline{\lambda_w^-})\Big]
\eea
Next, $\cL_2$ 
of Section~\ref{section2} gives contributions to the neutralino/chargino sectors:
\bea\label{ww4}
\cL_2 =\sum_{i=1}^8 T_{\cO_i}
\eea
The Susy part of this contribution contains one gaugino
 and one higgsino (charged or not), while its non-Susy one
 contains two higgsinos (charged or not):
\bea
T_{\cO_1}&=&
\alpha_{10}\sqrt 2 
\Big[
-(\vert h_1^0\vert^2+\vert h_1^-\vert^2)
\big[
g h_1^{0 *}\lambda_z \psi_1^0 +g_2\, (h_1^{0 *} \lambda_w^+ \psi_1^-+
h_1^{- *} \lambda_w^- \psi_1^0) - g h_1^{- *} \lambda_\gamma \psi_1^-
\big]
\nonumber\\
&-& (h_1^{0 *} \psi_1^0 + h_1^{- *}\psi_1^-)
\big[ g\lambda_z \vert h_1^0\vert^2+g_2\,(h_1^{0 *} \lambda_w^+ h_1^- +
h_1^{- *} \lambda_w^- h_1^0)- g \vert h_1^-\vert^2 \lambda_\gamma
\big]
\Big]
\nonumber\\
&-&
\alpha_{11}^* \,m_0 \,
(h_1^{0 *} {\psi_1^0}+ h_1^{- *}{\psi_1^-})^2
+h.c.
\eea
\bea
T_{\cO_2}&=&
\alpha_{20}\sqrt 2 
\Big[
-(\vert h_2^0\vert^2+\vert h_2^+\vert^2)
\big[-
g h_2^{0 *}\lambda_z \psi_2^0 +g_2\, (h_2^{0 *} \lambda_w^- \psi_2^+ +
h_2^{+ *} \lambda_w^+ \psi_2^0) + g h_2^{+ *} \lambda_\gamma \psi_2^+
\big]
\nonumber\\
&-& (h_2^{0 *} \psi_2^0 + h_2^{+ *}\psi_2^+)
\big[ - g\lambda_z \vert h_2^0\vert^2+g_2\,(h_2^{0 *} \lambda_w^- h_2^+ +
h_2^{+ *} \lambda_w^+ h_2^0)+  g \vert h_2^+\vert^2 \lambda_\gamma
\big]
\Big]
\nonumber\\
&-&
\alpha_{21}^* \,m_0 \,
(h_2^{0 *} {\psi_2^0}+ h_2^{+ *}{\psi_2^+})^2
+h.c.\eea
\bea
T_{\cO_3}&=&
\frac{\alpha_{30}}{\sqrt2}
\Big[
-(\vert h_1^0\vert^2+\vert h_1^-\vert^2)
\big[-
g h_2^{0 *}\lambda_z \psi_2^0 +g_2\, (h_2^{0 *} \lambda_w^- \psi_2^+ +
h_2^{+ *} \lambda_w^+ \psi_2^0) + g h_2^{+ *}\lambda_\gamma \psi_2^+
\big]\quad
\nonumber\\
&-&(h_1^{0 *} \psi_1^0 + h_1^{- *}\psi_1^-)
\big[ - g\lambda_z \vert h_2^0\vert^2+g_2\,(h_2^{0 *} \lambda_w^- h_2^+ +
h_2^{+ *} \lambda_w^+ h_2^0)+  g \vert h_2^+\vert^2 \lambda_\gamma
\big]
\nonumber\\
&-&(\vert h_2^0\vert^2+\vert h_2^+\vert^2)
\big[
g h_1^{0 *}\lambda_z \psi_1^0 +g_2\, (h_1^{0 *} \lambda_w^+ \psi_1^-+
h_1^{- *} \lambda_w^- \psi_1^0) - g h_1^{- *}\lambda_\gamma \psi_1^-
\big]
\nonumber\\
&-& (h_2^{0 *} \psi_2^0 + h_2^{+ *}\psi_2^+)
\big[ g\lambda_z \vert h_1^0\vert^2+g_2\,(h_1^{0 *} \lambda_w^+ h_1^- +
h_1^{- *} \lambda_w^- h_1^0)- g \vert h_1^-\vert^2 \lambda_\gamma
\big]
\nonumber\\
&-&\alpha_{31}^* m_0\,\,
(h_1^{0 *} \psi_1^0 + h_1^{- *}\psi_1^-)(h_2^{0 *} \psi_2^0 + h_2^{+ *}\psi_2^+)
+h.c.
\eea
\bea
 T_{\cO_4}&=&
-\alpha_{41}^* \,m_0\,(h_1^{0 *}\, h_2^{0 *}-h_1^{- *} h_2^{+ *})({\psi_1^0}\,\,
 {\psi_2^0}
-{\psi_1^{-}}\,\,{\psi_2^{+}})+h.c.\qquad\qquad\qquad
\qquad\qquad\qquad\,
\eea
\bea
T_{\cO_5}&=&
\frac{\alpha_{50}}{\sqrt 2}\,\,\Big[
(h_1^0\,h_2^0-h_1^- h_2^+ )
\big[
g h_1^{0 *}\lambda_z \psi_1^0 +g_2\, (h_1^{0 *} \lambda_w^+ \psi_1^-+
h_1^{- *} \lambda_w^- \psi_1^0) - g h_1^{- *} \lambda_\gamma \psi_1^-
+h.c.\big]
\nonumber\\
&+&\!\!\!
\big[
g\,\lambda_z\,\vert h_1^0\vert^2 +g_2 \,(h_1^{0 *} \lambda_w^+ h_1^- + h_1^{- *}
\lambda_w^- h_1^0)-g\lambda_\gamma\,\vert h_1^-\vert^2\big]
(\psi_1^0 \,h_2^0 +h_1^0 \psi_2^0 -\psi_1^- \,h_2^+ - h_1^- \,\psi_2^+)\Big]
\nonumber\\
&+&
\alpha_{51}^*\,m_0\,\Big[
(h_1^{0 *}\psi_1^0 +h_1^{- *}\,\psi_1^-)
(\psi_1^0 \,h_2^0 +h_1^0 \psi_2^0 -\psi_1^- \,h_2^+ - h_1^- \,\psi_2^+)
\nonumber\\
&+&
(\vert h_1^0\vert^2+\vert h_1^-\vert^2)(\psi_1^0\psi_2^0-\psi_1^-\psi_2^+)\Big]
+h.c.
\eea
\bea
T_{\cO_6}&=&
\frac{\alpha_{60}}{\sqrt 2}\,\,\Big[
(h_1^0 \,h_2^0-h_1^- \,h_2^+)
\big[-
g h_2^{0 *}\lambda_z \psi_2^0 +g_2\, (h_2^{0 *} \lambda_w^- \psi_2^+ +
h_2^{+ *} \lambda_w^+ \psi_2^0) + g h_2^{+ *} \lambda_\gamma \psi_2^++\! h.c
\big]
\nonumber\\
&+&\!\!\!\!\!
\big[
\!-g\lambda_z\vert h_2^0\vert^2\! +\!g_2 (h_2^{0 *} \lambda_w^- h_2^+\! +\! h_2^{+ *}
\lambda_w^+ h_2^0)+g\lambda_\gamma\,\vert h_2^+\vert^2\big]\,
(\psi_1^0 h_2^0 +h_1^0 \psi_2^0 -\psi_1^- h_2^+ - h_1^- \psi_2^+)\Big]
\nonumber\\
&+&
\alpha_{61}^*\,m_0\,\Big[
(h_2^{0 *}\psi_2^0 +h_2^{+ *}\,\psi_2^+)
(\psi_1^0 \,h_2^0 +h_1^0 \psi_2^0 -\psi_1^- \,h_2^+ - h_1^- \,\psi_2^+)
\nonumber\\
&+&
(\vert h_2^0\vert^2+\vert h_2^+\vert^2)(\psi_1^0\psi_2^0-\psi_1^-\psi_2^+)\Big]
+h.c.
\eea
\bea
\! T_{\cO_7}=-\frac{1}{4}
\alpha_{71}^w\,m_0\,(h_1^0\,h_2^0 -h_1^- h_2^+) (\lambda_w^3\lambda_w^3\!+\!
\lambda_w^+\lambda_w^-)
\!-\!\frac{1}{4}
\alpha_{71}^y\,m_0\,(h_1^0\,h_2^0 -h_1^- h_2^+)\,\lambda_y\lambda_y
\!+\!h.c.
\eea
\bea
T_{\cO_8}\!\! &=&\!\!\!
-2\alpha_{81}^*\,m_0
(h_1^0 \,h_2^0 -h_1^- h_2^+)
(\psi_1^0 \psi_2^0 -\psi_1^-\psi_2^+)+h.c.
\qquad\qquad\qquad\qquad\qquad\qquad\qquad\,\,\,
\eea
\bea
T_{\cK_0}=
-\zeta_{10}
\big[
2 (h_1^0 h_2^0\! - \! h_1^- h_2^+)
(\psi_1^0 \psi_2^0 \!-\!\psi_1^- \psi_2^+)
\!+\!
(\psi_1^0 \,h_2^0\! +\! h_1^0 \psi_2^0\! -\!
\psi_1^- \,h_2^+ \!-\! h_1^- \,\psi_2^+)^2
\big]
\!+\!h.c.\quad  
\eea
\bea
g\lambda_z & \equiv & g_2\lambda_w^3-g_1 \lambda_y\nonumber\\
g\lambda_\gamma & \equiv & g\lambda_w^3+g_1\lambda_y
\eea

\noindent
Finally, $\cL_3$ contains some four-chargino, four-neutralino as well as
two-chargino-two-neutralino interaction terms: 
\bea\label{ww5}
\cL_3\!\!\! &\supset & 
-\,\alpha_{10} (\overline{\psi_1^0}\psi_1^0 +\overline{\psi_1^-}\psi_1^-)^2
-\alpha_{20} (\overline{\psi_2^0}\psi_2^0 +\overline{\psi_2^+}\psi_2^+)^2
\nonumber\\
&-&\alpha_{30} (\overline{\psi_1^0}\psi_1^0 +\overline{\psi_1^-}\psi_1^-)
(\overline{\psi_2^0}\psi_2^0 +\overline{\psi_2^+}\psi_2^+)
+\alpha_{40}
(\overline{\psi_1^0}\overline{\psi_2^0}-\overline{\psi_1^-}\overline{\psi_2^+})
(\psi_1^0\psi_2^0-\psi_1^-\psi_2^+)\quad
\nonumber\\
&&\!\!\!\!\!\!\!\!\!\!\!\!\!
+\,\Big\{1/(2\sqrt 2)\alpha_{70}^w \big[
(\psi_1^0 h_2^0 \!-\! \psi_1^- h_2^+  \!+\! h_1^0\psi_2^0
\!- \! h_1^-\psi_2^+)\sigma^{\mu\nu} \lambda_w^a F_{w,\mu\nu}^a\!\!
\!+\!(\psi_1^0\psi_2^0\! -\! \psi_1^-\psi_2^+)(\lambda_w^3\lambda_w^3\!
+\!\lambda_w^+\lambda_w^-)\big]
\nonumber\\
&+&\!\!
1/(2\sqrt 2)\alpha_{70}^y \big[(\psi_1^0 h_2^0-\psi_1^- h_2^++h_1^0\psi_2^0
-h_1^-\psi_2^+)\sigma^{\mu\nu} \lambda_y F_{\mu\nu}
\!+\!(\psi_1^0\psi_2^0 -\psi_1^-\psi_2^+)
(\lambda_y\lambda_y)\big]\nonumber\\
&+&
h.c.\Big\}
\eea
where $a$ is an SU(2) index.

In applications not all operators are necessarily present. 
Depending on the cases considered, one can have only a subset of operators $\cO_i,\cK_0$ 
or just one of them,
in which case the neutralino/chargino Lagrangian corrections of order $1/M^2$ simplify considerably.
The Lagrangian in the neutralino/chargino sectors is then obtained, for each operator $\cO_j$,
by adding its contributions (identified by $\alpha_{jk}$), from 
eqs.(\ref{ww1}), (\ref{ww2}), (\ref{ww3}), (\ref{ww4}), (\ref{ww5}), plus the MSSM part.

\newpage
\def\theequation{C-\arabic{equation}}
\def\thesubsection{C}
\setcounter{equation}{0}
\subsection{Chargino mass corrections from effective operators.}\label{appendixC}

We use the notations:
\bea
m_w&=&\frac{g_2 v}{2};\quad 
\varphi=\big[m_2^2+\mu^2+2m_w^2\big]^2-4 \big[m_2 \mu+m_w^2 \sin2\beta\big]^2;
\quad
\alpha^r_{ij}=\frac{\alpha_{ij}+\alpha_{ij}^*}{2}
\nonumber\\
\tilde\lambda^{\pm}& \equiv &
\frac{1}{\sqrt{2}}(\lambda_w^1\mp i \lambda_w^2);\qquad 
\psi_1=\begin{pmatrix}
      \psi_1^0    \\
        \psi_-
\end{pmatrix},\quad \psi_2=\begin{pmatrix}
      \psi_+    \\
        \psi_2^0
\end{pmatrix};\,\,\,
\psi_1.\psi_2=\psi^0_1\psi^0_2-\psi_1^-\psi_2^+\,\,
\eea
The results for the chargino masses are:
\bea
m^2_{\tilde\chi_{1,2}^{+}}=m^2_{\tilde\chi_{1,2}^{+}, {\textrm{MSSM}}}
+\delta m^2_{\tilde\chi_{1,2}^{+}}
(\cK_0)+ \sum_{i}
\delta m^2_{\tilde\chi_{1,2}^{+}}(\cO_i)
\eea
where $i=1,..,8$ and $m_{\tilde\chi_{1,2}^{+}, 
{\textrm{ MSSM}}}^2=({1}/{2}) \big[m_2^2+\mu^2+2 m_w^2\mp
\sqrt{\varphi}\big]$ is the MSSM chargino mass with ``$-$" for
 the lighter chargino $\tilde{\chi}_{1}^{+}$ 
and ``$+$" for the heavier $\tilde{\chi}_{2}^{+}$. Also:
\bea
\label{mcmssm}
\delta m^2_{\tilde\chi_{1,2}^{+}}(\cO_1)&=&\pm\frac{1}{2 \sqrt{\varphi }}
\alpha_{10} \, v^2\cos^2\beta \big[-m_2^2 \mu^2+\mu^4\mp\mu^2 \sqrt{\varphi }
+2 m_w^2 \big(m_2^2+2 \mu^2\mp\sqrt{\varphi }\big) \cos^2\beta\nonumber\\
&+&4 m_w^4\cos^4\beta-8 m_2 \mu m_w^2\cos\beta \sin\beta+2 \mu^2 m_w^2 
\sin^2\beta-m_w^4 \sin^22\beta\big]
\nonumber\\
\delta m^2_ {\tilde\chi_{1,2}^{+}}(\cO_2)
&=&\delta m^2_ {\tilde\chi_{1,2}^{+}} (\cO_1)[\alpha_{10}
\rightarrow \alpha_{20},\beta\rightarrow{\pi/2}-\beta]
\nonumber\\
\delta m_ {\tilde\chi_{1,2}^{+}} ^2(\cO_3)
&=&\mp\frac{\alpha_{30} \, v^2 }{8 \sqrt{\varphi}}\big[2 m_2^2 \mu^2
-2 \mu^4-m_2^2 m_w^2-5 \mu^2 m_w^2\pm2 
\mu^2 \sqrt{\varphi }\pm m_w^2 \sqrt{\varphi }
\nonumber\\
&+&m_w^2 
\big(m_2^2+\mu^2\mp\sqrt{\varphi }\big)\cos4\beta+8 m_2 \mu m_w^2 \sin2\beta\big]
\nonumber\\
\delta m^2_ {\tilde\chi_{1,2}^{+}}(\cO_4)&=&
\pm\!\frac{\alpha_{41}^rm_0 \, v^2 \sin2\beta}{4 \sqrt{\varphi }}
\big[\mu \big(m_2^2\!-\!\mu^2\!-\!2 m_w^2\pm\sqrt{\varphi }\big)\!+\!
2 m_2 m_w^2 \sin2\beta\big]
\nonumber\\
&+&
2\delta m^2_{\cO_3}[\alpha_{30}\!\rightarrow \!\alpha_{40}]
\nonumber\\
\delta m^2_ {\tilde\chi_{1,2}^{+}}(\cO_5)&=&
\mp\frac{\alpha_{51}^r\, \, v^2\cos\beta}{2 \sqrt{\varphi }}\big[m_0 \mu 
\big(m_2^2-\mu^2-2 m_w^2\pm\sqrt{\varphi }\big)\cos\beta+4 m_0 m_2 m_w^2\cos^2\beta \sin\beta\big]
\nonumber\\
&\mp&\!\!\!\!
\frac{\alpha_{50}^r\, \, v^2\cos\beta}{2 \sqrt{\varphi }} 
\big[-2 m_2 \mu m_w^2\cos^3\beta+2 m_w^2 \big(2 
m_2^2+4 \mu^2-m_w^2\mp2 \sqrt{\varphi }\big)\cos^2\beta \sin\beta\nonumber\\
&+&4 m_w^4\cos^4\beta \sin\beta+2 m_w^4\cos^2\beta\cos2\beta
\sin\beta-2 \mu \sin\beta \big[\mu \big(m_2^2-
\mu^2\pm\sqrt{\varphi }\big)\nonumber\\
&+&m_w^2 \sin\beta (7 m_2\cos\beta-2 \mu \sin\beta)\big]\big]
\nonumber\\
\delta m^2_ {\tilde\chi_{1,2}^{+}}(\cO_6)
&=&\delta m^2_ {\tilde\chi_{1,2}^{+}}(\cO_5)[\alpha_{50}\rightarrow 
\alpha_{60},\alpha_{51}\rightarrow \alpha_{61},\beta\rightarrow{\pi/2}-\beta]
\nonumber\\
\delta m^2_ {\tilde\chi_{1,2}^{+}}(\cO_7^w)
&=&\!\!\!
\pm\frac{\alpha_{71w}^r \, v^2}{8 \sqrt{\varphi }}
\big[m_0 \mu m_w^2\!-\!m_0 \mu m_w^2\cos4\beta\!+\!m_0 m_2 
\big(-m_2^2\!+\!\mu^2\!-\!2 m_w^2\pm\sqrt{\varphi }\big) \sin2\beta\big]\nonumber
\\
&&\!\!\!\!\!\!\!\!\!\!\!\!\!\!\!\!\!\!\!
\pm\,\frac{\alpha_{70w}^r \, v^2}{8 \sqrt{\varphi }} \big[2 m_2
 \mu \big(m_2^2\!-\!\mu^2\!+\!4 m_w^2\!\mp\!
\sqrt{\varphi }\big)\pm2 \big(m_2^2\!+\!m_w^2\big) 
\sqrt{\varphi } \sin2\beta\!-\!2 m_w^4\cos4\beta \sin2\beta\nonumber
\\
&&\!\!\!\!\!\!\!\!\!\!\!\!\!\!\!\!\!\!
-\, 4 m_2 \mu m_w^2\cos4\beta-2 \big[m_2^2 (m_2-\mu) (m_2+\mu)
+3 \big(m_2^2+\mu^2\big) m_w^2+m_w^4\big] 
\sin2\beta\big]
\eea

\bea\label{mcmssmm}
\delta m^2_ {\tilde\chi_{1,2}^{+}} (\cO_8)&=&
\pm\frac{\alpha_{81}^r}{2\sqrt{\varphi }} m_0 v^2 \sin2\beta 
\big[\mu \big(m_2^2-\mu^2-2 m_w^2\pm\sqrt
{\varphi }\big)+2 m_2 m_w^2 \sin2\beta\big]\nonumber
\nonumber\\
\delta m^2_ {\tilde\chi_{1,2}^{+}}(\cK_0)&=&\pm\frac{v^2 \zeta_{10} 
\sin2\beta}{2\sqrt{\varphi}}
\big[\mu 
\big(m_2^2-\mu^2-2 m_w^2\pm\sqrt{\varphi }\big)+2 m_2 m_w^2 \sin2\beta\big]
\nonumber\\
&\pm&\frac{\, v^4 \zeta_{10}^2 \sin^22\beta}{8 \varphi ^{3/2}} 
\big[2 m_2^4 \mu^2-4 m_2^2 \mu^4+2 \mu^6-8 m_2^2 \mu^2 m_w^2+8 \mu^4 m_w^2\nonumber
\\
&+&4 m_2^2 m_w^4+8 \mu^2 m_w^4+m_2^2 \varphi -3 \mu^2 \varphi
 -2 m_w^2 \varphi \pm\varphi ^{3/2}-4 m_2^2 
m_w^4\cos4\beta\nonumber
\\
&+& 8 m_2\, \mu\, m_w^2 \big(m_2^2-\mu^2-2 m_w^2\big) \sin2\beta\big]
\eea
where the upper (lower) signs correspond to the lighter (heavier) 
chargino $\tilde\chi_{1}^{+}$ ($\tilde\chi_{2}^{+}$), respectively.

\end{document}